THE ROOTS OF PHYSICS STUDENTS' MOTIVATIONS: FEAR AND INTEGRITY

by

BEN VAN DUSEN

B.A., University of California at Berkeley, 2003

M.Ed., University of Oregon, 2004

A thesis submitted to the

Faculty of the Graduate School of the

University of Colorado Boulder in partial fulfillment

of the requirement for the degree of

Doctor of Philosophy

School of Education

2014

This thesis entitled:

The Roots of Physics Students' Motivations: Fear and Integrity

Written by Ben Van Dusen

has been approved for the School of Education
University of Colorado at Boulder

___________________________________
Dr. Valerie K. Otero, chair

___________________________________
Dr. Dan Liston

___________________________________
Dr. Joe Polman

___________________________________
Dr. Kris Gutiérrez

___________________________________
Dr. Noah Finkelstein

June 20, 2014

The final copy of this thesis has been examined by the signatories, and we find that both the content and the form meet acceptable presentation standards of scholarly work in the above mentioned discipline.

IRB Protocol: 0110.55

# ABSTRACT

Van Dusen, Ben (Ph.D., School of Education)
The Roots of Physics Students' Motivations: Fear and Integrity
Thesis directed by Associate Professor Valerie K. Otero

Too often, physics students are beset by feelings of failure and isolation rather than experiencing the creative joys of discovery that physics has to offer. This dissertation research was founded on the desire of a teacher to make physics class exciting and motivating to his students. This work explores how various aspects of learning environments interact with student motivation. This work uses qualitative and quantitative methods to explore how students are motivated to engage in physics and how they feel about themselves while engaging in physics. The collection of four studies in this dissertation culminates in a sociocultural perspective on motivation and identity. This perspective uses two extremes of how students experience physics as a lens for understanding motivation: fear and self-preservation versus integrity and self-expression. Rather than viewing motivation as a property of the student, or viewing students as inherently interested or disinterested in physics, the theoretical perspective on motivation and identity helps examine *features of the learning environments* that determine how students' experience themselves through physics class. This perspective highlights the importance of feeling a sense of belonging in the context of physics and the power that teachers have in shaping students' motivation through the construction of their classroom learning environments. Findings demonstrate how different ways that students experience themselves in physics class impact their performance and interest in physics. This dissertation concludes with a set of design principles that can foster integration and integrity among students in physics learning environments.

This thesis is dedicated to all students who have never had the opportunity to catch the spirit of science.

# ACKNOWLEDGEMENTS

I wish to convey my sincerest gratitude to Susie Nicholson-Dykstra for opening her classroom to me. Susie is a shining example of a teacher who continually striving to push the boundaries of their practice.

I also thank the National Science Foundation Grant DUE-0934921 and DUE-1340083, the Women Investing in the School of Education, and the Center for STEM Learning for their generous support of this work.

**Table of Contents**

# List of Tables

# List of Figures

**CHAPTER 1**

**INTRODUCTION**

The silence of our students is the same silence we have known in other settings: It is the silence of blacks in the presence of whites, of women in the presence of men, of the powerless in the presence of people with power. It is the silence of marginal people, people who have been told that their voice has no value, people who maintain silence in the presence of the enemy because in silence there is safety. Student silence is normally not the product of ignorance or indifference or cynicism. It is a silence born of fear. (Palmer, 1997, p. 2)

As a former high school physics teacher, I have seen students' apprehension and alienation from physics first hand. Each year, I saw some students make lasting connections to physics but the majority of the physics students appeared to only be engaging out of fear of failure and to look smart to their friends. The average student tried to hide from any situation in which he had to volunteer information and his facade of competence might be brought into question. This led me to wonder, what is it about traditional physics classroom environments that lead so many students, especially students from non-dominant backgrounds, to feel disengaged and bad at physics? All of my students, whether interested in physics or not, were very motivated and successful in other areas of their lives. I felt that if my students were highly motivated to engage in other activities, such as sports, the arts, and social networking, I should be able to foster that sort of interest in physics as well.

My dissertation research is designed to explore how to create learning environments in which students are motivated and feel good about themselves while engaging in physics. More over, it is critical to me that students do not feel good in *spite* physics, I want students to feel good *because* of physics. To this end, I have attempted to build and study physics learning contexts that integrate the cultural practices of students' peers and the physics classroom. My attempts to bridge students' peer cultural practices with the cultural practices of physics are based on the hunch that if students could just see themselves in the physics we present to them, they may be better able to identify with it. I believe that if students' can just engage with one or two interesting problems in physics, this can open up opportunities for them to find intrinsic enjoyment from doing it. My investigations and their findings are presented in four separate manuscripts. These manuscripts, their connections, and the body of work that they collectively represent will be explored in this dissertation.

The first manuscript (Van Dusen & Otero, 2012) describes a pilot study that examined how an urban, high school physics class responded to the inclusion of a classroom set of iPads and associated applications, such as screencasting. The study was exploratory in nature and was designed to identify how students integrated themselves, iPads, and physics classrooms. Ethnographic field notes and student survey responses led to the observation of three general trends in students interactions with iPads, screencasting, and physics: (1) students demonstrated increased *social status*, (2) students demonstrated opportunities to engage in *play*, and (3) students demonstrated a sense of *agency* in their work. It was hypothesized that students' experiences of these three constructs (social status, play, and agency) could lead to learning.

The second manuscript (Van Dusen & Otero, in review a) documents how screencasting on iPads could shape physics classroom cultural practices and students' performances. In this investigation we found that the practice of making screencasts was creating opportunities for students to engage in social interactions and exercise authorship within their physics assignments. We also found that students were creating more complete and correct physics solutions when using screencasts instead of their traditional notebooks. These findings led us to conjecture that the increases in students' social interactions and opportunities for authorship were leading to increases in motivations and improved problem solving performances.

The third manuscript examines the proposed links between social interactions, authorship, and motivation. To better understand this potential connection we drew on work done on Self-Determination Theory (Ryan & Deci, 2000; Ryan, 1995). Self-Determination Theory states that if one experiences competence (feeling that one can be successful), autonomy (feeling that one has choice in their actions), and relatedness (feeling connected to one's peers) when engaging in an activity, the activity gradually becomes integrated into one's identity and is internally motivating. Regression analysis of a survey in which students self-reported experiences of competence, autonomy, and relatedness showed a statistically significant link between positive affective experiences and two classroom outcomes: (1) class grades and (2) interests in learning physics. When performing an exploratory factor analysis, however, the three a priori determined constructs (competence, autonomy, and relatedness) were not shown to be empirically distinguishable from each other. The construct of relatedness was largely separable but competence and autonomy were blended into two emergent constructs. The two emergent constructs were primarily composed of the competency and autonomy questions that were either framed in a positive or negative social setting. These findings led us to speculate that feelings of social belonging are central in determining how students experience classroom environments.

The fourth manuscript applies a sociocultural lens to Self-Determination Theory in order to explore the role that students' sense of social belonging plays in shaping their motivations and identities. By integrating the work of from Ames (1992), Ross (2013), and Ross & Otero (2012), and Self Determination Theory (Ryan & Deci, 2000; Ryan, 1995) we created a model that outlines how a person's relationship to a social environment changes as she internalizes an activity. The two extremes of this model are associated with the motivational drivers of *fear and* *self-preservation* versus *integrity and self-expression*. We argue that integrity and self-expression are necessary for students to fully engage in the process scientific induction as intended by science educators and physicists. Feelings of fear and self-preservation, however, often emerge in classrooms that are based on correct answers, cookbook labs, and lecture. This theoretical model is described and applied to the data outlined in Van Dusen & Otero (in review a). We conclude that our model has practical implications for understanding the contextual nature of students' identities and motivations.

Combined, these papers serve as a guide to designing science learning environments that can meaningfully engage learners. The first and second papers provide examples of physics classroom settings that successfully fostered students' feelings of motivation and relatedness in their physics work. The third and fourth papers provide a theoretical lens to help understand how physics learning environments can promote students' engagement in physics from a place of integrity and self-expression rather than fear and self-preservation. Finally, these papers provide a means for thinking about how students' interactions with their learning environments can lead them to engage authentically, increase motivations, and internalize physics activities into their identities.

Perhaps most importantly, this work provides a set of tools that can be used to close the participation gap in physics. Rather than looking at some students as being inherently disinterested in physics, these findings help us examine *features of the learning environments* that put these students into the undesirable position of trying to protect their self-esteem. In physics classrooms, students' alienation is often due to the misalignment of the learning environments' with the students' more familiar cultural practices (Bang & Medin, 2010; Schwartz, Lederman, & Crawford, 2004; Traweek, 2009). By creating social spaces in learning environments in which these two sets of cultural practices can co-exist and blend, new cultural practices can emerge that are meaningful within both communities of practice. Through this process, we can provide students from *all* backgrounds opportunities to engage in physics classes from places of integrity. The fourth paper in this series provides a set of practical recommendations for building classroom environments that are more likely to promote integration and self-expression among students.

**CHAPTER 2**

**LITERATURE REVIEW**

Much of the United States' economic success has been attributed to its free public school and the education it provides (Berger & Fisher, 2013; NRC, 2007, 2010). In addition to sustaining and improving our nation's economy, education has been seen as a central tool for promoting equity among individuals. To quote Horace Mann, the 18$^{th}$ century educational reformer, "education then, beyond all other devices of human origin, is the great equalizer of the conditions of men, the balance-wheel of the social machinery" (H. Mann, 1848, p. 2). A quality education is supposed to provide every child with the skills they need to be successful.

Brown v Board of Education (Court, 1954) outlawed the segregation of children in public schools based on race. This ruling was designed to ensure that all children received an equal education. Despite this ruling, the intervening six decades have continued to produce widely varying educational outcomes for students of different races, genders, and economic statuses producing what is commonly referred to as the "achievement gap" (Ellison & Swanson, 2010; Miron & Urschel, 2008; NRC, 2010). These broader divisions in outcomes are no less severe in the science disciplines (NCES, 2009, 2011; NRC, 2011). Addressing this issue, Secretary of Education Arne Duncan has referred to education as, "the civil rights issue of our generation," stressing that, "every student should be given the tools they need to pursue careers as scientists, mathematicians, and engineers if they so choose" (Glickman, 2010).

The existence and persistence of these inequities have been well documented, the solution to them, however, is still a subject of significant discussion. It has been proposed that our focus should not be on creating *equal educational opportunities*, but rather on creating *educational opportunities of equal worth* (Howe, 1993). Focusing on the schools can lead one to believe that equality is gained through creating learning environments that are consistent across the nation. Simply providing all students equal access to educational opportunities, however, does not ensure that these opportunities are appropriate or useful for students with varying cultural resources, knowledge, and norms. If, as a society, we are to truly tackle the issues of inequality in our students' educations it will require a response that treats students as individuals and is dynamic enough to meet the unique needs each student.

In my dissertation I attempt to create a model by which physics learning environments can be created that meaningfully engage students from all backgrounds in the practices of science. Moreover, my goal is to create learning environments that engage students *through* engaging in science, rather than in *spite* of it.

**Scientific Induction/Inquiry**

The challenge of characterizing how physics learning environments interact with the hearts, minds, and social and cultural histories of students has been the subject of investigation since physics entered the high school curriculum in the late 1800s. During the "New Movement Among Physics Teachers" that began in 1906, university physics professors and physics high school teachers worked together to change the way in which students' experienced physics in the high school classroom (C. R. Mann & Twiss, 1910; C. R. Mann, 1909a, 1909b; Millikan & Gale, 1906; Twiss, 1920). In 1914, physicist Charles Mann concluded that the classroom had become a place in which teachers try to "impose [the scientific] order of thought on our pupils with the idea that we were thereby serving science." He went on to say that, "We have failed because the essence of the spirit we want is not of this sort. The essence of the scientific spirit is an emotional state, an attitude toward life and nature, a great instinctive and intuitive faith (Mann, 1914, p. 518)." Although physicists and physics teachers wanted to understand how to create an environment that fostered what Mann (1909) referred to as *the spirit of science,* what Ames (1992) referred to as mastery learning goals, or what we refer to as *integrity and self-expression,* the work of the early physicists did not quite achieve this—nor did the work of later physicists or science educators.

A number of national documents have attempted to outline a system through which students will come to embody the process of what Mann and Millikan referred to as "induction" and is now commonly referred to as "inquiry." In the 1993 the American Association for the Advancement of Science released their set of science standards titled *Project 2061 Benchmarks for Scientific Literacy* (AAAS, 1993). In the report, AAAS set a series of science literacy benchmarks for students to meet in the $2^{nd}$, $5^{th}$, $8^{th}$, and $12^{th}$ grades. In 1996 The National Research Council released another set of proposed standards titled the *National Science Education Standards* (NRC, 1996). This report defines inquiry in five parts and outlines how students should be inducted into the process of scientific inquiry. The latest national push for science standards is also from the National Research Council and is titled the *Next Generation Science Standards* (NRC, 2013). Much like *induction* before it, this document abandons the term *inquiry* in favor of *scientific practices*. While the preferred nomenclature has changed over time, the pursuit of helping students to engage in the practice of inducing principles using evidence from nature has remained constant over time.

A major goal of my research is determining methods by which physics curriculum can engage students and allow them to feel a sense of belonging. The large scale movement to make science students more engaged in their studies began as early as the 1906 with the creation of the "new movement" in physics education (Clemensen, 1933; Millikan, 1915; National Committee on Science Teaching, 1942; Noll, 1939). The movement set out to primarily teach physics topics through their direct applications to students' daily lives. In the intervening century, our understanding of why students engage in activities has expanded significantly.

I propose that a central reason for the ineffectiveness of the past century's reform efforts to increase participation in physics (National Science Foundation, 2010) is because they are based on a fatally flawed assumption. These reforms were largely designed to bring students into contact with the practices of physics, which is a critical step, but they incorrectly assumed that in doing so students would experience the same types of enjoyment and play that physicists find in the practices. While this approach has been shown to be effective in engaging a small subset of the student population, if we hope to engage a broader swath of our students we will have to create environments that can bridge the activities that students find meaningful and engaging with the practices that physicists find meaningful and engaging.

## Play

One method by which people come to feel a sense of belonging and engagement is through experiencing play. Play is often framed as an activity that children engage in and that once they grow up and become serious will stop engaging in. Far from being a childish activity however, play has the potential to be a highly productive activity. Adults can be in state of play in a variety of challenging activities, such as when engaging in sports, chess, or even science. In fact, play has been described as a key component of engaging in authentic physics (Hawkins, 1974). The framing of play as a non-productive activity is unfortunate because research from a range of fields, including anthropology, psychology and education, have indicated that play is an important mediator for learning and socializing (Rieber, 1996). Further, by creating environments in which students feel that the challenges are high, but that they are able to meet the challenges, leads students to enjoy participating, stretches their abilities, and increases their self-esteem (Csikszentmihalyi & LeFevre, 1989; Csikszentmihalyi, 1990).

While play has been characterized as being the opposite of work, this is a misconception. Rather then positioning play as the opposite of work, leisure is a more appropriate activity to put opposite of work. Play can occur during periods of work or leisure. Play has been generally defined as having four attributes: 1) it is usually voluntary; 2) it is intrinsically motivating, that is, it is pleasurable for its own sake and is not dependent on external rewards; 3) it involves some level of active, often physical, engagement; and 4) it is distinct from other behavior by having a make-believe quality (Rieber, 1996).

Play can be organized around four themes: play as progress, play as power, play as fantasy, and play as self (Rieber, 1996). The view of play as progress situates play an activity that mediates learning of things that are personally relevant or useful. Play as power positions refers to competitions in which adults are able assert dominance. The activity can be physical, such as wrestling, or mental, such as chess. Play as fantasy focuses on play's ability to create an environment in which one can engage in creative or imaginative thinking. This type of play creates a means by which one can be fantastical within a socially sanctioned activity. Play as self-positions play as a way to engage in activities that are of intrinsic worth and optimize one's life experiences.

Social media has emerged as a shared space that is conducive of people engaging in play, particularly around the formation of their digital identities (Kietzmann, Hermkens, McCarthy, &

Silvestre, 2011; National Science Foundation Task Force on Cyberlearning, 2008; Project Tomorrow, 2012). In the social media space it is common for people to create digital identities that are a blend of their existing identities and the identities that they wish to embody (Erikson, 1968). Kietzmann et al. (2011) discuss the creation of digital personas with fictitious attributes, such as names and professions. Central to a user's online experience is her identity development strategy that allows her to blend her real and virtual identities. Project Tomorrow (2012) highlights how digital devices allow for the creation of personalized learning environments, particularly when using a digital device that the user values outside of the classroom.

In the field of physics, play has been indicated as an important method for synthesizing physics concepts (Hasse, 2002). Although it is rarely included in the syllabus, physics faculty often reward students who engage in playing with physics concepts, equipment, and equations. Students' ability to use physics in the act of play has been hypothesized to be significant contributing factor to students' longevity in the discipline. Men have been observed to engage in play within the physics discipline at higher rates, which may be a contributing factor to women's lower representation in post graduate physics positions (Hasse, 2002). Though scientists may find inquiry to be a form of play, it is not clear how to create learning environments in which students engage in inquiry as a form of play.

**Modes of Physics Engagement**

Ames (1992) outlined a set of students' classroom motivational dispositions in terms of students' *goals*. She argued that students either engage in classroom activities through externally motivated goals associated with one's self worth, which she referred as *performance goals* or out of internally motivated goals of self-expression and inventiveness, which she referred to as *mastery goals*. Performance goals and mastery goals are closely related to Dweck's (Dweck & Leggett, 1988; Dweck, 2010) constructs of fixed and growth mindsets. Dweck defines fixed mindsets as being driven by a desire to reach a specific goal (e.g. getting an A on a test) and growth mindsets as being driven by a desire to improve (e.g. doing better on a test than the last time one took it). Ames used the idea of *performance* versus *mastery goals* to argue that it is the characteristics of classrooms, not characteristics of students, which increased the likelihood that students will engage in performance goals (protecting their self-worth) or mastery goals (self-expression and inventiveness). Using prior studies as examples, she demonstrates how these goals determine the quality of involvement of students in class, which greatly impacts students' efforts and outcomes. In our own terms, Ames (1992) showed that outcomes were largely attributable to whether the classroom context engendered an orientation toward *fear and self-preservation* or whether these contexts engendered an orientation toward *integrity and self-expression*—these are phrases I use throughout my dissertation. Ames emphasizes that the constructs of motivation and goals are determined by the nature of the context and how people see themselves in relation to that context and the people in it. I use Ames's work in relation to other work in the area of classic motivation theory (Ryan & Deci, 2000; Ryan, 1995) to develop a sociocultural perspective on identity and its relation to student performance in physics.

  Similarly, in an investigation of high school physics students' experiences Ross and Otero (2012) and Ross (2013) describe two competing narratives that high school students used to describe their past and present high school experiences in science. One narrative can be characterized as that of *fear of failure* and *preservation of self-esteem*. Students used terms such as "afraid, scared, judged, stupid, boring, gullible" and "looked down upon" when they talked about their experiences in science class. In contrast, when describing their experiences in a classroom environment that paid special attention to students development and defending of science ideas, students used terms consistent with *integrity* and *self-expression* such as "comfortable, interested, evidence, it's okay, legit, help each other, share," and "we have the answers."

Other sociocultural researchers have worked toward perspectives on learning using terms such as "agency," "identity," and "culture," in attempts to establish a theory of student learning as a function of sociocultural factors (Barton & Tan, 2010; Basu, Barton, Clairmont, & Locke, 2008; Brown, 2006; Holland, Lachichotte, Skinner, & Cain, 1998; Penuel & Wertsch, 1995). However, in each of these cases crucial terms that were used, such as "identity" and "agency," were not operationalized or no mechanism for their development was provided to the extent that these perspectives on identity could be easily utilized in other contexts.

**Social Spaces**

In trying to create physics learning environments that leverage students' cultural practices to engender a sense of connection and identification with physics, there are three social spaces that merit further examination: (1) Digital social networking community, (2) school science community, and (3) science community.

As previously discussed, social media is commonly used to engage in social behavior, play, and to try on different social identities (Kietzmann et al., 2011; National Research Council, 2010a; Project Tomorrow, 2012). A significant majority of K-16 students are currently engaged in some form of digital social networking (Project Tomorrow, 2012). Members of the millennial generation have been shown to spend 25 hours per week online, with eight of those hours spent on social media (WSL & Strategic Retail, 2014). Many of today's K-12 students have been raised in environments that are rich in social media and have come to feel that their digital persona is an important component of their identity (Buckingham, 2008; Kietzmann et al., 2011).

Every student in the United States is required to take science classes as a part of their K-12 education, but these classes largely fail to engage students (Aschbacher, Li, & Roth, 2009; Kadlec, Friedman, & Ott, 2007; Lee & Buxton, 2010). For most students there is little association between their school science classes and the types of creativity, social connection, and play that are found in social media (Project Tomorrow, 2012; M. J. Ross, 2013). In a report from Kadlec et al. (2007) they show that while students seem to be aware of the important of STEM disciplines in their lives, they find their classes irrelevant and unengaging and are not likely to pursue the subjects.

To a scientist, engaging in science is an opportunity to express their identity, be creative, and engage in play. Scientists embody what Mann referred to as the *spirit* of science, which is "an emotional state, an attitude toward life and nature" (Mann, 1914, p. 518). Within this social space, scientists feel connected to their community and experience a strong sense purpose and enjoyment (Lemke, 1990, 2001).

The relatively low percentage of college students who major in physics indicates that the high school science classes are largely failing to instill the spirit in students (National Science Foundation, 2010). The question then is how do we create learning environments that help students identify with physics the way they identify with their digital social networks. Put another way, how can we draw on the wealth of cultural practices and knowledge that students come into class with (Basu & Barton, 2007; Rosebery, Ogonowski, DiSchino, & Warren, 2010; Suarez & Otero, 2013) to enhance the learning of science and the development of a science identity? To better understand how these three social spaces might influence each other, I draw on the notion of boundary objects.

**Boundary objects**

In my work I focus on a specific tool, the iPad, and it's transformative potential in physics learning environments. The iPad's dynamic nature and appeal to diverse populations may make it a uniquely positioned learning tool. Specifically, I focus on the iPad as a tool within three distinct sets of cultural practices: (1) The iPad as a classroom tool—iPads are used to combine access to word processing and the class's website to create presentations, submit assignments, and provide peers feedback. (2) The iPad as a consumer and social media device—students use the iPad's camera, web browser, and Internet connectivity to engage in social activities, such as messaging friends, sharing pictures, and using social media. (3) The iPad as a physics computational device and research tool—physicists use iPads to facilitate the social creation of knowledge through the collection and sharing of data as well as engaging in peer review through its spreadsheet and PDF annotation applications. Figure 1 shows examples of iPad-facilitated activities in these three sets of cultural practices.

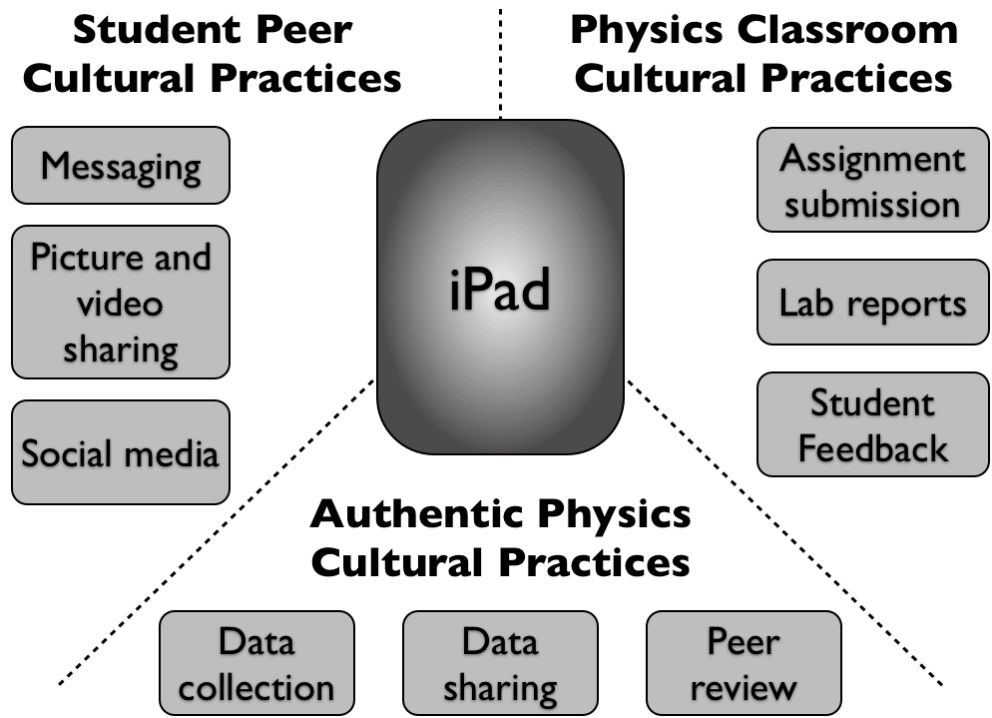

**Figure 1.** A model of the iPad as facilitating the blending of three different communities of cultural practices: students' peers, the physics classroom, and the larger physics community.

Because the iPad spans these sets of cultural practices it may act as a *boundary object* that facilitates the creation of new blended cultural practices (Star & Griesemer, 1989). Star and Griesemer define boundary objects as objects that adaptable across varying communities of practices yet robust enough to maintain an identity across them. Boundary objects achieve this balance by simultaneously being grounded in concrete, specific, and conventional uses while simultaneously being dynamic with abstract, general, and customizable uses. Buxton et al. (2005) discuss boundary objects' roles in acting as boundary spanners. Buxton specifically discusses boundary spanners ability to bridge the practices and discourses of varying groups, such as the discourse practices of students and their science classrooms. The concept of boundary objects has emerged from socio-cultural research on the role that shared tools, spaces, and concepts play in mediating social interactions (Akkerman & Bakker, 2011; Engeström,

Engeström, & Karkkainen, 1995; Leigh Star, 2010). In her seminal publication on boundary objects, Star (Star & Griesemer, 1989) examined how a natural history museum acted as a boundary object that bridged the cultural practices of researchers, conservationists, and trappers. While each group saw the museum as a means of preserving nature, they all had their own unique uses for it. The researchers used the museum as a means of data gathering. The conservationists saw the museum as a means of educating the public to the about the wonders of nature. The trappers used the museum as a means to share their collections. The museum brought together the three distinct groups and created a shared social space for them to interact in. Through the creation of shared social space, the museum acted as a boundary object that mediated these groups' cultural practices, facilitating the blending of their practices and the emergence of new, shared cultural practices.

The notion of boundary objects and the shared space they create is closely related to the notion of *third spaces* (Gutierrez, Baquedano-Lopez, & Tejeda, 1999; Gutiérrez, 2008; Gutierrez, 1997). Gutierrez defines *third space* as arising from environments that allows the discourse practices of various activity systems to come into coordination. When a classroom allows the classroom "script" and the students' "counterscripts" to coexist and blend, a new third space emerges that values and embodies aspects of both sets of discourse practices. Unlike third space, boundary objects focus on the role of tools in coordinating a range of cultural practices across time and space.

In my investigation, iPads have a slightly different role to play than the museum did in Star's work (Star & Griesemer, 1989). Rather than bringing multiple populations that embody varying cultural practices and spaces together, the iPads act as boundary objects to bring together multiple sets of cultural practices within a single population and space that changes across time. In my setting, the students take on the role of embodying several sets of differing cultural practices from different social spaces. Because they are in a physics classroom that values particular behaviors, the students embody the role of a physics student who uses the iPad to engage in behaviors such as creating and turning in assignments. Because the iPads offer access to peer interactions, the students also embody the cultural practices of their peers who engage in behaviors such as making jokes and digital social networking. In this way, the iPads are able to bring together multiple sets of social spaces and their associate cultural practices within a single set of participants, allowing their existing identities to blend and new identities to emerge.

Simply bringing these three spaces together and allowing them to interact, however, does not assure that students will begin to feel connected to physics and engage in its practices more fully. The creation of this space must also integrate aspects of what we know about human motivation and identity development. For this theoretical work, I draw on Self-Determination Theory.

**Self-Determination Theory**

Self-Determination Theory (SDT) considers identity and associated motivations as being attributes of an individual that slowly shift over time as the individual engages in activities (Deci & Eghrari, 1994; Ryan & Deci, 2000; Ryan, 1995). According to this perspective, given the appropriate conditions, an activity can be gradually integrated and internalized into one's identity. Through this internalization of an activity, one's choice to engage in an activity shifts from being externally motivated to being internally motivated. Through this shift, it can be said that an individual is motivated to engage in that activity.

According to Deci and Ryan (Ryan & Deci, 2000), internalization can be thought of as the assimilation of behaviors that were once external to the self. Through the process of internalization, individuals come to feel that what makes them engage in an activity (their locus of causality) moves from the external to the internal. Based on the study of organismic integration, SDT states that internalization occurs as an activity fulfills an individual's *basic psychological needs* (Deci & Ryan, 1991; Ryan & Deci, 2000). Organismic integration states that, like all natural processes, development through integration must be nurtured by the fulfillment of basic needs. These basic needs are defined as the, "nutriments or conditions that are essential to an entity's growth and integrity" (Ryan, 1995, p. 410). In the case of the human psyche, the basic nutriments for growth (or basic psychological needs) are feeling a sense of *competence, autonomy,* and *relatedness* (Reis, Sheldon, Gable, Roscoe, & Ryan, 2000; Ryan, 1995). When people engage in activities which provides them the experiences of competence, autonomy, and relatedness (instead of excessive controls, overwhelming challenges, and relational insecurity) they will be more likely to choose to engage in the same activities in the future.

SDT identifies six stages of internalization (*amotivation, external regulation, introjection, identification, integration,* and *intrinsic motivation*), shown in Figure 2. Each column in figure 2 represents one of the six stages of internalization, while the rows represent the corresponding *regulatory processes*, *mental and emotional processes*, *perceived locus of causality*, and *relative autonomy* for each stage. A person's development of intrinsic motivation is achieved through the internalization of an activity. As internalization of an activity occurs, one's regulatory process move from left to right through the six stages shown in Figure 2, making the activity more assimilated to one's self and increasingly intrinsically motivating.

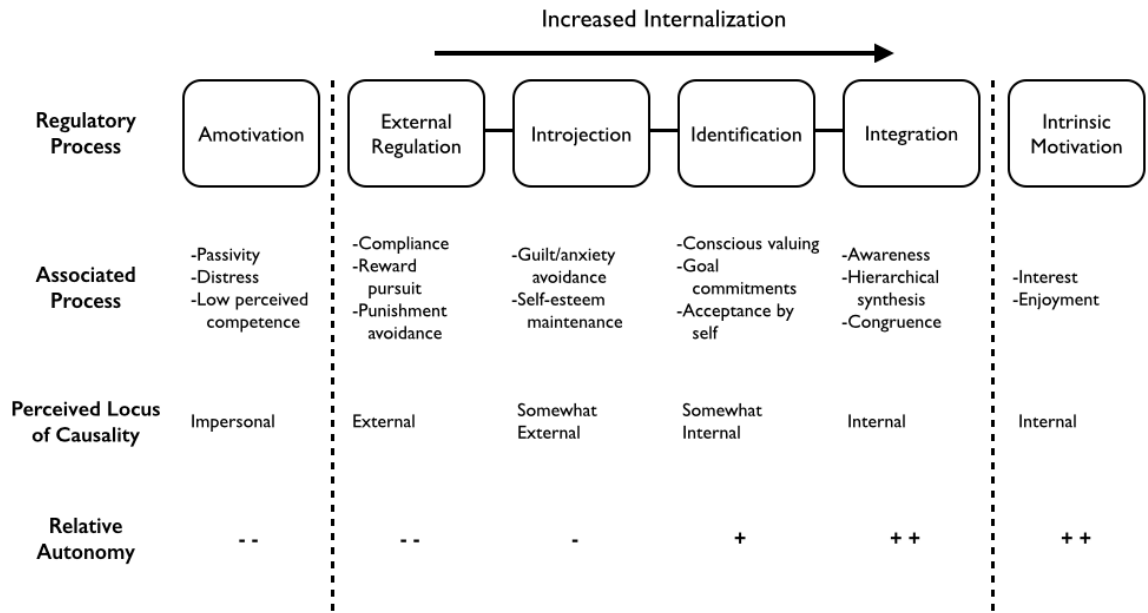

**Figure 2.** A guide to motivation in Self-Determination Theory (Ryan, 1995)

The extremes of the regulatory process scale (row 1) are *Amotivation* and *Intrinsic Motivation*. Amotivation actions are those that are seen as out of one's control. Intrinsic motivation, on the other hand, is the driver of actions that are done for their pure enjoyment and are perceived by the individual as being within their controls (Deci & Eghrari, 1994). Between the extremes of the regulatory scale, SDT identifies four regulatory process and their associated psychological mechanisms that drive peoples' actions (Deci & Eghrari, 1994; Ryan & Deci, 2000; Ryan, 1995). *External Regulation* is the first stage of internalization and is represented in the second column of Figure 2. Activities that undergo External Regulation are done out of compliance and hold little to no personal meaning to the individual. These types of activities require external coercion or reward. *Introjection* is the second stage of internalization and is represented in the third column. While Introjection is not entirely based on external motivation, it is driven by a sense of guilt or anxiety avoidance about potential judgment by others. *Identification* is the third stage of internalization and is represented in the fourth column. When in the Identification stage, motivation to engage in an activity is based on a feeling of acceptance and personal valuing of the activity. While this regulation process is perceived as being internally motivated and autonomously driven, it lacks integration with other parts of the self. The Identification stage can be thought of as the "trying on" stage, in which a person values an activity but has not yet fully embraced it. *Integration* represents the fourth stage of internalization and is represented in the fifth column. In the Integration stage of regulatory processes, various identifications are organized and brought into congruence with one's identity as a whole. Activities that have been assimilated to the Integration stage are those that we see as being central to our identities and we are intrinsically motivated to engage in.

Figure 2 also shows the proposed *associated processes*, *perceived locus of causality*, and *relative autonomy*. The *associated processes* (row 2) are example psychological mechanisms that drive people to engage in activities. The *perceived locus of causalities* (row 3) determines whether a subject's engagement in an activity is being driven externally or internally. The *relative autonomy* (row 4) is the extent to which people feel that they have control over whether or not they engage in activities.

I use Self-Determination Theory together with boundary objects to investigate how the iPad can mediate student motivation to engage in physics as play in school-based physics classes.

**Distributed Cognition**

To better understand the mediational role of technologies, such as the iPad, in the physics classroom I draw on the notion of Distributed Cognition (Hutchins, 1995, 1996). Distributed Cognition is a theoretical perspective that considers the active role of tools in the learning process. This perspective views the "cognitive unit" as a *socio-cultural cognitive system* encompassing the individual, surrounding people, available tools, configuration of the environment, and interactions among these things. Otero describes the socio-cultural cognitive system as:

> …students interacting with tools (such as laboratory apparatus and computer simulators), and students interacting with others and with tools are considered a cognitive system that generates learning. According to this perspective, each element of the system contributes to the cognitive product by sharing part of the cognitive load associated with a task (Otero, 2001, p. 1).

Distributed cognition considers the entire system sharing a part of the cognitive load (Hutchins, 1996). The individual may become more proficient at working within the socio-cognitive system to accomplish a task. Meanwhile, the roles of other components of the system (tools and interactions) shift systematically with changes in the role/behaviors of any one component, including changes in the way a single individual identifies with a tool in the context, or with the broader socio-cultural cognitive system (Otero, 2004). As such, in order to attend to student learning, it is important to monitor the changing use of tools and the environment as students perform tasks. In this study, I would view the context of the classroom as a socio-cultural cognitive system that generates learning. Within this system, the learning I am specifically interested in is students changing interactions with tools and each other, the changes in the way students interact with and identify with science, and the role science plays in students' visions of themselves today and in their futures.

As the name *socio-cultural cognitive system* would suggest, distributed cognition frames environmental systems socio-culturally while simultaneously examining the cognitive roles of each part of the system. Both individuals and tools play an important and active role within these systems, but to try to separate out any individual components from the system would strip away the most important facets of the system, the interactions.

Cognitive Scientist, Andy Clark (1997) uses scientists' discovery of how a sponge breathes as an analogy to understand the importance of viewing human cognition in context.

> The simple sponge, which feeds by filtering water, exploits the structure of its natural physical environment to reduce the amount of actual pumping it must perform: It orients itself so as to make use of ambient currents to aid its feeding. The trick is an obvious one, yet not until quite recently did biologists recognize it. The reason for this is revealing: Biologists have tended to focus solely on the individual organism as the locus of adaptive structure. They have treated the organism as if it could be understood independent of its physical world. In this respect, biologists have resembled those cognitive scientists who have sought only inner-cause explanations of cognitive phenomena (Clark, 1997, p. 46).

In this quotation, Clark highlights the importance of examining cognition in context. Scientists could not fully understand a sponge's ability to breath in isolation from its environment, nor could they understand it from the environment alone. It is only by examining the interactions between the sponge and its environment that the scientists were able to discover how a sponge is able to effectively breath. Similarly, it is only by examining the reflexive interactions between a person and her social contexts that the processes of learning and cognition can be understood.

**Technology**

There have been many other efforts to introduce technology in to the physics classroom. Two of the more transformative uses of technologies commonly found in physics classrooms are the use of computers to collect data and to run simulations. Over 20 years ago, Thornton and Sokoloff (1990, 1998) laid the groundwork for understanding how computers can be used to mediate physics students' lab experiences. By using digital probes students are able to collect data and quickly generate graphical representations of phenomena. In addition to showing significant learning gains when using Microcomputer-based Labs (MBL), Thornton and Sokoloff postulate that the primary reason for these gains are due to five MBL augmented classroom characteristics: (1) Students focus on the physical world, (2) immediate feedback is available, (3) collaboration is encouraged, (4) powerful tools reduce unnecessary drudgery, and (5) students understand the specific and familiar before moving to the more general and abstract (Thornton & Sokoloff, 1990).

The use of computers in physics learning was later expanded beyond the collection and analysis of lab data to include the manipulation of simulations. By creating simulations of real-world situations, projects such as Constructing Physics Understanding (Goldberg, 1997) and Physics Education Technology (Paul, Podolefsky, & Perkins, 2012; Podolefsky et al., 2009) provide students several new methods for understanding phenomena. These simulations afford students the opportunity to manipulate variables in situations that may be difficult or impossible to replicate in the lab. Some of the simulations also take processes that are normally invisible (e.g. the motion of gas molecules) and make them visible, thereby helping students visualize the mechanisms behind phenomena.

Further research in the technologies uses and effects in science learning environments have been a central component of several high profile national policy advising documents over the last decade. In their call to action, the National Research Council (2007) stressed the importance of both science and technology in supporting the continued functioning and growth of the U.S. economy. In a pair of response to the NRC's initial recommendations on STEM educational reforms, the Presidential Council of Advisors on Science and Technology (2010, 2012) detailed a set of recommendations on how K-12 and university settings could use technology to bring their practices into the 21st century and improve student outcomes. Specifically, the council identified six fields for improvement: (1) providing deeply digital course materials, (2) open-source modular course materials, (3) improved assessment systems, (4) personalized online tutoring systems, (5) automated systems and software to aid teachers, and (6) improved exchange of digital educational materials.

As an educational researcher and a former high school physics teacher that had integrated a number of newer technologies into my classes these recommendations resonated with me. It is my hope that this dissertation helps to answer highly applicable questions of how to design a science learning environment that engages students, as well as to add to our theoretical understanding of how people come to identify with and feel motivated to engage in physics.

Although it may be considered a finding of my investigations, my work has led to the creation of a model for student experience, identity, and motivation (Figure 3). Figure 3, shows how social relatedness (or levels of social integration) lead to different outcomes relevant to the basic psychological needs outlined by Deci & Ryan (1991). The constructs *competence* and *autonomy* will be expressed differently depending upon an individual's level of social integration. A context in which an individual feels highly integrated with their social environment (right side of figure) will lead to autonomy being experienced as self-expression/innovation, and competence being experienced as efforts for improvement of the self or environment. A context in which an individual feels like an outsider (left side of figure) will lead to autonomy being experienced as aloneness and alienation, and competence being experienced as efforts to preserve self-esteem. For example, students who are friends with their classmates and feel connections to the practices of their physics classrooms are willing to take risks and try out new ideas (autonomy) with the hope of better understanding physics (competence). Students who do not feel socially connected to other students, the teacher, or the goals and practices of physics are often afraid to express their ideas (competence), unlikely to take risks in problem solving, and often criticize the teacher and the content in efforts of not looking like the ones who are responsible for not fitting in (autonomy). Figure 3 illustrates that both competence and autonomy emerge in both types of settings, however ones integration within the social environment frames how these constructs are experienced and expressed. In order to have nurturing positive feelings of competence and autonomy emerge from a system students must feel some connections to their social environment (peers and the cultural practices within the system). The details behind and implications of this model are further explored in Chapter 6 of this dissertation.

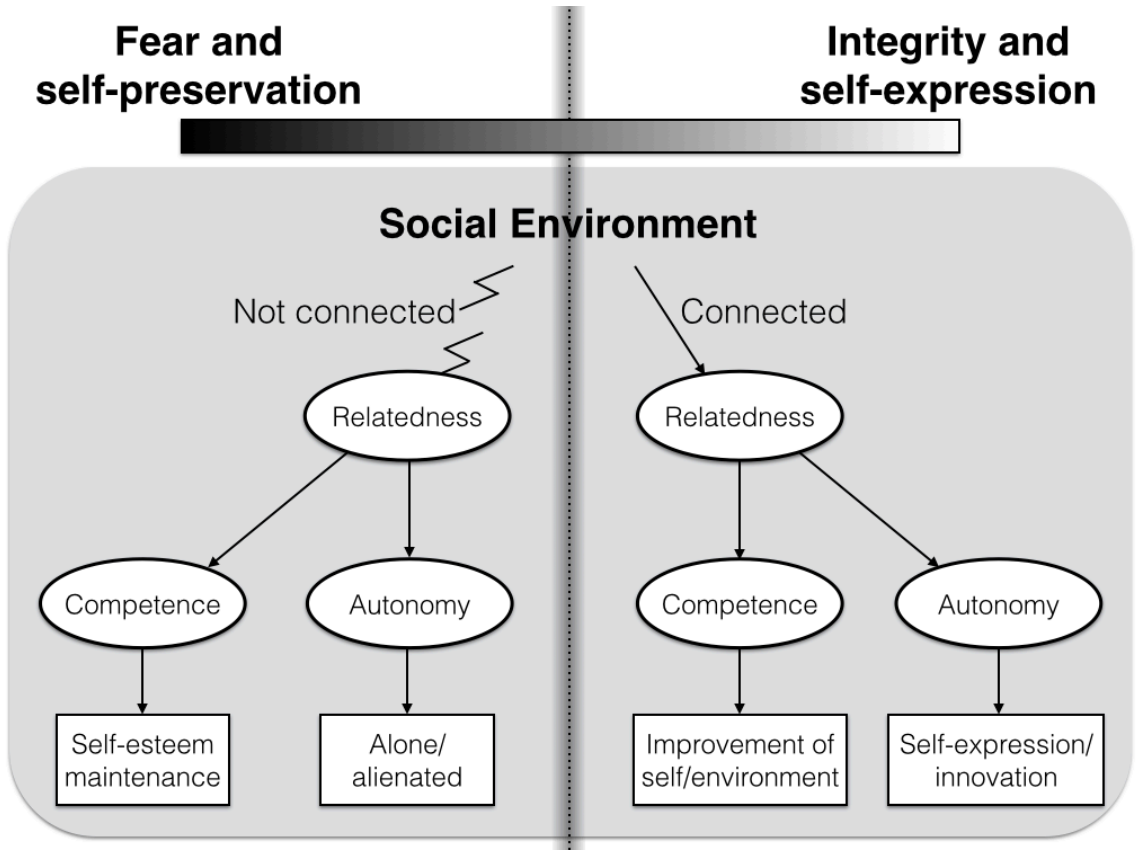

**Figure 3.** A toy model of the how "basic psychological needs" are expressed in context.

# CHAPTER 3

# INFLUENCING STUDENTS' RELATIONSHIPS WITH PHYSICS THROUGH CULTURALLY RELEVANT TOOLS

Ben Van Dusen and Valerie Otero

*School of Education, University of Colorado, Boulder, 80309, USA*

**Abstract**. This study investigates how an urban, high school physics class responded to the inclusion of a classroom set of iPads and associated applications, such as screencasting. The participatory roles of students and the expressions of their relationships to physics were examined. Findings suggest that iPad technology altered classroom norms and student relationships to include increased student agency and use of evidence. Findings also suggest that the iPad provided a connection between physics, social status, and play. Videos, observations, interviews, and survey responses were analyzed to provide insight into the nature of these changes.

## Introduction

On January 12th, 2012, two high school physics students took the initiative to leave the classroom with an iPad to shoot a video in their car. The video showed them starting to drive, placing a coffee cup on their dashboard, and then quickly accelerating to a stop. The video clearly shows that the coffee cup kept moving forward even after the car had come to rest. The students inserted this video into their digital lab report and showed it to their classmates.

The above example struck us as a particularly intriguing interaction. How can the actions of the students be explained? What compelled the students to stretch the bounds of the classroom in ways that were never formally sanctioned by the teacher in order collect and log data from a relevant experiment that they designed?

The iPad Enhanced Active Learning (iPEAL) project was designed specifically to explore the effects iPads in high school physics classrooms. Like the personal computers (PCs) before them, iPads have been hyped as a "magical" product that will revolutionize education. PCs introduced new ways of collecting and analyzing classroom data through probeware [1], video-based motion analysis [2], and introducing model-like evidence through simulations [3]. Such PC-based activities have changed the role of data collection and analysis in classroom physics. At the same time, specific populations of students remain largely underrepresented in university and college physics. We hypothesize that in order to capture the natural curiosity of *all* students, and to introduce them to the richness that physics inquiry could bring to their lives, the classroom environment must be shifted significantly to facilitate the gradual process of personal identification with physics.

By examining changes in student activities and peer-interactions, we investigate the questions: (1) In what ways do iPads change student interactions with physics, if at all? (2) In what ways do iPads mediate student relationships to physics, if at all?

**Research Context**

This research was conducted in 5 high school physics classes (4 regular and 1 Advanced Placement) in an urban area. The school is primarily composed of students who have been traditionally underserved and are underrepresented in science. While the student enrollment of the courses varied throughout the year, there were approximately 140 students at any given time. At the start of the school year 73% of the students were juniors and 27% were seniors.

The five classes shared a single set of 38 iPads. This allowed each student to have an iPad that was unique to them during class, but was shared with four other students throughout the day.

A third-year teacher with a background in biology, including a Ph.D. in biochemistry, taught all of the classes. Like most high school teachers of physics, she did not have a physics or physics education degree [4]. She is a Streamline to Mastery [5] teacher, engaged in NSF-funded teacher-driven professional development.

Students engaged in iPad-supported activities that were intended to supplement traditional physics assignments. For example, students created screencasts of their textbook problem solutions. Through screencast technology, they created a video of the iPad screen as they recorded think-aloud audio while solving physics problems using the stylus. This allowed students to record and play back their dynamic problem solutions. These screencasts were later made available to other students.

Theoretical framework

We take a critical perspective, where we assume that high school student physics experiences are too often wrought with fear and failure rather than being enjoyable, empowering, and personally meaningful. Our research is based on the assumption that in order for learning to occur, the learner must be engaged in an activity that is personally meaningful. Further, for an activity to be personally meaningful, it must produce, or be produced by, positive experiences. We focus on the iPad as a tool that could potentially *mediate* a change from negative to positive experiences in physics. From these assumptions we have created a tentative model (Fig. 1) in which positive experiences and personally meaningful activities are both reflexive and necessary for creating an environment in which learning may occur.

While similar models have been described in the literature [6], we were drawn to this type of model through our first year of observations in the iPad learning environment. The construct of "personal meaning" was salient in the social context yielding potential for helping us understand the experiences that students were having with physics via the iPad. During the second-year of observation and interviews we intend to further articulate this model in terms of the specific ways in which students engage with the iPad, physics, or both.

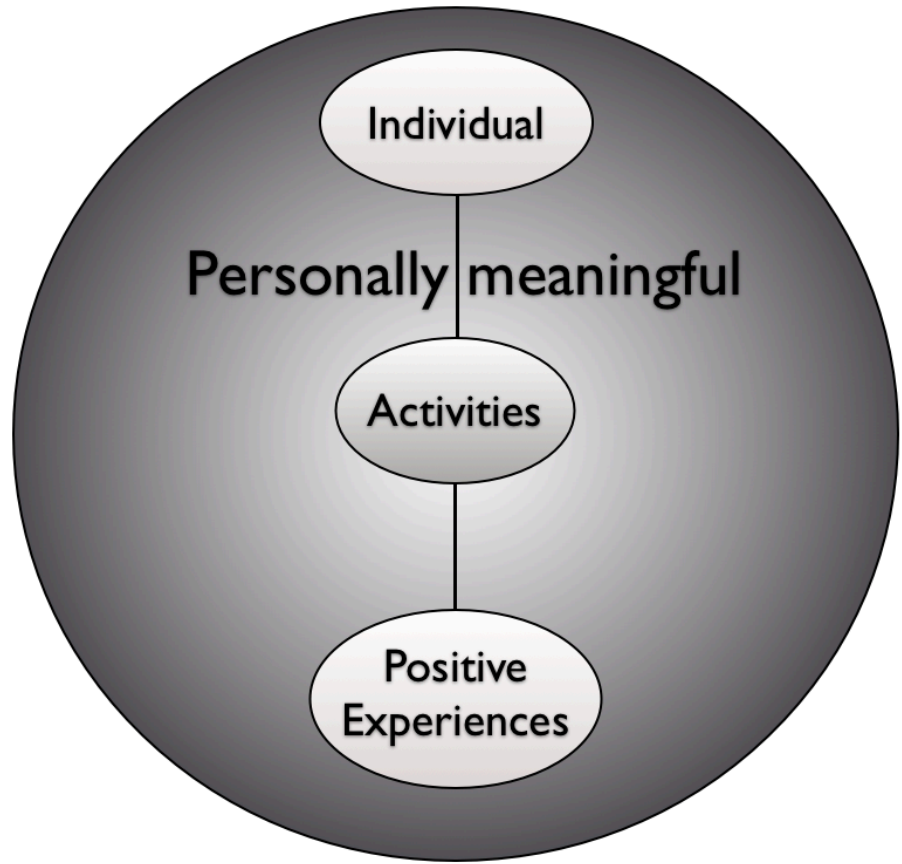

**Figure 1**. A simple model of personal meaning.

## Methods

Field-notes, video recordings, artifacts, student surveys, and student interviews were collected. The interviews were typically administered the same week as the surveys and included a similar set of question in order to collect expanded answers from a smaller set of students. Findings from the interviews and surveys were triangulated with the field-notes from the weekly observations. Classroom videos were used as references to supplement field-notes. We used a generative coding methodology to discover any themes that emerged from the data. Eleven themes were found, four of which are relevant to the construct of personal meaning. The use of iPads in this context: 1) facilitated student use of evidence, 2) facilitated student play, 3) increased student agency, and 4) appeared to impact student social status. These findings along with exemplars are described in the section below.

**Findings**

**Finding 1**: The iPads facilitated students' use of evidence. This was accomplished by making it easy to complete tasks that were previously either difficult or impossible. By facilitating data collection, analysis, and collaboration, the iPads allowed students to draw their own conclusions based on evidence, rather than relying on the book or the teacher to provide solutions.

The shift of authority from the teacher and textbook to evidence was most apparent during labs. For example, the iPad altered the task of evaluating the results of a sound lab. The original lab required students to cut PVC piping to predetermined lengths in order to produce different harmonic frequencies. Students then listened to the pitch of the pipe and were to determine if they had created the intended pitch. Only the teacher and one of the students in the observed classes had the musical training to identify the pitch classes by ear. The rest of the students were unable to aurally identify if their pipes were producing the correct frequency and relied on the teacher's assessment. With the introduction of the iPad, students used several applications to identify the pitches of the PVC experiment. One application displayed the frequency the pipes were emitting numerically, another produced varying pitches for students to reference and match, and the last application identified the pitch class and reflected the tuning variations on a virtual dial. With these three applications, the iPad *transformed the task* of asking the teacher to evaluate the sound into a different task in which students could use a mathematical, audio-matching, or visual-spatial model to assess the sound. There were many situations such as this, where tasks were transformed from teacher-as-authority to evidence-as-authority. These transformations were tractable for students and allowed them to reason with evidence instead of deferring to an external authority.

Finding 2: The iPad facilitated student play. Even before the iPads were actually implemented, students demonstrated excitement to use them in class. During the first two months of the school year, when students did not yet have the iPads, they regularly inquired about when the iPads would be arriving. Once the iPads were implemented, students used the time before class and during transitions to explore the iPads. Typical student-initiated activities included taking pictures of friends, setting the background image, exploring simulations, and playing games. Unlike previous years, student began to regularly go into the physics classroom outside of class time to work on physics projects. When asked how much they enjoyed doing iPad work versus traditional work, students articulated a strong preference for iPad work (Fig. 2). The students who found the work to be less enjoyable on the iPad, expressed a preference for laptops.

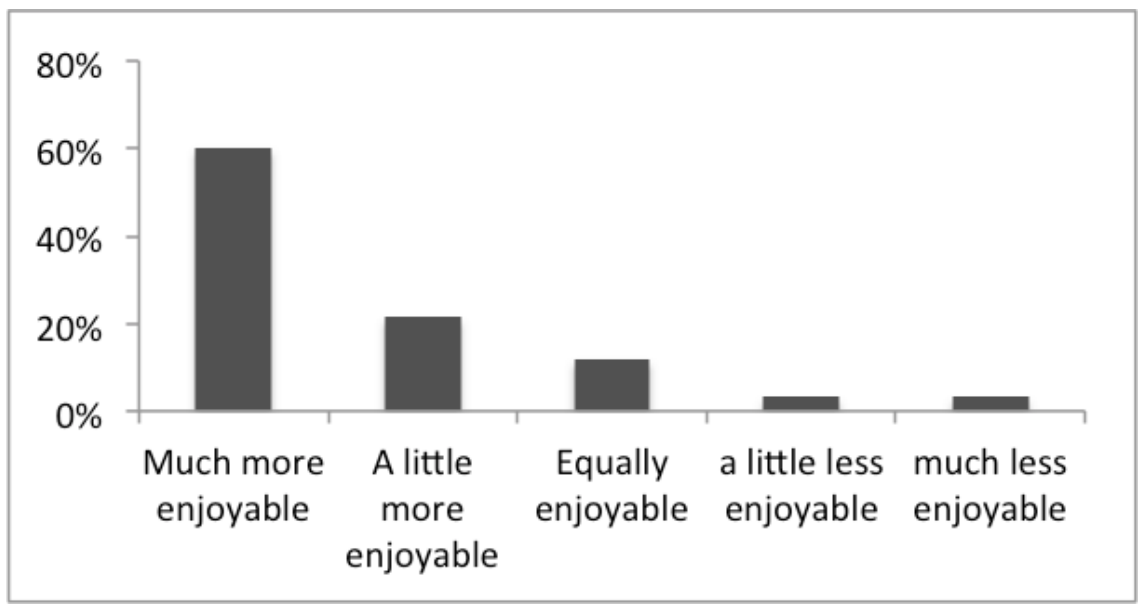

**Figure 2**. Comparison of iPad to traditional work.

**Finding 3**: The iPad increased student agency. Through specific activities made possible by the iPad, such as students' creation of screencasts, some responsibilities were shifted from the teacher to the student. As described earlier, students used screencast technology to record a verbal explanation with video documentation of the steps in their problem solutions and then shared them with other students. When the AP students were asked whether the teacher, themselves, or the class determined what steps should be shown in their work, students were more likely to reference the teacher for traditional lab work and themselves for screencasts (Fig. 3), none said the class.

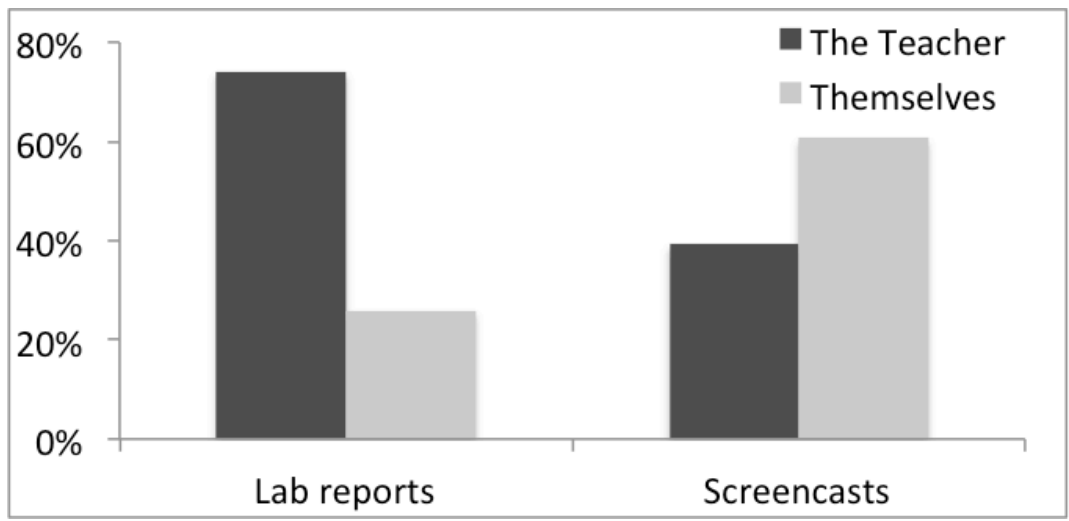

**Figure 3**. Who determines the steps to be shown.

Students were aware of and able to vocalize the differences in personal agency in determining steps in physics labs. When asked if screencasts helped them learn, one AP student answered, "You can learn from visuals and reading, but teaching, writing, and teaching yourself again is a very effective way. I think it's the most effective way because you think that you're going to give someone a lesson and you test your own knowledge. You don't have anyone telling you, you test yourself" (Manuel, 5/4/12). Another AP student answered, "Trying to explain to someone else is like being a teacher" (Julia, 5/4/12). We infer from this set of data that the iPad has the potential to increase student agency more broadly through activities such as self-guided screencasts.

**Finding 4**: The iPad appeared to impact student social status. We have some evidence that has led us to further investigate this claim. It was common for students in non-physics classes to express a desire to be involved with the iPad classroom environment. When physics students were asked what they enjoyed about using the iPad, one student said, "I do really like the photobooth, it's really cool. I can send pictures to my email and put them up [on Facebook]. Then that's another way for people to know that we have iPads in our class. They're like, 'how did you do that?' I'm like, 'oh we have iPads in our physics class,' and they're like, 'what?!'" (Sally, 1/13/12).

Using the iPads was important to the students. For example, when they learned that there was not enough money to purchase AppleCare, cases, and styluses for the iPads, the students offered to raise the money or pay for the equipment themselves.

The effect of the iPads reached beyond the class to raise the status of the school itself. The physics classes had more iPads than the rest of the district combined. Once the iPads were implemented, school visitors were brought to observe the physics classes on a regular basis. One prospective student was overheard by the teacher saying that he thought the school was "ghetto" and was not going to go there until he heard that it had "the physics classes with iPads."

Data collection to support the claim of social status has been challenging, but we intend to continue to investigate this and our other claims throughout the upcoming year.

For research purposes, we have analytically separated play, status, and agency but some findings were difficult to categorize. For example, when asked what percentage of their digital assignments (e.g. screencasts and digital lab reports) and analogue assignments (e.g. book problems and handwritten labs) they shared with fellow physics students, AP and regular students jointly expressed that they shared more of their digital assignments (Fig. 4). We were uncertain how this finding should be classified; it could represent any number of things ranging from agency and status to teacher-driven requirements. We report it here because it reveals a possible shift in the classroom collaborative environment.

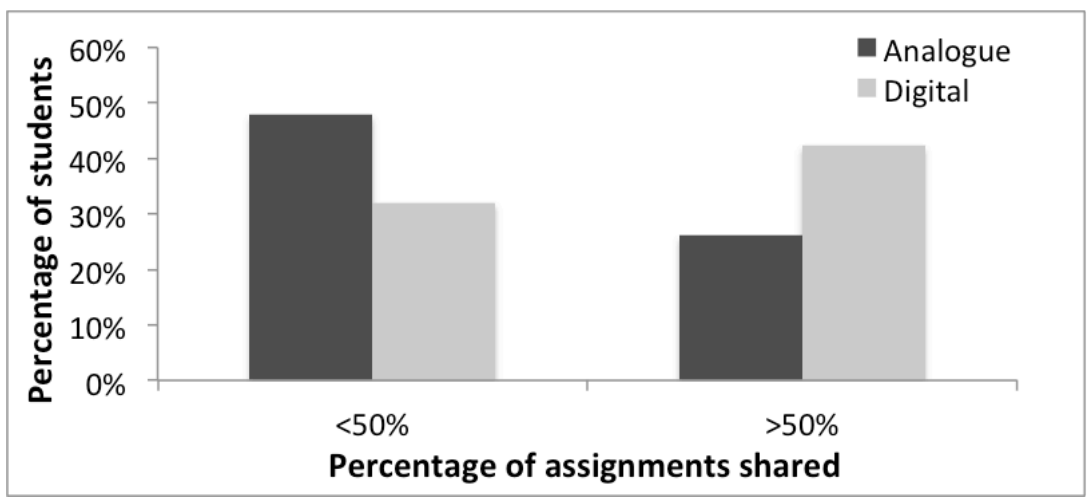

**Figure 4**. Percentage of assignments shared with fellow physics students.

**Theoretical implications**

Based on our initial findings, we have further operationalized the construct of positive experiences to include social status, play, and agency (Fig. 5). Using this lens we can investigate the potential role of the iPad in physics learning. We postulate that the iPad created a "bridge" connecting students, via social status, play, and agency to physics.

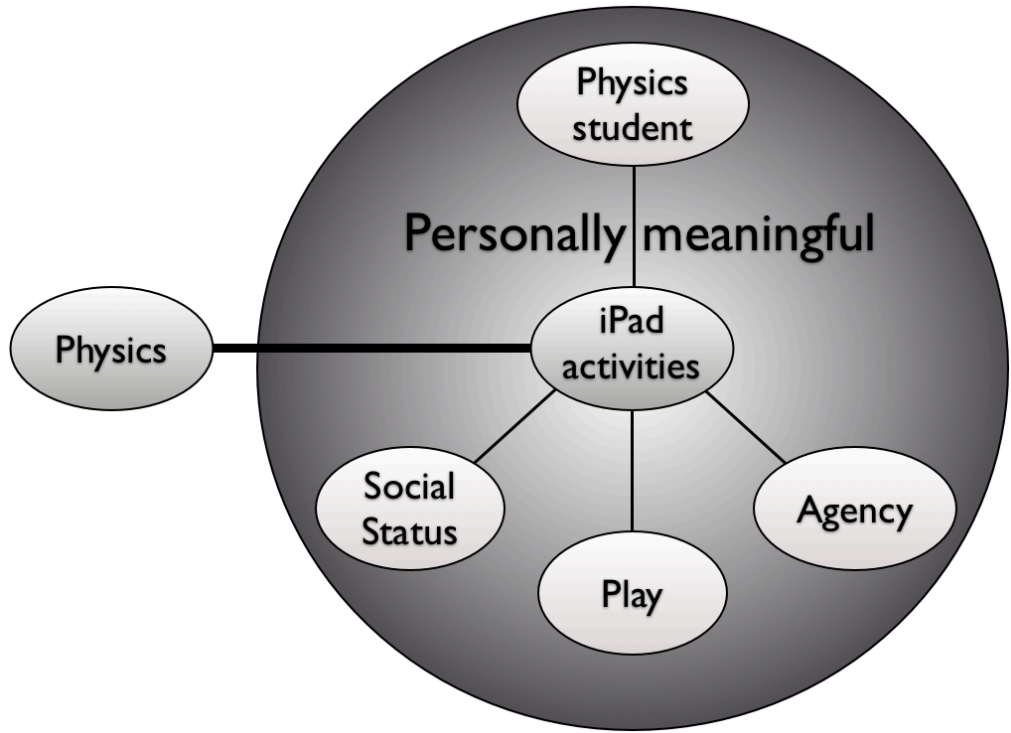

**Figure 5**. Personal meaning connects students to physics.

We propose a model in which the iPad may have the potential to mediate a positive relationship between students and physics. Based on our current data analysis, we suggest that agency, status, and play afforded by the iPad can serve as mechanisms for reshaping students' relationships with physics.

Although this argument hedges on classical conditioning/behaviorist models of learning appearing in the literature as early as 1910 [7], it may have some value in helping to connect what a student internally experiences in the physics classroom to the external experiences that are provided in the class environment. These experiences may produce fear and anxiety or they may feel empowering and exciting to the student. We provided some evidence that a mediating artifact such as the iPad could serve to facilitate a positive relationship with physics through personally meaningful activities that engender a sense of play, status, and agency. Our model goes further in positing that the "bridge" between physics and the iPad could eventually be switched off and student's sense of personal engagement in physics would remain (Fig. 6). In such cases, the iPad has *mediated* the student's positive relationship with physics. It follows that the student would be more likely to continue to engage in future studies of physics.

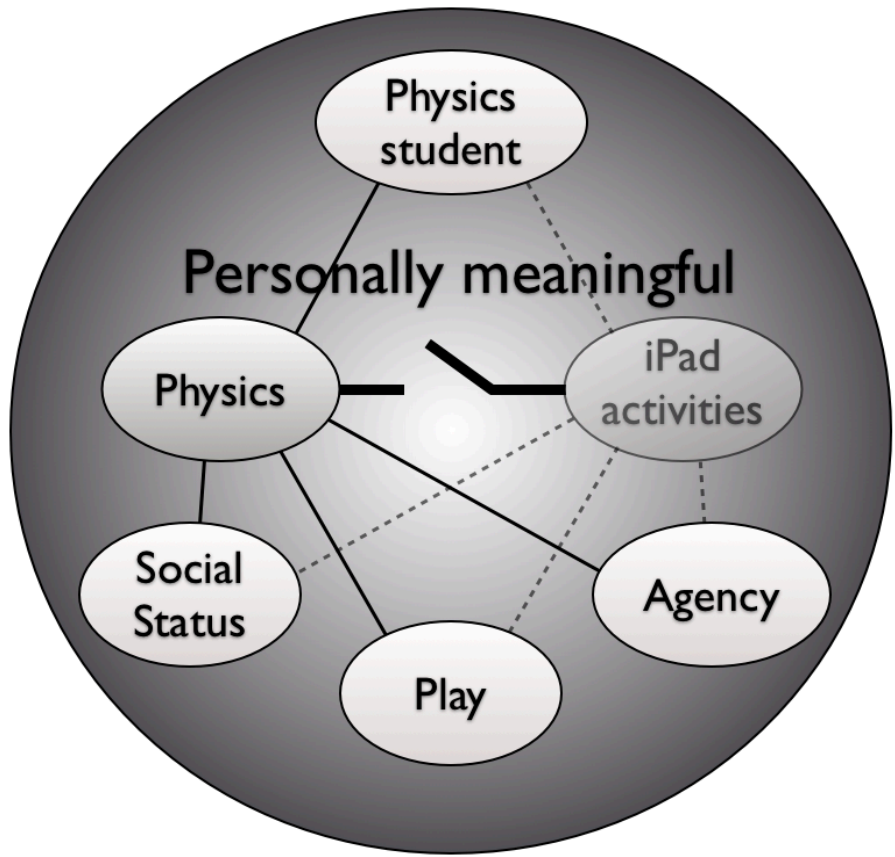

**Figure 6**. Shifted student relationship with physics.

It is unknown if any of the students in the study reported here fell into this category. However, the opening example of students videotaping and testing Newton's first law provides some indication that such a transformation could be taking place. We will continue to test this model in the future, collecting data that is specifically relevant to the categories that were revealed through the current study.

**Acknowledgements**

We thank Susie Dykstra for creating, and allowing us to study, her iPad physics enhanced environment, and the WISE and NSF (DUE #934921) grants.

# CHAPTER 4

# BLURRING THE BOUNDARY BETWEEN PHYSICS PROBLEM SOLIVNG AND STUDENTS' PEER CULTURAL PRACTICES

Ben Van Dusen and Valerie Otero

*School of Education, University of Colorado, Boulder, 80309, USA*
This study investigates differences in AP physics students' solutions to problems when creating them in traditional pencil and paper notebooks and in screencasts. The students' notebook and screencast solutions were examined for structural differences in their problem solving procedures and for correctness. These findings were combined with the examination of students' self-reports of behaviors and experiences while creating their solutions. Findings show that, within this classroom environment, students created more complete solutions to traditional physics problems and were more likely to get the correct answer when using screencasts versus notebooks. Student surveys show that they felt more socially connected, an increased sense of authorship, and less frustration when creating solutions in their screencasts. Our findings suggest that the improvements in student work and increased social interactions were associated with the iPad-based screencasts acting as *boundary objects*, giving students the opportunities to creatively incorporate personally meaningful practices into their physics assignments.41

**PACS:** 01.40.EK, 01.40.FK, 01.50.HT

Van Dusen, B., & Otero, V. (in review a). Blurring the Boundary Between Physics Problem Solving and Students' Peer Cultural Practices. *Physical Review Special Topics - Physics Education Research*.

**Introduction**

Our research is founded on the idea that in order to capture students' natural curiosity and introduce the richness that scientific inquiry can bring to their lives, physics classroom environments must be made engaging and motivating. We assume that certain tools hold meaning both within students' peer cultural practices and the physics classroom. Such tools can transform students' experiences by serving as bridges that can bring students' personal lives into the physics classroom and physics into their personal lives (Buxton et al., 2005). We hypothesize that such tools can lead to the rearrangement of classroom social practices, which can result in students feeling more engaged and motivated to do physics. The iPad is an example of an object that has meaning in many youth cultural practices as well as physics classrooms.

In this analysis we focus on student behavior surrounding a single iPad-facilitated practice, creating screencasts. Screencasting apps simultaneously record a user's voice and the their dynamic interactions with the iPad's screen and combine them to create a video (Van Dusen, 2013). In this study, students used screencasts to create tutorials for their peers, showing them how to solve traditional physics problems. Findings suggest that the iPad acted as a *boundary object* that facilitated the blending of the cultural practices of the students, the physics class, and the physics community. The merging of these practices acted to generate an environment in which students were motivated to engage in the creation of traditional physics assignment solutions.

## Theoretical framework

Students' physics experiences are too often wrought with fear, failure, and rote memorization rather than being enjoyable, empowering, and motivating (Belleau, Ross, & Otero, 2012; M. J. Ross, 2013). We investigate the extent to which a learning environment can produce physics experiences that are engaging and motivating. The iPad's dynamic nature and appeal to diverse populations may make it a uniquely positioned learning tool. In this work we focus on the iPad as a tool within three distinct sets of cultural practices: (1) The iPad as a classroom tool—iPads are used to combine access to word processing and the class's website to create presentations, submit assignments, and provide peers feedback. (2) The iPad as a consumer device—students use the iPad's camera, web browser, and Internet connectivity to engage in social activities, such as messaging friends, sharing pictures, and using social media. (3) The iPad as a physics computational device and research tool—physicists use iPads to facilitate the social creation of knowledge through the collection and sharing of data as well as engaging in peer review through its spreadsheet and PDF annotation applications. Figure 1 shows examples of iPad-facilitated activities in these three sets of cultural practices.

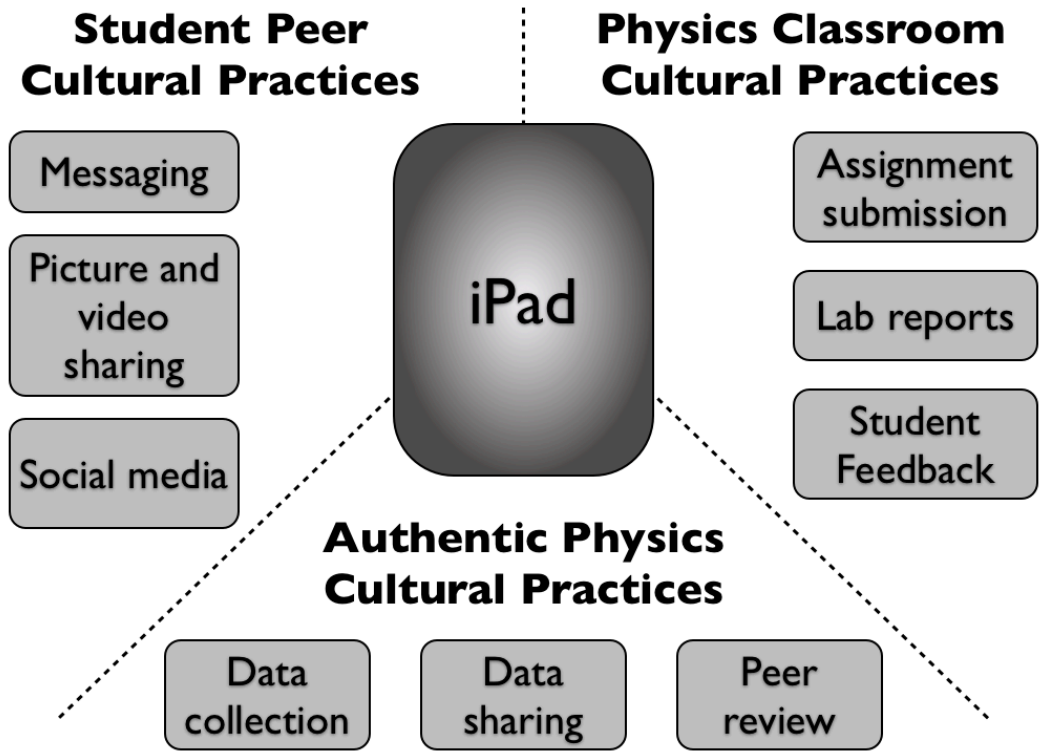

**FIGURE 1.** A model of the iPad as facilitating the blending of three different communities of cultural practices: students' peers, the physics classroom, and the larger physics community.

Because the iPad spans these sets of cultural practices it may act as a *boundary object* that facilitates the creation of new blended cultural practices (Buxton et al., 2005). Boundary objects are objects with meaning in multiple sets of cultural practices (Star & Griesemer, 1989). The concept of boundary objects has emerged from socio-cultural research on the role that shared tools, spaces, and concepts play in mediating social interactions (Akkerman & Bakker, 2011; Engeström et al., 1995; Leigh Star, 2010). In her seminal publication on boundary objects, Star (Star & Griesemer, 1989) examined how a natural history museum acted as a boundary object that bridged the cultural practices of researchers, conservationists, and trappers. While each group saw the museum as a means of preserving nature, they all had their own unique uses for it. The researchers used the museum as a means of data gathering. The conservationists saw the museum as a means of educating the public to the about the wonders of nature. The trappers used the museum as a means to share their collections. The museum brought together the three distinct groups and created a shared space for them to interact in. Through the creation of shared space, the museum acted as a boundary object that mediated these groups' cultural practices, facilitating the blending of their practices and the emergence of new, shared cultural practices.

We view learning in the classroom as a social practice involving the interactions among students, the teacher, tools, and environment. In order to understand the mediational role that tools, such as the iPad, play in shaping student experiences in physics classes, we use the lens of distributed cognition (Hutchins, 1995, 1996). Through this perspective, we view the classroom as a *socio-cultural cognitive system* encompassing the individual, surrounding people, available tools, configuration of the environment, and interactions among these things (Fig. 2). These systems are highly dynamic and their components are interdependent. A useful analogy is an electric circuit—a change in any one component of the system results in overall changes in the system. Similarly, when any component of a socio-cultural cognitive system changes (e.g. addition of an iPad or a student learns) the properties of the socio-cultural cognitive system undergoes change. In order to examine student learning, it is important to monitor the changing use of tools as students perform tasks. By viewing the context of the classroom as a socio-cultural cognitive system, we are not only interested in students' interactions with the teacher and with traditional paper and pencil tests, but we are also interested in how students interact with each other and with other tools. The nature of students' interactions with various tools often depends upon students' motivation to engage with them.

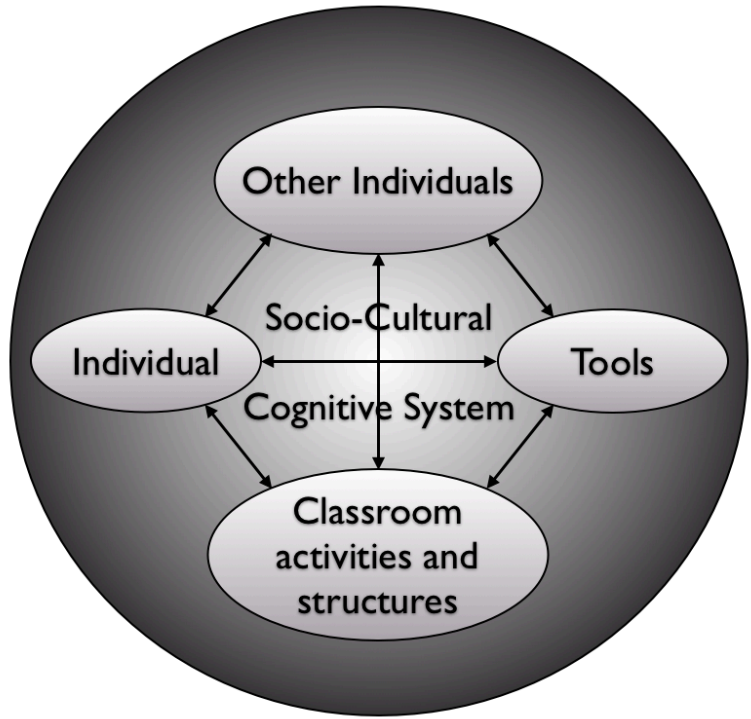

**FIGURE 2.** A model of the components of and interactions within a socio-cultural cognitive system.

Most of the literature on motivation frames motivation as a property of the individual (Maslow, 1943; NRC, 2004; Ryan & Deci, 2000). We view motivation as a dynamic, flexible, and emergent characteristic of interactions within a socio-cultural cognitive system. We define motivation as emerging from interactions within the system, as is evidenced by varying levels of student participation. Just as the interactions within a socio-cultural cognitive system can be reorganized with the introduction of a tool, the motivations that emerge from interactions within the system undergo similar shifts. Viewing the iPad as a boundary object, we hypothesize that its introduction can rearrange the classroom social practices to make physics more engaging and motivating for students.

## Research Questions

In our work we examine in what ways the introduction of iPads transforms classroom practices and how this influences student motivation, if at all. In order to understand these complex interactions, we investigated the following questions: (1) In what ways did AP physics students' performances on traditional notebook and screencast problem solutions differ, if at all? (2) What differences were there, if any, in students' behaviors and self-reported experiences when creating notebook and screencast solutions? (3) In what ways can we account for any differences that may exist in students' performances on notebook and screencast problem solutions?

## Setting

Data for this research project were collected in a high school AP physics class. The teacher has a background in biology, including a Ph.D. in biochemistry. Like the majority of US high school teachers of physics [8], the teacher did not have a degree in physics. The teacher in this study worked in a public high school located in an urban community that was primarily composed of students from non-dominant backgrounds. Table 1 shows the school demographics.

**TABLE 1**. School demographics

| Ethnicity | | | | Free or reduced lunch | ESL/FEP | IEP |
|---|---|---|---|---|---|---|
| Hispanic | White | Asian | Afr. Am. | | | |
| 56% | 32% | 8% | 3% | 41% | 49% | 11% |

The teacher was recruited for the research project because of her role as a Streamline to Mastery [9] teacher. Streamline to Mastery is an NSF-funded teacher-researcher community that collaborates with our university to engage in educational research as a mechanism for professional development. To be eligible for the Streamline to Mastery program, teachers must work in a school district that has been identified as largely serving students from non-dominant backgrounds, have a masters degree, and undergo an application process that focuses on their desire to improve their teaching practices through self investigation. The cornerstone of the Streamline to Mastery program is using research as a mechanism for generating principles about effective teaching and learning. For her research, the teacher in this study examined how providing her high school students opportunities to act as Learning Assistants (Otero, Pollock, & Finkelstein, 2010) for elementary school students affected their content learning (Nicholson-Dykstra, Van Dusen, & Otero, 2013). This experience led the teacher to continue to focus on creating teaching-to-learn opportunities for her students and motivated the creation of the screencasting activities.

During the first year of our collaboration, we worked with the teacher to jointly raise grant money to purchase a classroom set of thirty-eight iPads. The iPads were shared between the teacher's five classes. This allowed students to have their specific iPads during class, but four other classes used them throughout the school day. The teacher taught AP Physics, Biology, and Bio-Medical classes. This study focuses only the AP Physics class. The AP physics students typically had the opportunity to take the iPads home, provided they brought them back for the first class of the next day. The school experienced a significant transient student population, so class sizes were not constant. However, the AP physics class retained the majority of the 30 students who began the year in the class. Of those 30 students, this study examines the work of the 27 students who completed the class. While most of the students had access to the Internet at home, very few had access to an iPad outside of their physics class. The students were explicitly made aware that their class was being studied, in part, to better understand how the iPad could be used to facilitate science learning. They were also informed that the findings from their classes would help direct the future use of iPads in science classes.

Throughout the year, the students engaged in a variety of iPad-supported activities that were intended to supplement traditional AP physics assignments. For example, students used their iPads to collect videos of projectiles and digitally analyze their motion in order to create position, velocity, and acceleration graphs.

In this study we focus on a single iPad-supported activity that the teacher initiated—the students' creation of screencasts. Screencasting is a technology that captures a video of the iPad's screen while using the microphone to capture and merge audio to that file. This technology allows students to create various types of dynamic presentations. Within the physics class, the students' screencasting assignment prompts were to create a tutorial teaching how to solve a problem of their choice from a specific worksheet or chapter from their book. How these tutorials were to be created and what they should look like was not specified by the teacher and was left for the students to determine. Typical screencasts consisted of students talking through their thought process for solving a problem while writing out the solutions in real-time or while showing specific parts of a solution they had written down prior to recording the screencast (Van Dusen, 2013). Figure 3 shows screen-captures from the start and end of a student's screencast solving a question about electrostatics. The students posted their screencasts on the class's Edmodo™ site (a Facebook-like social media site created specifically for education). After the screencasts were posted online, students were often prompted by the teacher to view several other students' work and to provide their peers feedback on the quality and usefulness of their screencasts. In addition to the students regularly using the Edmodo site to provide written and verbal feedback to each other on their screencasts, three times throughout the year the teacher orchestrated a whole-class discussion about how the screencasting was going and how students felt the process could be made better. During the first of these whole-class discussions, the teacher prompted the students to co-create a rubric by which to evaluate the quality of their screencasts. While the students regularly gave each other feedback using the rubric, the teacher never formally graded any of the students' screencasts. The teacher's expectations and grading scheme were made explicit to the students at the start of the school year and were continually reinforced through the teacher's instruction and grading.

Start of screencast solution | End of screencast solution

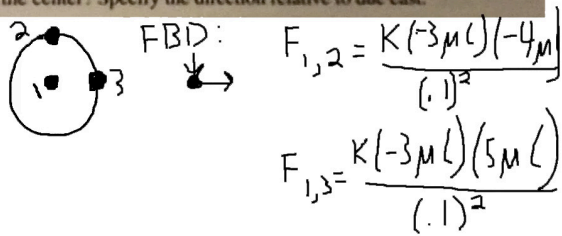 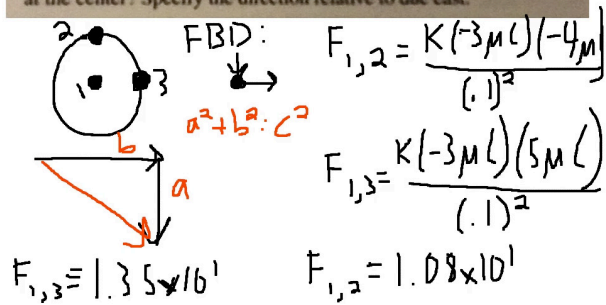

**FIGURE 3.** Screen captures from a student's solution to an electrostatics problem (Van Dusen, 2013).

The student screencast solutions are compared to the solutions that students created in their traditional pen and paper notebooks. The students were regularly given problems either from the back of their textbook or from worksheets to complete in their notebooks. In all of the problem solving assignments, the students were instructed to follow the AP guidelines for showing their work. This grading expectation was so commonly used in the class that students would refer to it as their "Bible" for solving problems. Like the screencasts, the teacher never formally graded these solutions. After the homework was completed, each group of students would be randomly assigned a solution from the assignment to share with the rest of the class. After each group shared their solution using a whiteboard, the other groups had the opportunity to ask questions and provide them feedback on any mistakes they may have made. As with the screencasts, students were periodically asked how the process of making notebook solutions was going and how it could be improved.

## Methods

### Data Collection

During the weekly observations, a variety of data were collected. For this study, we focus on the AP physics students' screencasts (N=96), traditional pencil and paper notebook problems (N=454), and survey responses (N=19). Table 2 shows the relevant data sources from the AP physics class in this investigation.

**TABLE 2**. Data sources

| Data type | Description |
|---|---|
| Artifacts | Students' screencast (N=96) and notebook solutions (N=454) |
| Student Surveys | Students' surveys about their experiences with screencasts and notebooks (N=19) |

Screencasts were collected digitally throughout the year, and the pencil and paper notebooks were collected at the end of the year. Students regularly used their notebooks until the very end of the school year, so students had to volunteer to come back after the semester ended to hand in their notebooks. We received no notebooks from the seniors (N=11), who got out of school a week before classes officially ended. Ultimately, eight of the sixteen juniors' notebooks were collected, analyzed, and compared to the student screencasts. In the eight notebooks, we examined the problems (N=454) from the kinematics, momentum, and electrostatics units. Only some of the notebook problems (N=337) contained sufficient information to score the students' answers for the correctness. All of the screencasts (N=96) had sufficient information to be able to score the students' final answers for correctness.

**Data Analysis**

For this study, students' screencasts and notebooks were scored using two methods. The first method scored students' final answers for correctness. The second method scored students' solutions for completeness using a rubric that mirrored the teacher's rubric for grading their solutions (which itself mirrored the AP test's grading guidelines). The grading rubric was aligned to the teacher's performance goals for students' problem solving procedures. A limitation of this rubric is that it only allowed for the assessment of quantitative problems, so conceptual questions were not included in this portion of the analysis. Our scoring rubric, much like the AP scoring rubrics, focused on the procedural steps to solving a problem. The rubric uses a scoring system in which students get 1 point for each step of a problem solution (when applicable). There were 5 steps possible in each question: 1) writing down the equation, 2) solving the equation (before substituting in values), 3) substituting in values, 4) getting a final answer, and 5) using correct units throughout. Figure 4 shows an example of a student's screencast solution with each of the 5 procedural steps annotated. To control for media affordances, the screencasts were scored based only on what the students wrote down, no points were associated with the students' verbal explanations.

**Scoring Rubric**

Step 1: Write down the equation — 1 point

Step 2: Solve the equation — 1 point

Step 3: Plug in the numbers — 1 point

Step 4: Solve for the final answer — 1 point

Step 5: Correct units used throughout — 1 point

**Student work**

5. A 15 000 kg railroad car traveling at 2.45 m/s couples with a 12 500 kg car which is at rest. What is the final velocity of the two cars?

$$P = P'$$
$$m_1 v_1 + m_2 v_2 = m_1 v_1' + m_2 v_2'$$
$$m_1 v_1 = (m_1 + m_2) v'$$
$$v' = \frac{m_1 v_1}{(m_1 + m_2)}$$
$$v' = \frac{(15000 kg)(2.45 m/s)}{(15000 kg + 12500 kg)}$$
$$v' = 1.34 m/s$$

**FIGURE 4.** An example screencast solution with the scoring rubric applied.

In order to have a representative sample of students' problem solving performances throughout the year, work from the first (kinematics), middle (momentum), and last (electrostatics) units were scored. Student work in these three units included a series of notebook problems (N=454) and screencasts (N=96). Table 3 shows the number of screencast and notebook solutions that were scored for completeness and correctness, for each unit. The number of problems students completed varied by unit. This variance comes from a combination of the differences in the number of problems assigned per unit and students' problem completion rates.

**TABLE 3.** The number of solutions scored for completeness and correctness, by unit.

|  |  | Kinematics (N) | Momentum (N) | Electrostatics (N) | Total (N) |
|---|---|---|---|---|---|
| Notebook | Completeness | 343 | 55 | 56 | 454 |
|  | Correctness | 319 | 6 | 12 | 337 |
| Screencast | Completeness | 31 | 18 | 47 | 96 |
|  | Correctness | 31 | 18 | 47 | 96 |

To ensure an even weighting between units (despite the variation in the number of problems), each student was given up to two scores for completeness of their solutions and for correctness of their answer (one for notebook problems and one for screencast problems) for each unit. Each student's scores of completeness and correctness consist of the average notebook and the average screencast scores for the unit. The unit averages for each individual student were then averaged to provide a representative score for the entire class's performance on notebook and screencast solutions. Figure 5 shows each of the steps carried out in creating an average score for student notebook solutions. The notebook and screencast scores were statistically analyzed using an independent samples t-test. Table 4 shows the full set of students' average unit scores on notebook and screencast completeness. While the majority of both the completeness and correctness scores are for the kinematics sections, we see similar trends in all three units.

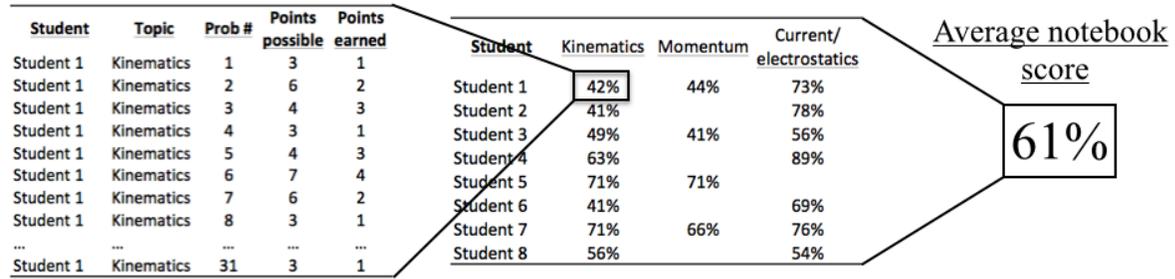

**FIGURE 5.** Calculating the completeness score of the average notebook solution.

**TABLE 4.** Students' unit averages for notebook and screencast problem completeness and correctness.

| Medium | Student | Scoring type | Kinematics | Momentum | Electrostatics | Average |
|---|---|---|---|---|---|---|
| Notebooks | Student 1 | Completeness | 43% | 44% | 73% | 54% |
| | | Correctness | 72% | - | - | 72% |
| | Student 2 | Completeness | 41% | - | 78% | 59% |
| | | Correctness | 40% | - | - | 40% |
| | Student 3 | Completeness | 49% | 41% | 56% | 49% |
| | | Correctness | 61% | - | - | 61% |
| | Student 4 | Completeness | 63% | - | 89% | 76% |
| | | Correctness | 47% | - | - | 47% |
| | Student 5 | Completeness | 69% | 71% | 96% | 70% |
| | | Correctness | 75% | 0% | 40% | 38% |
| | Student 6 | Completeness | 41% | - | 69% | 55% |
| | | Correctness | 27% | - | - | 27% |
| | Student 7 | Completeness | 71% | 66% | 76% | 71% |
| | | Correctness | 78% | - | - | 78% |
| | Student 8 | Completeness | 56% | - | 54% | 55% |
| | | Correctness | 51% | - | 29% | 40% |
| Notebooks | **Average** | Completeness | **54%** | **55%** | **74%** | **61%** |
| | | Correctness | **56%** | **0%** | **34%** | **50%** |
| Screencasts | **Average** | Completeness | **88%** | **82%** | **77%** | **81%** |
| | | Correctness | **90%** | **72%** | **83%** | **83%** |

To account for the fact that only 30% of the students are represented in the notebook problem data, the overall class performance of students who did and did not turn in their notebooks were compared. The first and second semester grades were averaged for students who did and did not turn in their notebooks. The difference in the average of the semester grades for each group was also statistically analyzed using an independent samples t-test. Figure 6 shows the average semester grades for students who did (83%) and did not (77%) turn in their notebooks.

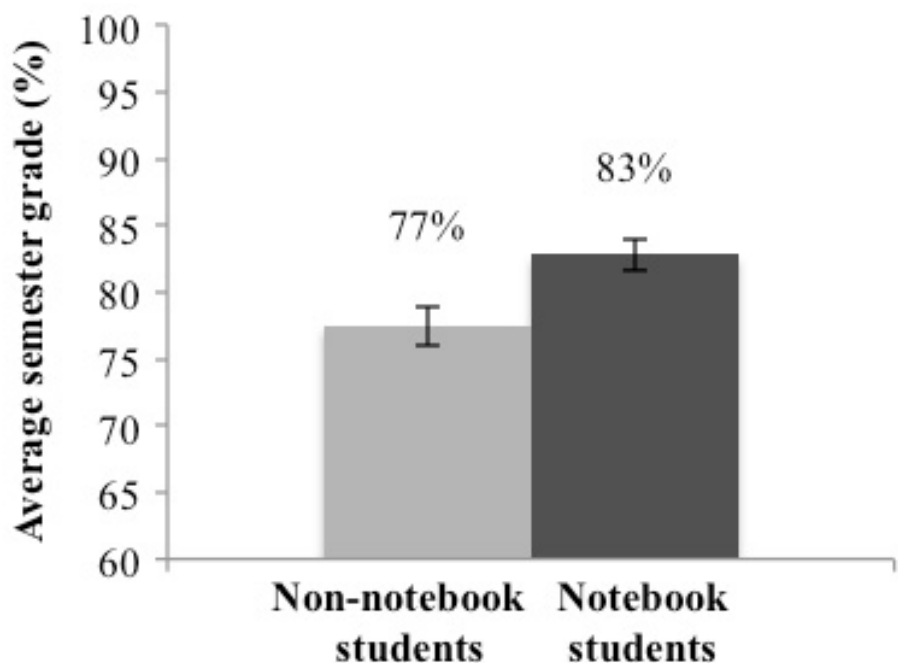

**FIGURE 6.** Average semester grades for students who did and did not turn in notebooks (with standard errors).

Our analysis shows that the students who turned in their notebooks performed, on average, 6% better in the class than their peers who did not turn in their notebooks. An independent samples t-test shows the difference in the students' average semester grades to be statistically significant (p=0.034). The full results of this analysis are reported in the findings section.

A second part of this study examines students' experiences when creating problem solutions. A survey asking students to answer nine likert-style questions pertaining to their experiences was administered shortly before the end of the year. The questions asked students whether various feelings or thoughts were more likely to occur when they were solving a notebook problem or creating a screencast, or whether there was no difference. The questions addressed students' social engagement, feelings of authorship, affective issues, and effort levels (actual questions are shown in table 6). Student responses of "much more likely" and "somewhat more likely" were collapsed into a single category.

Finally, in order to classify the types of behaviors students exhibited while creating their screencasts, eighty-five screencasts, spanning two years of student work, were examined using open (generative) coding that attempted to capture any commonalities among screencasts. This led to the creation of twenty-one screencasting codes, each representing a visible or auditory activity students engaged in during their screencast. These codes were then clustered by commonalities, which led to the emergence of two categories of codes: 1) technological moves and 2) social moves. The full list of codes, their categories, and examples of them can be seen in Table 5.

**TABLE 5**. Screencasting categories, codes, and examples

| Category | Code | Example |
|---|---|---|
| Technology Moves | Picture | A student inserts a picture of their hand doing the right-hand rule |
| | Pen | A student uses the pen to write out an equation |
| | Typing | A student types up the question |
| | Scaling | After completing part A of a problem, a student shrinks its size to make room for part B |
| | Pointer | A student uses the pointer function to direct the viewer's attention to the free body diagram |
| | Multi-Take | After completing part A of a problem, a student stops recording and picks up recording part B at a later time |
| | Multi-Page | After completing part A of a problem, a student uses a fresh page to complete part B |
| | Multi-Color | A student draws the acceleration vectors in one color and the velocity vectors in another color |
| | Moving Parts | After drawing a diagram, a student moves it to the side of the screen |
| | Large Space | After completing part A of a problem, a student moves their screen to a new area to complete part B |
| | Erase | A student accidentally writes down the wrong number and then erases it |

| Category (cont.) | Code (cont.) | Example (cont.) |
|---|---|---|
| Social Moves | Clarify | After explaining why the car had positive acceleration a student goes back and explains it in another way |
| | Apologize | After sloppily writing down an equation, a student apologizes to the audience |
| | Directly Address Audience | A student tells their listener that they hope that they listener can follow their work |
| | Friends included | A student brings in a friend to help them do a funny voice |
| | Group Pronouns | A student talks about the answer that "we" get for a problem |
| | Humor | A student uses their "Darth Vader" voice to read a problem |
| | Introduction | A student starts their screencast by saying their name and thanking the audience for listening |
| | Offer Advice | In addition to explaining how they did the work, a student advises their listeners how to solve another problem |
| | Tag Line | A student ends their screencast by thanking their listeners |
| | Extra Details | A student takes a circular motion problem and adds in the detail that it's actually a space hamster orbiting earth |

This screencast coding scheme was applied to the same set of 96 screencasts from the kinematics, momentum, and electrostatics units, on which the grading rubric was used. If any of the technological moves or social moves were observed in a screencast, the entire screencast was coded with that code. To test for reliability in the coding, 10% of the kinematic screencasts were independently coded by an external education researcher. Because the coding scheme is relatively simple, the external coding was a 100% match with coding performed by the authors. Findings from the coded screencasts are reported below.

Limitations to this study include: (1) Student notebook collection occurred through a self-selection process, which could lead to a bias in the notebook data. To attend to potential differences in the performance of the students who turned in their notebooks and the rest of the class we examined the students' semester grades. This examination showed us that the students who turned in their notebooks were, on average, higher performers in the class. (2) The teacher provided the students with a notebook grading rubric, but co-generated the screencasting grading rubric with the students. It could be that the process of co-generating the rubric allowed students to feel more engaged in the process and to better understand the expectations for creating screencast solutions and that this lead to better screencasting performance. This does not appear to the case, however, since the kinematics screencasts were created prior to the creation of the class' screencasting rubric and the screencast problems still showed significantly more complete solutions (88%) than the kinematics notebook problems (54%).

**Findings**

*Notebook and Screencast scores*

Data show that students were significantly more likely to answer problems correctly in their screencasts than their notebooks. Figure 7 shows that students answered 83% of their screencast problems correctly, compared to 50% of notebook problems answered correctly. These data show that when solving problems in screencasts the students' answers were 33% more likely to be correct than when solving them in their notebooks. An independent sample t-test also shows the average improvement of three letter-grades is a statistically significant difference (p<0.001).

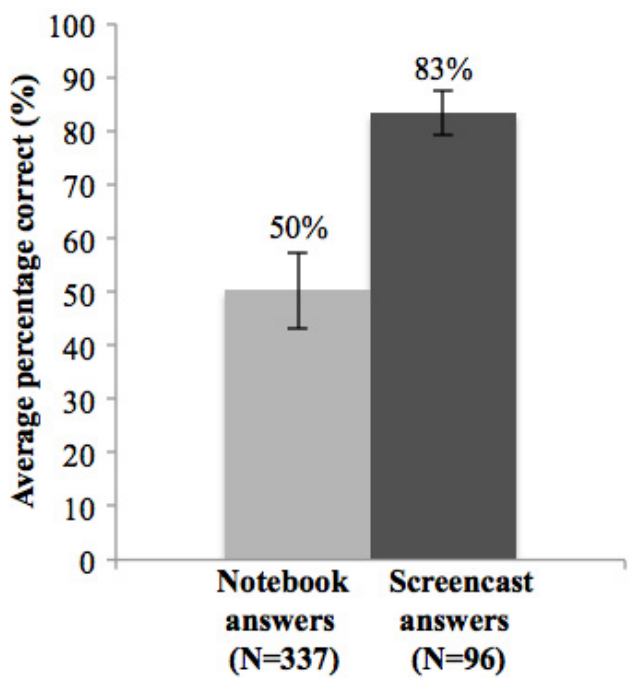

**FIGURE 7.** Students' average percentage of correct solutions on notebook and screencast problems.

The 5-point grading rubric, outlined in Figure 4, was used to score students' problem solving procedures to traditional physics problems. Results showed that the students' scored higher on their screencast solutions than their notebook solutions. Figure 8 shows the students' average unit scores and standard errors for their notebook and screencast solutions. Students showed 61% of the steps involved in solving problems in their notebooks, as compared to 81% in their screencasts.

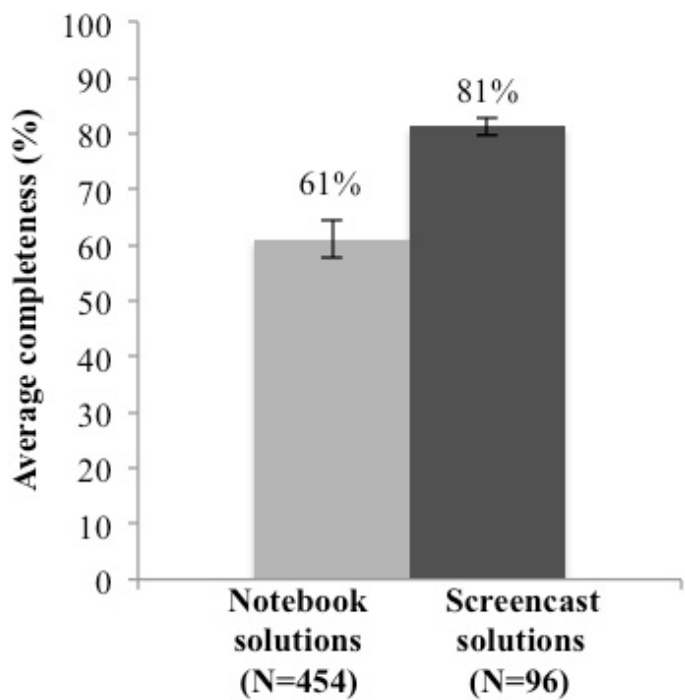

**FIGURE 8.** Students' average percentage of solution completeness on notebook and screencast problems.

These data show that when solving problems in screencasts students' solutions were 21% more complete then when solving problems in their notebooks. An independent sample t-test also shows the average improvement of two letter grades is a statistically significant difference ($p<0.001$).

*Student surveys*

The survey, which asked students to compare the processes of creating solutions to traditional physics problems in their notebooks versus screencasts, was examined. Table 6 shows the students' responses to questions about creating solutions in their notebooks and screencasts.

The questions have been categorized to indicate the four main themes: social interactions, feelings of authorship, affect and issues of effort.

**TABLE 6.** Students' survey responses. (N=19)

| Topic | # | Question | Screencasts | No difference | Notebooks |
|---|---|---|---|---|---|
| Social | 1 | I give my friends more feedback on their work when using: | 37% | 58% | 5% |
| | 2 | I more often think about what my friends will think of my work when using: | 53% | 42% | 5% |
| | 3 | I include more references to things I like when using: | 42% | 53% | 5% |
| | 4 | I include more jokes when using: | 63% | 37% | 0% |
| Authorship | 5 | I feel like I can be more creative when using: | 63% | 32% | 5% |
| | 6 | I put more things in my own words when using: | 26% | 68% | 5% |
| Affect | 7 | I am more excited about my work when using: | 53% | 26% | 21% |
| | 8 | I feel more frustrated when using: | 16% | 37% | 47% |
| Effort | 9 | I am more likely to do as little work as I can to complete the assignment when using: | 16% | 53% | 32% |

These data show that while many students reported no difference (47%) in social behaviors when using screencasts or notebooks, a larger portion of the students reported that they are more likely to engage in various social behaviors when using screencasts (49%) as compared to 4% of the students who report engaging in more social behaviors when doing notebook-based work rather than screencast-based work. Specifically, students reported higher rates of social interactions (giving friends feedback, making socially relevant references, and including jokes) as well as preparing for social interactions (thinking about what friends will think) when creating screencast solutions than when creating notebook solutions.

Students' responses to questions of authorship show a similar pattern to those about social behavior. For questions of student authorship, 45% of the students report increased feelings of authorship when using screencasts, 50% report no difference, and 5% report increased feelings of authorship when using notebooks. More students associated screencasts, rather than notebooks, with opportunities to be creative and to put things in their own words. The data also show that students associated more positive affect with screencasts than with notebooks. Specifically, the most common student responses were that they were more excited about their work when using screencasts (53%) and more frustrated when using notebooks (47%). Finally, the survey data show that students report putting forth more effort when creating screencasts than when solving problems in their notebooks. While approximately half of the students reported no difference (53%), twice as many students report doing as little work as possible to complete an assignment in their notebooks (32%) compared to creating screencasts (16%).

We propose that students' association of screencasting with positive social interactions, opportunities for authorship, and affective experiences is associated with putting forth more effort in the creation of physics solutions on screencasts. We interpret this increase in effort as associated with increased motivation, which helps to explain why students created more complete solutions in their screencasts than in their notebooks.

*Screencast coding*

To verify students' self reports of social interactions and opportunities for authorship when making screencasts, we examined the screencasts themselves. The students' screencasts were coded for the types of social and technological behaviors that the students exhibited in them. As previously described, if a student demonstrated any of the behaviors in the screencast coding scheme, the entire screencast was given the associated code. Figure 9 shows the percentage of screencasts in which each of the social moves appeared. Each screencast included, on average, 1.8 social moves. As figure 8 shows, the most common social move was using group pronouns (e.g. us, we, and our) when explaining the steps to solving their physics problem. For example, the problem was "our question," each step was something that "we do," and the results were "our results."

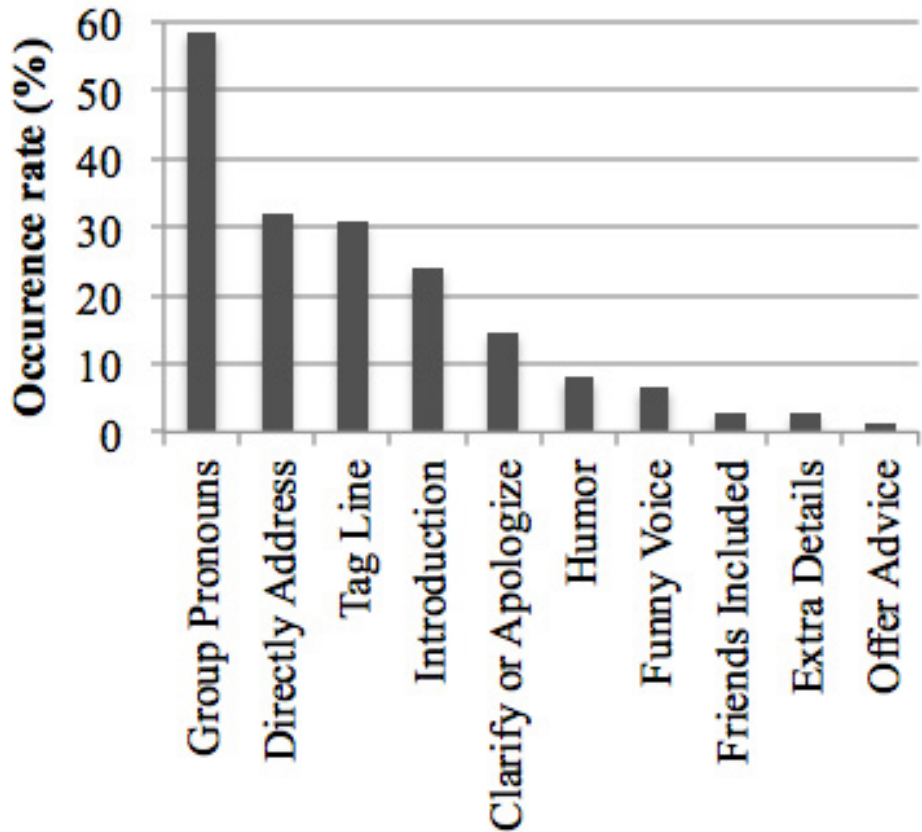

**FIGURE 9.** Percentage of screencasts that exhibit social moves.

The students' use of group pronouns suggests that, while they created their screencasts individually, they saw the process as a social interaction. These data also show that students integrated a range of social activities, such as directly addressing their audience, including jokes, and using funny voices, into their problem solutions. Within each of these codes, students exhibited a range of behaviors. For example, some students incorporated humor into their screencasts by altering the question to have humorous components while others made popular references. It should be noted that students' use of group pronouns may have originated as mimicry of the teacher's practice of using group pronouns when solving problems in front of the class.

The screencasts were also coded for the technological tools that students used in them. While students used, on average, 3.2 technological tools per screencast, Figure 10 shows that the majority of the tools were used in around 20% of the screencasts.

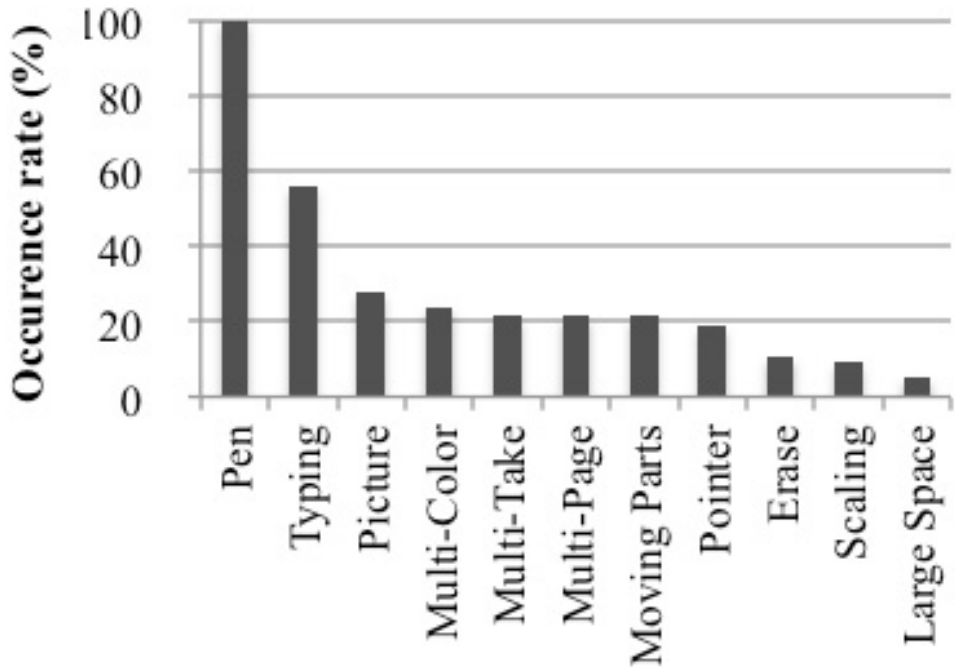

**FIGURE 10.** Percentage of screencasts that exhibit technological moves.

With the exception of using the pen (which was used in all of the screencasts) and the typing functions (which was used in just over 50% of the screencasts), students were selective in choosing which tools to use in their screencasts. These data support the students' self-reports that the screencasting allowed them the opportunity to exert authorship and creativity in their work.

**Discussion**

Our findings are nicely summarized by a student's response to a survey question that asked what the most important difference was between doing screencast problems and notebook problems:

*I think creativity in solving the problems can be important because it shows what you can do in your own way and it may be unique as well. Screencasts open doors to bringing that creativity out and letting other people see that.* (Salina, 5/14/2013)

This student proposes that screencasts were different from notebooks because they allowed students to be creative in solving problems and because they facilitated the sharing of this creativity with the world. This claim of increased creativity and social engagement is substantiated in both the students' self-reports and in the analysis of the screencasts themselves. We infer through this student's talk of how screencasts "open doors" to "bring creativity out" that the student associates positive affective experiences with screencasts, which is also supported in the students' self-reported data. We suggest that the social involvement, opportunity for self-authorship, and positive affective experiences are associated with motivation and resulted in students' placing more effort into screencasts than notebooks. This extra effort then resulted in greater diligence in following the steps to creating a complete problem solution and ultimately getting the correct answer.

We conclude that students' positive associations with screencasting emerged through the combining of socially meaningful behaviors and traditional physics classroom activities. Screencasts enabled meaningful inclusion of students' peer cultural practices (e.g. making jokes or popular references) into the physics classroom cultural practices (e.g. solving problems and turning in assignments). It was through the blending of these practices that new cultural practices emerged that were relevant both to the students as well as to the teacher and her goals for the physics class (e.g. sharing screencasts solutions with peers and increased effort). We hold that screencasting mediated the blending of the students' personal selves with physics in a way that created a new meaning for physics homework. This was accomplished through screencasting and the iPad serving as a boundary object that connected students' peer cultural practices with the physics classroom cultural practices. Figure 11 shows examples of student peer cultural practices (making jokes, engaging with peers, making social references, and social networking) and physics classroom cultural practices (writing equations, solving problems, submitting homework, and evaluating work) that the iPad brought into coordination. By serving as a bridge between these two sets of cultural practices, screencasts created a space in which the blending and emergence of new practices could occur. We suspect that this type of blending of the self with subject matter illustrates how identity is developed.

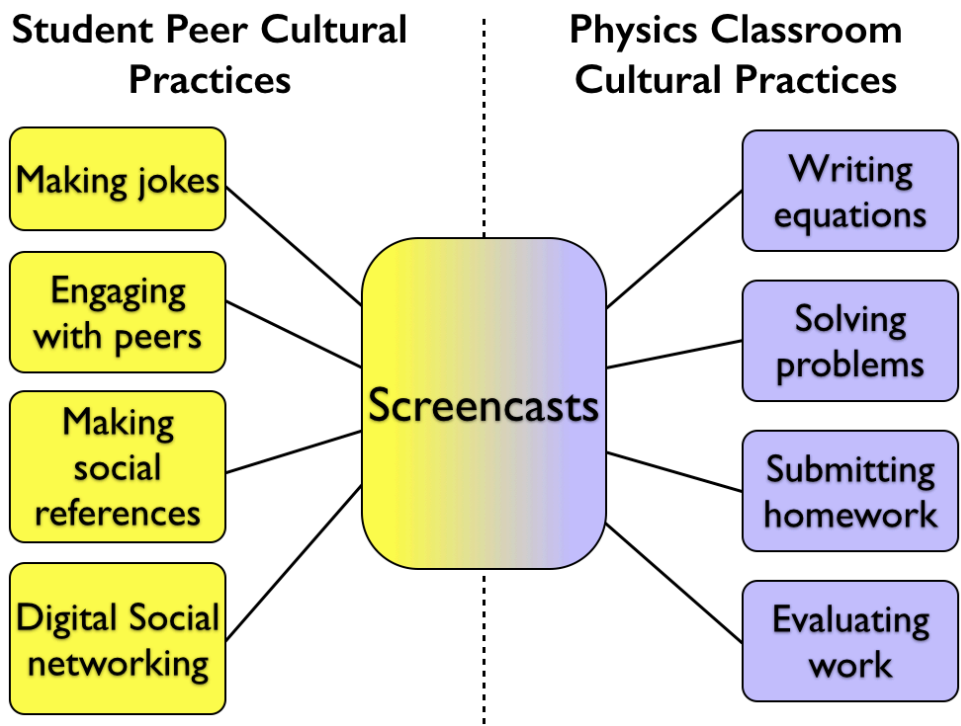

**FIGURE 11.** Screencasts acting as a boundary object.

It should be noted that, it may not have been the screencasts that were the critical boundary objects in this system. It could be that another boundary object, such as the screencasting rubric that the teacher co-generated with the students, mediated this blending of cultural practices. As previously stated, the kinematics problems, which were created prior to the screencasting rubric, showed similar trends as the momentum and electrostatics problems, which were created after the screencasting rubric. This leads us to believe that the co-generation of the screencasting rubric was not the central mediating activity in this class. We do, however, believe that the effects of students' co-creation of rubrics warrant further investigation.

While screencasts have meaning in the cultural practices of both the students' peers and the physics classroom, simply creating screencasts may not be enough to mediate the blending of these practices. Our environments (socio-cultural cognitive systems) mediate the ways that we engage with the tools and people around us. We do not have the evidence required to identify all of the aspects of the classroom environment that acted to shape students' interactions with screencasts, but our observations have led us to identify several features of the AP physics class that we believe may have played a role in facilitating the success of screencasts on iPads in acting as boundary object. These factors include: (1) the visibility of us as researchers in the classroom, (2) the co-creation of a screencasting rubric, (3) the peer-review system for examining screencasts, and (4) the public nature of the assignments. As part of our research, students were explicitly told that the researchers were studying how to effectively use iPads in physics classes, that they should help figure out how to best use them, and that what was learned from their class would effect how iPads were used in other classes. While students were allowed to explore how to create screencasts, the teacher implemented a set of practices that scaffolded the students' experiences. One of these experiences was prompting the students to create a rubric by which they could evaluate the quality of their own and peers' screencasts. The teacher also told students that they were to upload their screencasts to the course website, where students were encouraged to view each other's work and provide feedback.

Our observations have also led us to identify several features of classroom tools and the students' interactions with them that may have contributed to making screencasts on iPads effective boundary objects. These tools include: (1) wireless Internet, (2) the course website, (3) and the iPads themselves. The wireless Internet and the course website acted as complimentary technologies that allowed students to easily share their screencasts with fellow students (a non-trivial task with traditional notebook problems). The course website, which is an Edmodo™ site, was designed to be reminiscent of Facebook and incorporates many social networking elements. The screencasts were made on iPads, objects that are status symbols within our students' peer cultural practices (Van Dusen & Otero, 2012). While screencasts can be made on desktop computers, the iPads provided students the affordance of easily writing out problems and drawing diagrams. Screencasts also have affordances that notebooks lack, such as the ability to easily incorporate dynamic features, such as digital pictures and moving components.

**Implications**

It is important to note that the authors are not advocating specifically for the integration of screencasts or iPads into physics classroom environments. What made these particular activities effective was not simply the tools. Rather it was the tools, students, contexts, and interactions between them that created a classroom environment (a sociocultural-cognitive system) in which the activities mediated the blending of practices and subsequent improvement in student performance. It was within this rich environment, with all of its intricacies, that these positive outcomes emerged.

What we have shown in this study is one example of how tools can be used to reorganize students' social practices and classroom power structures to create more engaging and motivating physics experiences. What we advocate is for physics instructors to be thoughtful in their integration of tools (e.g. computers, lab equipment, or the curriculum itself) into the learning environment such that they value students' peer cultural practices and create experiences that are meaningful and engaging rather than frustrating and belittling. It is through the emergence of these types of positive experiences that we can begin to foster long-term physics trajectories in our students. Until students feel that physics is engaging and meaningful to them, no amount of improvement in student conceptual learning will create the STEM workforce or scientifically literate population that our world needs going into the future.

**Aknowledgements**

We would like to thank Susie Nicholson-Dykstra for sharing her classroom and teaching expertise with us. We would also like to thank CU Boulder's Women Investing in the School of Education and STEM Learning Center for funding our research.

# CHAPTER 5

# STUDENT ENGAMENT AND BELONING IN PHYSICS CLASS

Ben Van Dusen and Valerie Otero

*School of Education, University of Colorado, Boulder, 80309, USA*

This study investigates the notion of *basic psychological needs*. Particularly, how competence, autonomy, and relatedness affect physics students' classroom outcomes. The three constructs were examined using a modified version of the Basic Needs Satisfaction at Work Survey. The exploratory factor analysis of our survey results showed that three factors emerged. The emergent constructs did not map onto the *a priori* constructs (competence, autonomy, and relatedness). Our literature review showed that the majority of the studies that utilize the constructs of competence, autonomy, and relatedness assumed them to be independent of the each other. Our finding that competence, autonomy, and relatedness are not independent is consistent with other studies that examine these constructs using exploratory factor analysis. Findings also show that students' self-reports that indicate a blend of competence and autonomy are predictive of higher semester final grades and increases in interest in learning physics. We conclude that the constructs may not be separable because they are a function of a person's dynamic, iterative, and feedback-dependent interactions within their environment.

**PACS**: 01.40.ek, 01.40.Fk, 01.40.Ha

## Introduction

Our study investigates what types of learning environments produce physics experiences that are motivating to students. We focus on how tools that bridge the cultural practices of students' peers and the cultural practices of the physics classroom mediate experiences in a high school AP physics class. We believe that these tools can transform students' experiences, leading to shifts in students' levels of engagement and motivation in physics, ultimately resulting in better classroom performance.

Our pilot work utilized a grounded approach to identify classroom constructs that motivate students to engage in physics. In our qualitative examination we identified three types of experiences that appeared to influence student classroom motivation, which we labeled *play*, *agency*, and *social belonging* (Van Dusen & Otero, 2012). The emergence of these three constructs led us to examine the literature base for theoretical perspectives that use similar constructs. *Self-Determination Theory* identifies three constructs (competence, autonomy, and relatedness) that are said to be responsible for students' motivations (Ryan, 1995). Figure 1 compares our constructs to those identified in Self-Determination Theory. Both our own qualitative study and research done in the field of Self-Determination Theory identify a blend of positive inter- and intra- personal experiences that appear to be positively related to feeling intrinsically motivated to engage in an activity.

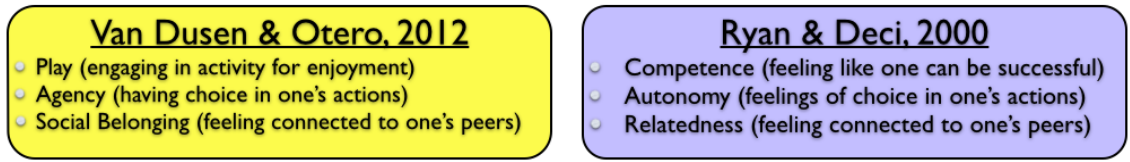

**Figure 1.** The central constructs from two studies on motivation.

We were pleased to see some similarities in the findings from our small study and those from work done on Self-Determination Theory. Because Self-Determination Theory has a large corpus of literature supporting it, we decided to leverage their work going forward. In this study we examine the constructs of competence, autonomy, and relatedness and their effects on high school AP physics students' semester grades and interest in learning physics.

## Conceptual Framework

Self-Determination Theory (SDT) considers identity and associated motivations as being attributes of an individual that slowly shift over time as the individual engages in activities (Deci & Eghrari, 1994; Ryan & Deci, 2000; Ryan, 1995). According to this perspective, given the appropriate conditions, an activity can be gradually integrated and internalized into one's identity. Through this internalization of an activity, one's choice to engage in an activity shifts from being externally motivated to being internally motivated. Through this shift, it can be said that an individual is motivated to engage in that activity.

According to Deci and Ryan (Ryan & Deci, 2000), internalization can be thought of as the assimilation of behaviors that were once external to the self. Through the process of internalization, individuals come to feel that what makes them engage in an activity (their locus of causality) moves from the external to the internal. Based on the study of organismic integration, SDT states that internalization occurs as an activity fulfills an individual's *basic psychological needs* (Deci & Ryan, 1991; Ryan & Deci, 2000). Organismic integration states that, like all natural processes, development through integration must be nurtured by the fulfillment of basic needs. These basic needs are defined as the, "nutriments or conditions that are essential to an entity's growth and integrity" (Ryan, 1995, p. 410). In the case of the human psyche, the basic nutriments for growth (or basic psychological needs) are feeling a sense of *competence, autonomy,* and *relatedness* (Reis, Sheldon, Gable, Roscoe, & Ryan, 2000; Ryan, 1995). When people engage in activities which provides them the experiences of competence, autonomy, and relatedness (instead of excessive controls, overwhelming challenges, and relational insecurity) they will be more likely to choose to engage in the same activities in the future.

SDT identifies six stages of internalization (*amotivation, external regulation, introjection, identification, integration,* and *intrinsic motivation*), shown in Figure 2. Each column in figure 2 represents one of the six stages of internalization, while the rows represent the corresponding *regulatory processes*, *mental and emotional processes*, *perceived locus of causality*, and *relative autonomy* for each stage. A person's development of intrinsic motivation is achieved through the internalization of an activity. As internalization of an activity occurs, one's regulatory process move from left to right through the six stages shown in Figure 2, making the activity more assimilated to one's self and increasingly intrinsically motivating.

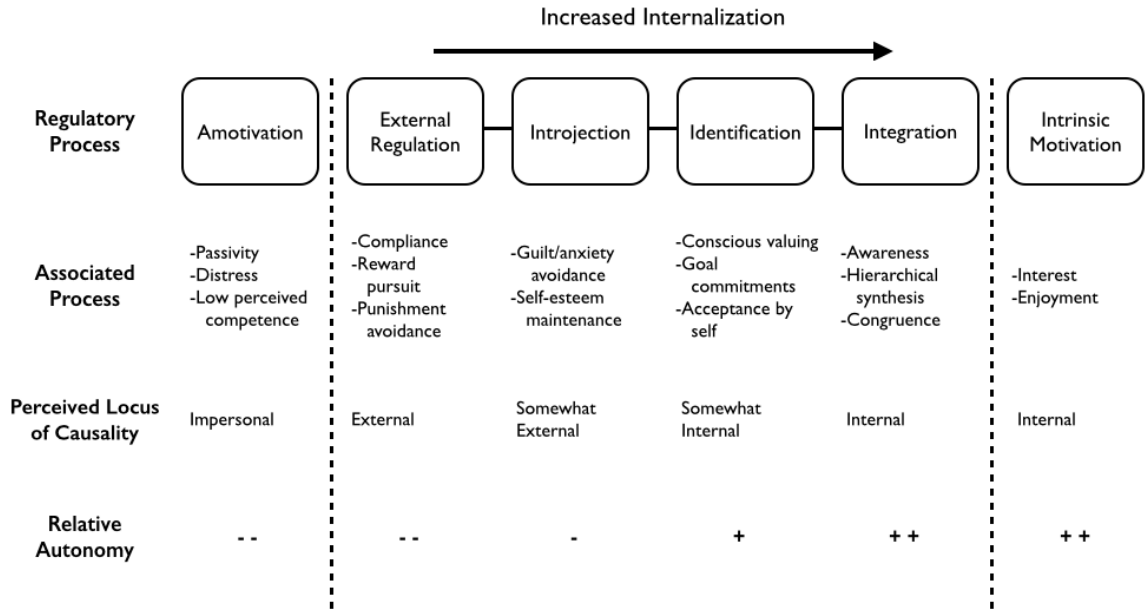

**Figure 2.** A guide to motivation in Self-Determination Theory (Ryan, 1995)

The extremes of the regulatory process scale (row 1) are *Amotivation* and *Intrinsic Motivation*. Amotivation actions are those that are seen as out of one's control. Intrinsic motivation, on the other hand, is the driver of actions that are done for their pure enjoyment and are perceived by the individual as being within their controls (Deci & Eghrari, 1994). Between the extremes of the regulatory scale, SDT identifies four regulatory process and their associated psychological mechanisms that drive peoples' actions (Deci & Eghrari, 1994; Ryan & Deci, 2000; Ryan, 1995). *External Regulation* is the first stage of internalization and is represented in the second column of Figure 2. Activities that undergo External Regulation are done out of compliance and hold little to no personal meaning to the individual. These types of activities require external coercion or reward. *Introjection* is the second stage of internalization and is represented in the third column. While Introjection is not entirely based on external motivation, it is driven by a sense of guilt or anxiety avoidance about potential judgment by others. *Identification* is the third stage of internalization and is represented in the fourth column. When in the Identification stage, motivation to engage in an activity is based on a feeling of acceptance and personal valuing of the activity. While this regulation process is perceived as being internally motivated and autonomously driven, it lacks integration with other parts of the self. The Identification stage can be thought of as the "trying on" stage, in which a person values an activity but has not yet fully embraced it. *Integration* represents the fourth stage of internalization and is represented in the fifth column. In the Integration stage of regulatory processes, various identifications are organized and brought into congruence with one's identity as a whole. Activities that have been assimilated to the Integration stage are those that we see as being central to our identities and we are intrinsically motivated to engage in.

Figure 2 also shows the proposed *associated processes*, *perceived locus of causality*, and *relative autonomy*. The *associated processes* (row 2) are example psychological mechanisms that drive people to engage in activities. The *perceived locus of causalities* (row 3) determines whether a subject's engagement in an activity is being driven externally or internally. The *relative autonomy* (row 4) is the extent to which people feel that they have control over whether or not they engage in activities.

This perspective provides a model for how one's identity co-develops with one's motivations. It is our goal to utilize this perspective to better understanding the role that competence, autonomy, and relatedness play in developing an identity and motivation in the physics classroom.

**Research Questions**

In this paper we use SDT to examine the ways in which classroom experiences influence student motivation. In order to understand these complex interactions, we investigated the following questions: (1) To what extent do students' feelings of competence, autonomy, and relatedness in physics class lead to improved classroom outcomes, if at all? (2) To what extent are the constructs of competence, autonomy, and relatedness distinguishable from each other, if at all?

## Literature Review

Surveys have been a central tool in SDT research, therefor the field has created a variety of surveys designed to assess constructs central to SDT, such as a person's basic psychological needs, intrinsic motivation, and self-regulation. To understand these surveys and their uses, we reviewed the literature that is most central to SDT. To narrow our search, we selected our literature from a website that is dedicated to SDT and is designed to act as a clearing house of information on SDT (Rochester, 2014). The website, titled "Self-Determination Theory: and approach to human motivation & personality" is hosted by the University of Rochester, which is home to Edward Deci and Richard Ryan (the original developers of SDT). The Self-Determination Theory website has a page dedicated to surveys. The page lists 17 categories of surveys, each of which has a subpage with several surveys from the category listed. For example, on the Basic Psychological Needs Scale page, there are four surveys listed: (1) The Basic Need Satisfaction at Work Scale (Baard, Deci, & Ryan, 2004; Kasser, Davey, & Ryan, 1992), (2) the Basic Need Satisfaction in Relationships Scale (La Guardia, Ryan, Couchman, & Deci, 2000), (3) the basic Need Satisfaction in Life Scale (Gagné, 2003), and (4) the Basic Need Satisfaction in Physical Education Classes Scale. Each of the items in these surveys are a priori associated with either competence, autonomy, or relatedness and utilize likert-style answers ranging from 1 (not at all true) to 7 (very true). To calculate a score for each construct, the score for each construct's associated items are averaged after the responses to the negatively framed questions are inverted.

The Basic Psychological Needs at Work Scale (Kasser et al., 1992) is the oldest and most commonly used of the basic psychological needs scales (Rochester, 2014). The development of the three more recent scales (listed in the previous paragraph) all leverage the questions from the Basic Need Satisfaction at Work Scale. Table 1 shows a complete list of the items on the Basic Need Satisfaction at Work Scale and their associated constructs. The Basic Need Satisfaction at Work Scale originated from an investigation of discrepancies between employee and supervisors perceptions of the workplace (Kasser et al., 1992). As a part of the investigation, the Work Motivation Form (the original Basic Needs Satisfaction at Work Scale) was created.

**Table 1.** The Basic Needs Satisfaction at Work Scale items and constructs.

| Item # | Construct | Framing | Question |
|---|---|---|---|
| 1 | Autonomy | + | I feel like I can make a lot of inputs to deciding how my job gets done. |
| 2 | Relatedness | + | I really like the people I work with. |
| 3 | Competence | - | I do not feel very competent when I am at work. |
| 4 | Competence | + | People at work tell me I am good at what I do. |
| 5 | Autonomy | - | I feel pressured at work. |
| 6 | Relatedness | + | I get along with people at work. |
| 7 | Relatedness | - | I pretty much keep to myself when I am at work. |
| 8 | Autonomy | + | I am free to express my ideas and opinions on the job. |
| 9 | Relatedness | + | I consider the people I work with to be my friends. |
| 10 | Competence | + | I have been able to learn interesting new skills on my job. |
| 11 | Autonomy | - | When I am at work, I have to do what I am told. |
| 12 | Competence | + | Most days I feel a sense of accomplishment from working. |
| 13 | Autonomy | + | My feelings are taken into consideration at work. |
| 14 | Competence | - | On my job I do not get much of a chance to show how capable I am. |
| 15 | Relatedness | + | People at work care about me. |
| 16 | Relatedness | - | There are not many people at work that I am close to. |
| 17 | Autonomy | + | I feel like I can pretty much be myself at work. |
| 18 | Relatedness | - | The people I work with do not seem to like me much. |
| 19 | Competence | - | When I am working I often do not feel very capable. |
| 20 | Autonomy | - | There is not much opportunity for me to decide for myself how to go about my work. |
| 21 | Relatedness | + | People at work are pretty friendly towards me. |

In the creation of the original Basic Needs Satisfaction at Work Scale, the reliability argument for how items were mapped onto constructs was created through a review of previous literature on SDT. The survey's internal reliability was tested through the calculation of a Cronbach's alpha. While each item was associated with one of the competence, autonomy, and relatedness constructs, the constructs' sub-scores were not designed to be used independently. As Kasser states, "a total-motivation score was computed by averaging the scores on the three motivational subscales, i.e., autonomy, competence, and relatedness (1992, p. 179)." This total-motivation score was then used to predict workplace outcomes. As our literature review will show, later investigations claim to measure competence, autonomy, and relatedness independently. These investigations most often utilized confirmatory factor analyses to support their conclusions (E. L. Deci et al., 2001; Ilardi, Leone, Kasser, & Ryan, 1993).

We examined the literature to determine how surveys that were designed to measure competence, autonomy, and relatedness created their reliability arguments. The primary methods for creating reliability arguments were through reviews of previous literature and factor analyses. Factor analysis is a statistical method used to link a set of variables with underlying constructs. There are two types of factor analyses, exploratory and confirmatory. Exploratory factor analysis is used to reduce the number of variables into as few underlying constructs as possible. In doing so, exploratory factor analysis makes no a priori assumptions about how variables will load onto underlying constructs. Confirmatory factor analysis uses structural equation analysis to test whether a hypothesized model that is based on a priori constructs can account for observed data.

As stated earlier, papers for our literature examination were selected from the Self-Determination Theory website (Rochester, 2014). The SDT website allows users to browse papers which are cross listed under the 23 topical areas. To identify papers of a similar nature to ours, we narrowed our selection to the publication hosted under the header of "Education." Within the Education header, there are articles classified as "overviews" and "research reports." Because we are interested in the methods used in SDT studies, we further narrowed our selection to the "research reports." Of the "research reports" approximately half of the publications were hosted on the SDT website, while the other half either linked to the journal websites or to the authors' email addresses. For reasons of convenience, we further limited our selection to the publications that were hosted on the SDT website (n=101). In order to make the number of papers to review more manageable, we placed the papers in chronological order and reviewed every other one (n=51). This gave us a final set of fifty-one semi-random papers that examine educational settings and are evenly spread across time. The full list of articles reviewed is in the appendix.

We categorized each of the publications into one of three types of classifications: (1) Type I, publications that measured competence, autonomy, and/or relatedness separately and performed a confirmatory factor analysis that included these constructs. (2) Type II, publications that measured competence, autonomy, and/or relatedness separately and performed an exploratory factor analysis that included these constructs. (3) Type III, publications that did not attempt to measure competence, autonomy, and/or relatedness separately. Table 2 shows the number of studies that fell into each classification and their defining features.

**Table 2.** The three publication classifications that emerged from our literature review.

| Classification | Separate measurement of Competence, Autonomy, and/or Relatedness | Validity Argument | | Examples |
| --- | --- | --- | --- | --- |
| | | Factor Analysis | Literature Review | |
| Type I (n=22) | Yes | Confirmatory | Yes | Williams et al., 1997; Wallhead & Ntoumanis, 2004 |
| Type II (n=3) | Yes | Exploratory | Yes | Katz et al., 2009; Niemiec et al., 2006 |
| Type III (n=26) | No | N/A | N/A | Wormington et al., 2011; Valas & Sovik, 1993 |

It was common for both Type I and II publications to use a significant number of surveys, often creating new survey instruments that draw on items from several other studies. For example, Chen and Jang (2010) used selections of ten separate surveys to measure their participants competence, autonomy, relatedness, motivation, engagement, achievement, perceived learning, and course satisfaction. Only one of the ten surveys used in Chen and Jang's study was a previously existing survey, the other nine surveys were created either from scratch or by altering previously existing surveys. Table 3 shows a complete list of constructs, instruments, the development process, and statistical analysis used in Chen and Jang's (2010) publication.

**Table 3.** Survey instruments used in Chen and Jang's publication (2010).

| Construct | Instruments | Development process | Statistical analysis |
|---|---|---|---|
| Competence 1 | Self created | Two open-ended questions were coded and used to develop a 15 item instrument. | Cronbach's Alpha |
| Competence 2 | Intrinsic Motivation Inventory (McAuley et al., 1989) | Only the perceived competency subscale was used and the items were reworded to say "in this online course" instead of "at this task." | Cronbach's Alpha |
| Autonomy 1 | Learning Climate Questionnaire (William & Deci, 1996) | Of the original 15 questions, the 9 items that are "most tied to the autonomy construct" or those that include concrete actions of instructors. | Cronbach's Alpha |
| Autonomy 2 | PE-modified Learning climate questionnaire (Standage et al., 2005) | Where the survey asked about "PE class" they replaced the words "online course." | Cronbach's Alpha |
| Relatedness | Sense of Community (South, 2006) | Nine items were used from the trust, interactivity, and shared values subscales. | Cronbach's Alpha |
| Motivation | Academic Motivation Scale (Vallerand et al., 1992) | Problems were reworded to fit the context, such as replacing "a college of education" with "this online class." | Cronbach's Alpha & Exploratory Factor Analysis |

| Construct (cont.) | Instruments (cont.) | Development process (cont.) | Statistical analysis (cont.) |
|---|---|---|---|
| Engagement | Self created & objective measures | One self-report question and a count of the number of times students accessed an online resource were used. | None |
| Achievement | Self created & objective measures | One self-report question and the students' final grades were used. | None |
| Perceived learning | Perceived Learning Scale (Alavi, 1994) | This instrument was used in its entirety without modification. | Cronbach's Alpha |
| Course satisfaction | Online Course Satisfaction Survey (Hao, 2004) | The 10 items were adapted to fit the research context. | Cronbach's Alpha |

After using Cronbach's alpha scores to assess the reliability of the majority of the surveys, the students' responses were used to assess the fit of six different a priori models through confirmatory factor analyses. Each of these models proposed a set of relationships between the measured constructs and one of six outcome variables (hours per week studying, number of website hits, expected grade, final grade, perceived learning, and course satisfaction). The structural equation modeling provides a partial correlation matrix, path coefficients, and the overall model fit. The path coefficients are used to specify the relationship between constructs and the overall model fit is used to determine the plausibility of the model fitting the data. The model predicting the students' final grades failed to show a statistically significant connection to any of the predictor variables and the models predicting perceived learning and expected grade failed to meet the overall model fit cut-offs. Chen and Jang concluded that while their data affirmed the fit of four of their a priori models, additional investigation should be done to test the validity of alternative models (2010).

Publications that were classified as Type I assumed the independence of the competence, autonomy, and relatedness constructs without performing any type of analysis to test whether the factors should be collapsed into a single construct of basic psychological needs. Because of their use of confirmatory factor analysis for the creation of a reliability argument, Chen and Jan's article was classified as a Type I publication. Of the publications that measured at least one of the constructs competence, autonomy, and relatedness independently (Type I & II), the majority (88%) of them were classified as Type I publications.

Publications classified as Type II were those that measured competence, autonomy, and relatedness separately and performed exploratory factor analyses on these constructs. Of the three publications in our literature review that were classified as Type II, only one of them reported the constructs of competence, autonomy, and relatedness emerging from their exploratory factor analysis (Ommundsen & Kvalø, 2007). The other two Type II studies reported that the three constructs did not emerge independently, but rather that a single basic psychological needs construct emerged (Katz, Kaplan, & Gueta, 2009; Niemiec et al., 2006). The findings of these two exploratory factor analyses are consistent with the indented use of the original Basic Needs Satisfaction at Work Scale survey.

This analysis illustrates the extent to which studies that utilize SDT adapt surveys to create and test models. It also shows that it is a common untested assumption that competence, autonomy, and relatedness are independent constructs. Of the three studies in our literature review that examine the independence of these three constructs, all of them utilized exploratory factor analysis and two of them did not find the three basic psychological needs factors to emerge independently. When modifying a survey (a common practice in SDT studies), one nullifies any existing validity and reliability arguments that are associated with the original survey. Our literature review suggests that it is unclear how independent the constructs of competence, autonomy, and relatedness are and led to our second research question.

## Methods and Data

*Setting*

Data for this study were collected in a high school AP physics class. The teacher has a background in biology, including a Ph.D. in biochemistry. Like the majority of US high school teachers of physics, the teacher did not have a degree in physics (Hodapp, Hehn, & Hein, 2009). The teacher in this study worked in a public high school located in an urban community that was primarily composed of students from non-dominant backgrounds. Table 4 shows the school demographics.

**Table 4**. School demographics

| Ethnicity | | | | Free or reduced lunch | ESL/FEP | IEP |
|---|---|---|---|---|---|---|
| Hispanic | White | Asian | Afr. Am. | | | |
| 56% | 32% | 8% | 3% | 41% | 49% | 11% |

The teacher was recruited for the research project because of her role as a Streamline to Mastery teacher (M. Ross, Van Dusen, & Otero, 2014; M. Ross, Van Dusen, Sherman, & Otero, 2011; Van Dusen, Ross, & Otero, 2012). Streamline to Mastery is an NSF-funded teacher-researcher community that collaborates with our university to engage in educational research as a mechanism for professional development. To be eligible for the Streamline to Mastery program, teachers must have a masters degree, work in a school district that has been identified as largely serving students from non-dominant backgrounds, and undergo an application process that focuses on their desire to improve their teaching practices through self investigation. The cornerstone of the Streamline to Mastery program is using research as a mechanism for generating principles about effective teaching and learning. For her research, the teacher in this study examined how providing her high school students opportunities to act as Learning Assistants (Otero et al., 2010) for elementary school students affected their content learning (Nicholson-Dykstra et al., 2013). This experience led the teacher to continue to focus on creating teaching-to-learn opportunities for her students.

The school experienced a significant transient student population, so class sizes were not constant. However, the AP physics class retained the majority of the 30 students that began the year in the class. Of those 30 students, this study examines the work of the 27 students that completed the class.

*Data*

Our analysis utilized two types of data sources: outcome variables and environment (mediating) variables.

*(1) Outcome variables*. Our first outcome variable is the students' perceived changes in interest in learning physics. Their perceived changes in interest in learning physics were measured by asking the students how their current interests in learning physics compared to their interests in learning physics at the start of the course. Students' responses (n=22) were coded as either increasing, staying constant, or decreasing.

Our second outcome variable is the students' course grades. Students' course grades (n=27) were calculated by averaging each individual student's final score in the class at the end of the first and second semester.

(2) *Environment (mediating) variables.* Our mediating variables were designed to reflect student experiences of competence, autonomy, and relatedness. Student competence, autonomy, and relatedness were measured using a survey that was a modified version of the Basic Need Satisfaction at Work Scale (Baard et al., 2004; E. L. Deci et al., 2001; Ilardi et al., 1993) that was outlined in the literature review. Our modification of the survey was developed to reflect the fulfillment of students' basic psychological needs while engaged in classroom physics activities rather than work activities. Table 5 shows the questions from the Basic Need Satisfaction at Work Scale and our modified survey as well as the hypothesized construct (competence, autonomy, or relatedness) associated with each item on the original survey (Baard et al., 2004; E. L. Deci et al., 2001; Ilardi et al., 1993).

See the appendix for descriptive statistics of students' survey results (Table 11) and reports of increasing or decreasing interest in learning physics (Table 12).

**Table 5.** Survey items from the Basic Needs Satisfaction at Work Scale and our modified survey.

| Item # | Const. | Basic Needs Satisfaction at Work Scale | Modified Survey for Physics Classroom |
|---|---|---|---|
| 1 | Autonomy | I feel like I can make a lot of inputs to deciding how my job gets done. | I felt like I was free to decide how to do the problem(s). |
| 2 | Relatedness | I really like the people I work with. | I got to positively interact with my peers. |
| 3 | Competence | I do not feel very competent when I am at work. | I felt like I was not good at doing the problem(s). |
| 4 | Competence | People at work tell me I am good at what I do. | People told me I was good at what I was doing. |
| 5 | Autonomy | I feel pressured at work. | I felt pressured to do the problem(s). |
| 6 | Relatedness | I get along with people at work. | I got along with the people I talked with. |

| Item # (cont.) | Const. (cont.) | Basic Needs Satisfaction at Work Scale (cont.) | Modified Survey for Physics Classroom (cont.) |
|---|---|---|---|
| 7 | Relatedness | I pretty much keep to myself when I am at work. | I pretty much kept to myself and did not have a lot of social contact. |
| 8 | Autonomy | I am free to express my ideas and opinions on the job. | I generally felt free to express my ideas and opinions. |
| 9 | Relatedness | I consider the people I work with to be my friends. | I consider the people I interacted with to be my friends. |
| 10 | Competence | I have been able to learn interesting new skills on my job. | I was able to learn interesting new skills. |
| 11 | Autonomy | When I am at work, I have to do what I am told. | I had to do what I was told. |
| 12 | Competence | Most days I feel a sense of accomplishment from working. | I felt a sense of accomplishment from what I did. |
| 13 | Autonomy | My feelings are taken into consideration at work. | The people I interacted with took my feelings into consideration. |
| 14 | Competence | On my job I do not get much of a chance to show how capable I am. | I did not get a chance to show how capable I am. |
| 15 | Relatedness | People at work care about me. | People I interacted with care about me. |
| 16 | Relatedness | There are not many people at work that I am close to. | I did not get to interact with people that I'm close to. |
| 17 | Autonomy | I feel like I can pretty much be myself at work. | I felt like I could be myself. |
| 18 | Relatedness | The people I work with do not seem to like me much. | The people I interacted with do not seem to like me much. |
| 19 | Competence | When I am working I often do not feel very capable. | I did not feel very capable. |

| Item # (cont.) | Const. (cont.) | Basic Needs Satisfaction at Work Scale (cont.) | Modified Survey for Physics Classroom (cont.) |
|---|---|---|---|
| 20 | Autonomy | There is not much opportunity for me to decide for myself how to go about my work. | There was not much opportunity for me to decide for myself how to do the problem(s). |
| 21 | Relatedness | People at work are pretty friendly towards me. | People were generally pretty friendly towards me. |

*Relevant Classroom Activities*

Students were asked to complete the modified survey after two different teacher-initiated physics problem solving activities that took place during the second semester. The first activity was the completion of physics problems on whiteboards. When creating solutions on whiteboards each student would work with a group of four or five peers to jointly create a problem solution that would be shared with the rest of the class. The second activity was the students' creation of physics problem solutions using screencasts (Nicholson-Dykstra et al., 2013; Van Dusen & Otero, 2012). Screencasting is a technology that captures a video of the iPad's screen while using the microphone to capture and merge audio to that file. Each student worked individually to create a screencast that solved a problem from the back of the physics book or a worksheet. The screencast solutions were uploaded to the class's website where they could be viewed by fellow students.

*Analysis Part I*

*Determining Survey Constructs.* To test for the existence of underlying constructs, students' survey results were analyzed using exploratory factor analysis. Principal axis factoring was used to reduce the number of variables in our modified survey. The number of responses from the two administrations of the survey totaled to thirty-eight. Due to our small sample size (n=38), we only used questions that were shown to have a loading value above 0.6 in order to ensure the strength of our factors (de Winter, Dodou, & Wieringa, 2009). This led to the removal of five questions that did not have sufficiently strong loading values for any individual factor. Due to the potential interrelated nature of the constructs (competence, autonomy, and relatedness) direct oblimin rotation was used (de Winter et al., 2009). The oblimin rotation does not assume that the constructs are orthogonal and therefor allows the factors to be correlated. Three factors emerged from our analysis that represent the students' self reported feelings when creating physics problem solutions. Cronbach's alphas were calculated for each factor to assess their internal reliability.

The factors from our exploratory factor analysis were used to create a set of factor scores for each student. These scores were calculated using a combination of the students' survey responses and the factor loading values for each item. The students' factor scores represent the number of standard deviations each student is above or below the class mean on each factor.

*Analysis Part II*

*Testing Causal Links.* We use multiple linear regression (MLR) to build models that predict either continuous (semester grades) or categorical (interest in learning physics) outcomes through the analysis of continuous input variables (factors analysis scores). Our initial MLR models used the three factors that emerged from the factor analysis as well as the interactions between them (Factor1xFactor2, Factor1xFactor3, Factor2xFactor3, and Factor1xFactor2xFactor3) as the input variables. Our models were further refined using a backward elimination technique. We engaged in an iterative process in which the factors with the highest p-values were removed and the models were rerun. This process was repeated until all of the remaining factors, if any, had a p-value less than 0.05. Each of these models met the assumptions of linearity, independence of errors, homoscedasticity of errors, and normality of error distribution.

Once the final models were completed, we focused on the impact that each input variable had on the outcome variable by examining the input variables' effect sizes (odds ratios) of the coefficients ($\beta$). When using the expression for linear regression to describe the probability function, the odds ratio can be expressed as $e^\beta$, which gives the relative effect of a single variable to the prediction of an outcome variable. For example, consider our factors where each student's score represents how many standard deviations the student scored above or below the class average for that construct. If a factor's $\beta=0.395$, then it's odds ratio is $e^{0.395}=1.48$. This means that for every standard deviation a student scores higher on that factor, we can predict that their final grade in the class will be 48% higher than the class average. All of our statistical analyses were completed using SPSS v20.

*Model*

The results of Part I and II of our analyses were combined to create a model of how classroom activities lead to improved outcomes. Figure 3 shows our Input-Environment-Outcome (IEO) model that was expanded using our analysis results. In this model, the Input factors are the classroom activities, the Environment (mediating) factors are the constructs that emerge from the exploratory factor analysis of our modified basic psychological needs survey, and the Outcome factors are the classroom outcomes.

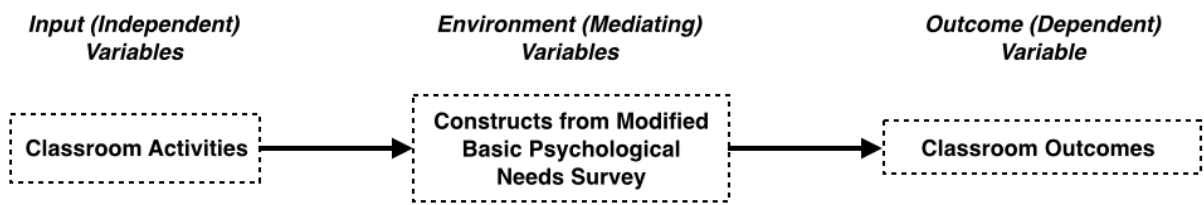

**Figure 3.** Our initial IEO model of student physics classroom experiences.

*Limitations*

A limitation in our analysis is our sample size. As previously stated, in order to account for our limited sample size and ensure the strength of our factors, we used a higher cut off (0.6) for the loading value of each item. With this constraint in place, we met the required assumptions for factor strength (de Winter et al., 2009). Our higher loading values, levels of communality, and lack of cross-loading lead us to conclude that despite our smaller sample size, there are useful conclusions that can be drawn from this data that are relevant to the education community.

An additional limitation of our analysis is the potential conflation of our environment variables with our outcome variables. For example, rather than being driven by the constructs measured in our modified basic psychological needs survey, it may be that our outcome variables (semester grades and changes in interest in learning physics) are simply an extension of these constructs. In our multiple linear regression analyses we presuppose that students' semester grades and changes in interests in learning physics are outcomes that are mediated by the fulfillment of students' basic psychological needs.

Finally, in addition to making small changes to the wording of the questions in our survey, our likert-style scales answers of "not at all true" and "very true" ranged from one to five instead of one to seven, like the original survey.

## Results

**Findings Part I**

*Exploratory Factor analysis of Modified Survey*

We examined the twenty-one questions from the modified Basic Needs Satisfaction in General survey for underlying constructs using Exploratory Factor Analysis (EFA). Three factors emerged from this analysis. As described in our methods, because of our smaller sample size (n=38) we used a factor loading cut-off score of 0.6. Using the principle axis factoring extraction method, we removed five questions (7, 8, 11, 17, and 18) which did not load strongly enough onto any of the three factors. With these question removed the EFA was re-run with the sixteen remaining questions. In the second, and final, EFA, sampling accuracy was confirmed with a Kaiser-Meyer-Olkin (KMO) score of .707 and the correlations between individual questions were also confirmed through a significant Bartlett's test of sphericity, $c^2(91) =328.772$, $p<.0001$. Each construct has at least four items that load at or above the 0.6 thresholds. Only one item (#13) cross-loaded onto more than one factor. The average communality score for the items is 0.707, indicating that over 70% of the variance in our variables are accounted for in our three factors. Together the three factors accounted for 63.9% of the variance. Internal reliability is an estimate of how consistent a set of questions are and is typically measured using a Cronbach's alpha score. It is commonly accepted that an alpha>0.7 is adequate for internal reliability (Cortina, 1993). Our three factors yielded alpha scores of 0.832, 0.728, and 0.800 respectively. The factor analysis results are presented in Table 6.

**Table 6.** Factor analysis results (bold indicates factor loading values above 0.6).

| Item # | Construct | Factor 1 | Factor 2 | Factor 3 | Communality | Question wording |
|---|---|---|---|---|---|---|
| 6 | Relatedness | **0.894** | -0.166 | -0.210 | 0.824 | I got along with the people I talked with |
| 9 | Relatedness | **0.874** | 0.070 | -0.306 | 0.780 | I consider the people I interacted with to be my friends |
| 2 | Relatedness | **0.859** | -0.037 | -0.312 | 0.741 | I got to positively interact with my peers |
| 21 | Relatedness | **0.806** | -0.057 | -0.441 | 0.761 | People were generally pretty friendly towards me |
| 16 | Relatedness | **0.727** | 0.363 | -0.492 | 0.735 | I did not get to interact with people that I'm close to |
| 10 | Competence | **0.660** | 0.036 | -0.130 | 0.464 | I was able to learn interesting new skills |
| 19 | Competence | 0.068 | **0.784** | -0.012 | 0.661 | I did not feel very capable |
| 3 | Competence | -0.052 | **0.777** | -0.152 | 0.733 | I felt like I was not good at doing the problem(s) |
| 4 | Autonomy | -0.207 | **0.671** | -0.080 | 0.684 | I felt pressured to do the problem(s) |
| 15 | Competence | -0.078 | **0.640** | 0.019 | 0.639 | I did not get a chance to show how capable I am |
| 1 | Autonomy | 0.069 | **0.632** | -0.095 | 0.741 | I felt like I was free to decide how to do the problem(s) |

| Item # (cont.) | Construct (cont.) | Factor 1 (cont.) | Factor 2 (cont.) | Factor 3 (cont.) | Communality (cont.) | Question wording (cont.) |
|---|---|---|---|---|---|---|
| 20 | Autonomy | 0.182 | **0.625** | -0.240 | 0.513 | There was not much opportunity for me to decide for myself how to do the problem(s) |
| 14 | Relatedness | 0.476 | -0.077 | **-0.834** | 0.784 | The people I interacted with took my feelings into consideration |
| 12 | Relatedness | 0.530 | -0.208 | **-0.792** | 0.805 | People I interacted with care about me |
| 5 | Competence | 0.196 | 0.184 | **-0.791** | 0.641 | People told me I was good at what I was doing |
| 13 | Competence | 0.009 | **0.602** | **-0.707** | 0.799 | I felt a sense of accomplishment from what I did |
| Eigenvalue | | 5.209 | 3.602 | 2.786 | | |
| Cronbach's Alpha | | 0.832 | 0.728 | 0.800 | | |
| Variance | | 32% | 22% | 10% | | Total: 63.9% |

Structure Matrix

Extraction Method: Principal axis factoring.

Rotation Method: Oblimin with Kaiser Normalization.

    Table 6 shows that each of the three emergent factors has a blend of competence, autonomy, and relatedness. These blends can be seen by matching the a priori constructs listed in the second column of Table 6 with the bolded loading values for the factors shown in column 3-5. Specifically, Factor 1 is primarily relatedness with some competence, Factor 2 is a blend of competence and autonomy, and Factor 3 is a blend of competence and relatedness.

The three factors that emerge from this analysis allow us to begin to flesh out our Input-Environment-Outcome (I-E-O) model, shown in Figure 4. In this model, the input variables are our two classroom activities (creating solutions to physics problems on screencasts and whiteboards) and the environment variables are the three factors that emerged from the analysis of the students' survey results. In the next section we will use regression analysis to explore how the environment variables (student feelings) relate to the outcome variables (classroom outcomes).

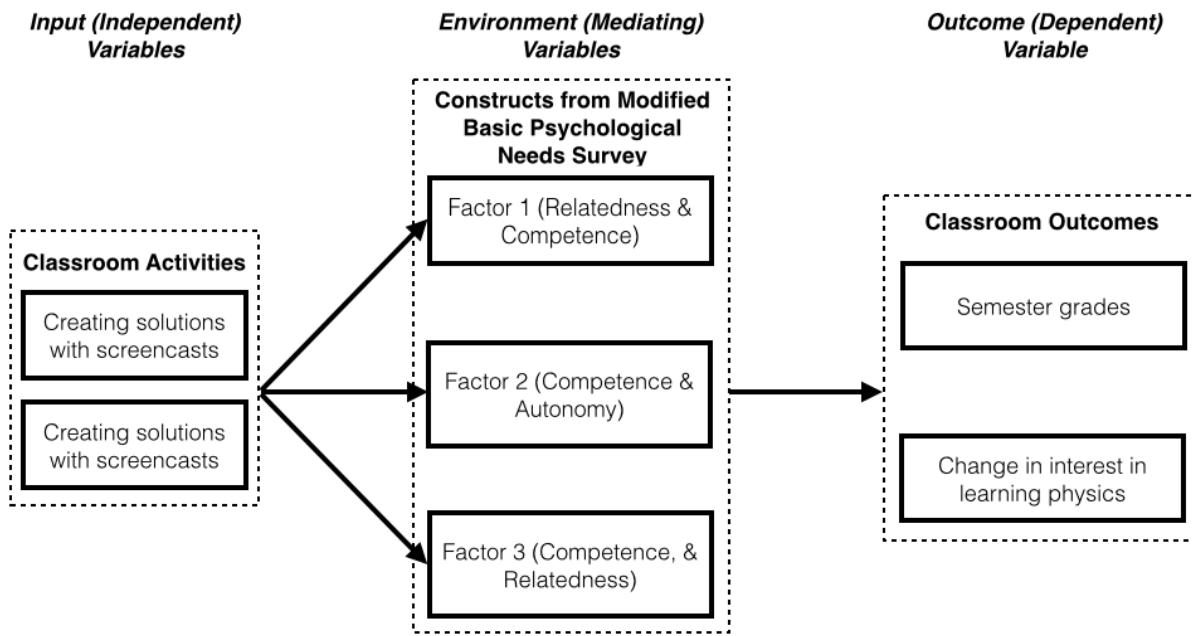

**Figure 4.** Our I-E-O model with the identified input, environment, and outcome variables.

**Findings Part II**

*Regression analysis*

Multiple linear regression (MLR) was used to test the relationship of the three factors to the classroom outcome items (semester grades and changes in student interest in learning physics). In order to identify the factors involved in each of the outcome variables, a two-step procedure was followed. The first step was to perform the MLR with each of the three factors identified in our exploratory factor analysis as well as their interactions. If not all of the factors had a p<0.05 then the second step was to remove the factor with the highest p-value and re-run the analysis. This procedure was repeated until all of the remaining factors had a p<0.05.

*Semester grades.*

Our first MLR examined the influence of all three factors and their interaction effects on the students' semester grades. Table 7 shows the results of our initial regression analysis for semester grades.

**Table 7.** Initial regression model of our emergent factors' effects on student semester grades.

|  | β | S.E. | T-value | p-value | Exp(β) | Odds ratio |
|---|---|---|---|---|---|---|
| Factor 1 (R & C) | -0.06 | 2.415 | -0.24 | 0.813 | 0.942 | 1.062 |
| Factor 2 (C & A) | 0.296 | 2.283 | 1.086 | 0.288 | 1.344 | 1.344 |
| Factor 3 (A & R) | 0.093 | 1.763 | 0.502 | 0.62 | 1.097 | 1.097 |
| Factor 1 x Factor 2 | -0.061 | 1.837 | -0.222 | 0.827 | 0.941 | 1.063 |
| Factor 1 x Factor 3 | 0.042 | 2.669 | 0.134 | 0.894 | 1.043 | 1.043 |
| Factor 2 x Factor 3 | -0.19 | 2.361 | -0.713 | 0.483 | 0.827 | 1.209 |
| Factor 1 x Factor 2 x Factor 3 | 0.4 | 1.687 | 1.251 | 0.223 | 1.492 | 1.492 |

Our initial regression model (Table 7) was refined by removing the factor with the largest p-value (Factor 1 x Factor 2) and rerunning the model. This process was repeated six times such that all of the remaining factors had p-values<0.05. Table 8 shows the final MLR model that emerged from the repeated process of eliminating factors with high p-values.

**Table 8.** Final regression model of our emergent factors' effects on student semester grades.

|  | β | S.E. | T-value | p-value | Exp(β) | Odds ratio |
|---|---|---|---|---|---|---|
| Factor 2 (C & A) | 0.395 | 0.153 | 2.58 | 0.014 | 1.48 | 1.48 |

This model shows that students with higher Factor 2 (competence & autonomy) scores were significantly more likely to have higher semester final grade. Specifically, analyses indicated that students who posted a one standard deviation higher Factor 2 score had, on average, 47% higher semester final grades.

*Changes in Interest in Learning Physics*

Our second MLR examined the influence of all three factors and their interaction effects on the changes in student interest in learning physics. Table 9 shows the results of our initial regression analysis for changes in student interest in learning physics.

**Table 9.** Initial regression model of our emergent factors' effects on student interest in learning physics.

|  | β | S.E. | T-value | p-value | Exp(β) | Odds ratio |
|---|---|---|---|---|---|---|
| Factor 1 (R & C) | -0.202 | 2.415 | -0.852 | 0.401 | 0.817 | 1.224 |
| Factor 2 (C & A) | 0.524 | 2.283 | 2.334 | 0.027 | 1.689 | 1.689 |
| Factor 3 (A & R) | 0.183 | 1.763 | 1.058 | 0.298 | 1.201 | 1.201 |
| Factor 1 x Factor 2 | 0.121 | 1.837 | 0.509 | 0.614 | 1.129 | 1.129 |
| Factor 1 x Factor 3 | -0.156 | 2.669 | -0.538 | 0.594 | 0.856 | 1.169 |
| Factor 2 x Factor 3 | -0.008 | 2.361 | -0.035 | 0.972 | 0.992 | 1.008 |
| Factor 1 x Factor 2 x Factor 3 | -0.234 | 1.687 | -0.926 | 0.362 | 0.791 | 1.264 |

This model was refined using the same technique as the previous MLR model. After six iterations of removing the factors with the largest p-values a final model emerged with all p-values<0.05. Table 10 shows the final MLR model that emerged from the repeated refinement process.

**Table 10.** Final regression model of our emergent factors' effects on student interest in learning physics.

|  | β | S.E. | T-value | p-value | Exp(β) | Odds ratio |
|---|---|---|---|---|---|---|
| Factor 2 (C & A) | 0.556 | 0.152 | 3.665 | 0.001 | 1.74 | 1.74 |

This model shows that students with higher Factor 2 (competence & autonomy) scores were significantly more likely to have increases in their interest in learning physics. Specifically, analyses indicated that students who reported a one standard deviation higher Factor 2 score had a 74% increase in their likelihood of reporting an improvement in their interest in learning physics.

This analysis shows that while three factors relating to students experiences were present when students engaged in solving problems, only one of those factors was a statistically significant predictor of classroom outcomes (semester grades and interest in learning physics). As Figure 5 shows, higher Factor 2 scores where positively correlated with both student semester grades and changes in student interest in learning physics. The implications of our findings and model are discussed in the following section.

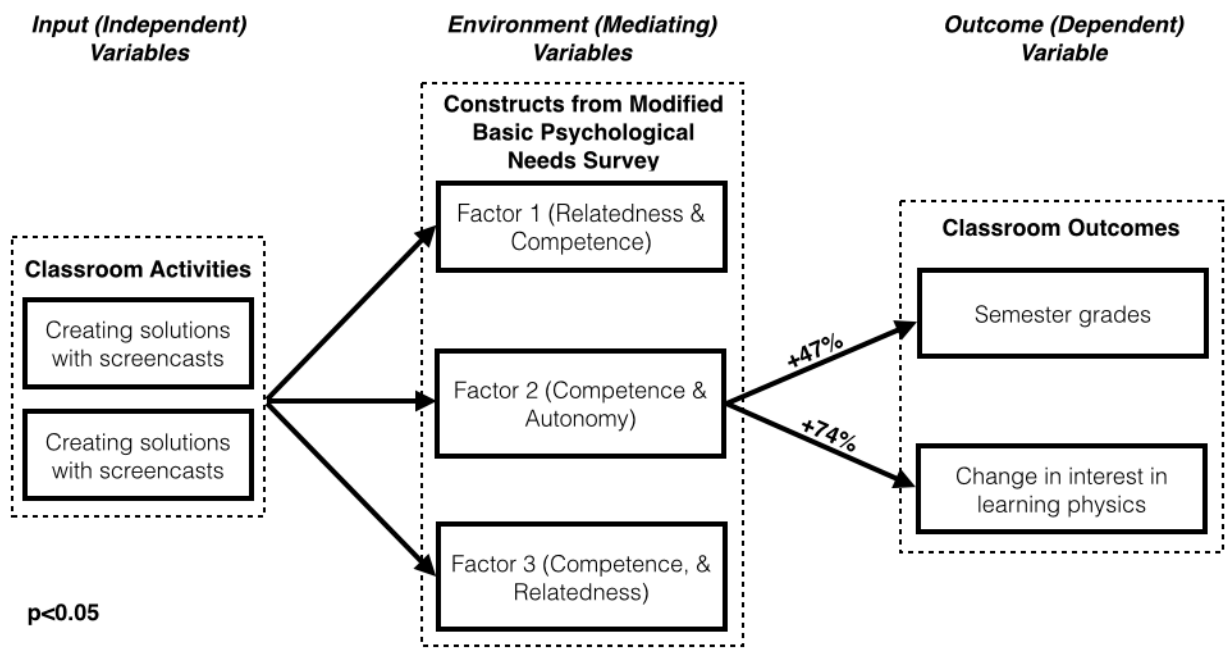

**Figure 5.** Our final I-E-O model with regression analysis data.

Although our exploratory factor analysis did not find the constructs of competence, autonomy, and relatedness to be distinguishable from one another, the items that were developed to measure these constructs were indeed found to load onto three distinct factors. Factor 1 was found to be a blend of (our modified) items that were *a priori* assigned to the constructs of relatedness and competence by the Basic Need Satisfaction at Work Scale. Factor 2 was found to be a blend of the items assigned to the constructs of competence and autonomy, and Factor 3 was found to be a blend of the items assigned to the constructs of competence and relatedness. Furthermore, our regression analysis showed students' Factor 2 scores to be predictive of both students' final semester grades *and* students' increased interests in learning physics.

**Discussion**

The finding that factor 2 predicted classroom grades and self-reported interest outcomes leads to the conclusion that some blend of competence and autonomy, or perhaps a different construct altogether, is related to both students' interest and performance. However, by observing the specific items that loaded onto factor 2, we see that a majority of them were stated in the negative, for example, "I did not feel very capable" and "I felt pressured to do the problem." The only positively stated item that loaded onto factor 2 reads, "I felt that I was free to decide how to do the problem." While our data is not sufficient to determine if this clustering of negatively framed items is a random occurrence, one can imagine potential causes for this grouping. For example, it could be that what matters is not the presence of positive feelings of competence and autonomy, but rather the absence of negative feelings of competence and autonomy.

Also a comparison of the items that are associated with relatedness that loaded onto Factor 1 to those that loaded onto Factor 3 reveals that those that loaded onto factor 1 tend to attribute external outcomes and contexts to the self. The items associated with relatedness that loaded onto Factor 3 tend to attribute external outcomes to others. This is particularly interesting because the theoretical perspective of Self-Determination Theory is founded on a model of identity and motivation that posits that people progress from an external to an internal perceived locus of causality. This observation has led us to question the specific constructs of competence, autonomy, and relatedness. Perhaps, the affective constructs that are operating in classroom contexts are somewhat or entirely different from those theorized in the literature.

Our original emergent constructs of play, agency, and social belonging were derived from a small study of high school physics classroom interactions involving the iPad. Our original iPad intervention was driven by our desire to create a classroom environment in which students felt good about themselves rather than being in survival mode (Ames, 1992). Findings from our factor analysis yielded three factors, the Factor 1 is comprised of items that both addressed the survey respondent in a relatively positive way *and* these questions attributed external outcomes to the self. Factor 2 was comprised of questions that addressed the survey respondent in a relatively negative way and were a mix of internal and external attribution. The items that loaded onto Factor 3 seem to address the survey respondent in neither a positive or negative way, but they also seemed to attribute external outcomes to others. Our analysis did not support the hypothesis that (a) competence, autonomy, and relatedness can be measured independently, assuming that the items appropriately measure these constructs or (b) that the items that were chosen to represent these constructs did, in fact, measure these constructs. These finding are consistent the findings of the majority of the studies in our literature review that were classified as Type II (having performed exploratory factor analysis).

Though the constructs of competence, autonomy, and relatedness were not distinguishable, it is notable that these highly identity based constructs, which address individuals' feelings about themselves in the world, impact learning outcomes. Our analyses shows that Factor 2, which is a combination of competence and autonomy items, predicted learning outcomes as measured by semester grade and interest in learning physics.

The constructs of competence, autonomy, and relatedness (or closely related variations on them) have emerged as the central focus of a range of research agendas, including our qualitative studies (Van Dusen & Otero, 2011, 2014), work on Self-Determination Theory (Gagne & Deci, 2005; R Ryan & Deci, 2000), science identity (Carlone & Johnson, 2007), and flow (Csikszentmihalyi & LeFevre, 1989; Csikszentmihalyi, 1990). All of these types of investigations examine people's beliefs about themself in an environment. We propose that beliefs about one's competence, autonomy, and relatedness are part of a dynamic feedback loop within the interactions of a person and his or her surroundings. The environment and a person in that environment can both change rather dramatically as the individual interprets this feedback.

The dynamic, integrated nature of a person's interactions with an environment may explain why the constructs of competence, autonomy, and relatedness appear to be intertwined. For example, a person's feelings of competence in a physics class may emerge from their assessment of their own mastery of physics content knowledge in comparison to their fellow classmates. In this way, a person's belief about their personal ability to perform a task (competence) may be inseparable from his or her relationships with their peers (relatedness).

The entanglement of these constructs is consistent with Vygotsky's views of learning. Similar to Self-Determination Theory's envisioning of the internalization of activities into one's identity, Vygotsky postulated that all learning was the internalization of social activities (L S Vygotsky, 1978). Using this lens, it is impossible to reflect one's levels of competence and autonomy without framing them in a larger social context. While Self-Determination Theory has traditionally taken a cognitive perspective on human behavior and motivation, we believe that it may be productive to examine the tenets of Self-Determination Theory using a more socio-cultural lens.

**Future Work**

    Future work includes using the students' interest and performance variables as inputs for each other. This would help determine in what ways students' perceptions of increased interests and test performances drive one another.

    Future work also includes performing a qualitative validity analysis of students' responses to specific survey items to determine how the questions are being interpreted. By performing and analyzing the results of student think-alouds, we hope to glean what themes are associated with the items that load onto the three emergent factors from our exploratory factor analysis. This will help us better understand these constructs as well as potentially providing insight into the widely sought after construct of identity.

**Appendix:**

Literature review articles

[1]   C. Benware and E. Deci, "Quality of learning with an active versus passive motivational set," *Am. Educ. Res. J.*, vol. 21, no. 4, pp. 755–765, 1984.

[2]   R. J. Vallerand, M. R. Blais, N. M. Briere, and E. T. Luc, "Construction et validation de l'echelle de motivation en education ( EME )," *Can. J. Behav. Sci.*, vol. 21, no. 3, 1989.

[3]   R. M. Ryan, J. P. Connell, and R. W. Plant, "Emotions in nondirected text learning," *Learn. Individ. Differ.*, vol. 2, no. 1, pp. 1–17, Jan. 1990.

[4]   R. Ryan and C. Powelson, "Autonomy and relatedness as fundamental to motivation and education.," *J. Exp. Educ.*, vol. 60, no. 1, pp. 49–66, 1991.

**Table 11.** Modified Basic Needs Satisfaction survey items averages.

| Item # | A prior construct | Average score |
|--------|-------------------|---------------|
| 3      | Competence        | 2.61          |
| 5      | Competence        | 2.63          |
| 10     | Competence        | 3.45          |
| 13     | Competence        | 3.87          |
| 15     | Competence        | 2.29          |
| 19     | Competence        | 2.45          |
| **Ave.** | **Competence**  | **2.88**      |
| 1      | Autonomy          | 4.11          |
| 4      | Autonomy          | 2.37          |
| 8      | Autonomy          | 4.16          |
| 11     | Autonomy          | 3.45          |
| 14     | Autonomy          | 3.61          |
| 17     | Autonomy          | 4.34          |
| 20     | Autonomy          | 2.18          |
| **Ave.** | **Autonomy**    | **3.46**      |
| 2      | Relatedness       | 4.13          |
| 6      | Relatedness       | 4.24          |
| 7      | Relatedness       | 1.95          |
| 9      | Relatedness       | 4.16          |
| 12     | Relatedness       | 3.68          |
| 16     | Relatedness       | 2.11          |
| 18     | Relatedness       | 1.45          |
| 21     | Relatedness       | 4.47          |
| **Ave.** | **Relatedness** | **3.27**      |
| **Ave.** | **Total**       | **3.22**      |

**Table 12.** Student reports of changes in their interest in learning physics (n=22).

|  | Increase | No change | Decrease |
|---|---|---|---|
| Interest in learning physics (n=22) | 27% | 36% | 36% |

**CHAPTER 6**

**THE ROOTS OF PHYSICS STUDENTS' MOTIVATIONS: FEAR AND INTEGRITY**

Ben Van Dusen and Valerie Otero

*School of Education, University of Colorado, Boulder, 80309, USA*

This study utilizes a sociocultural interpretation of Self-Determination Theory to better understand the role that learning contexts play in creating student motivation, engagement, and identity. By drawing on previous work that examined motivation we develop a model that describes how students' senses of belonging in social settings can transform their goals and their experiences. We use the ideas of fear and integrity to understand students' motivations to engage in activities. A student's sense of connection and belonging, or not, in a social setting also a drives whether she experiences competence as self-esteem maintenance or self-improvement and autonomy as alienation or innovation. These findings stress the importance of feeling a sense of belonging and that this is achieved through alignment of the goals and practices of the individual and an activity. This model is applied to three classroom examples to illustrate how feelings of social connection and isolation can be exhibited in a physics classroom setting. We conclude by discussing physics learning environment design principles that foster feelings of connection for *all* students.
**PACS**: 01.40.ek, 01.40.Fk, 01.40.Ha

## Introduction

Our work is driven by a desire to create physics learning environments that are motivating and engaging for high school students. All too often high school physic classes leave students feeling uninterested and disconnected from the practices of science. In order to create positive student learning experiences, we have examined how to construct physics learning environments that foster students' feelings of personal investment and motivation. In an investigation of high school physics students' experiences, Ross and Otero (2012) and Ross (2013) describe two competing narratives that high school students used to describe their past and present high school experiences in science. One narrative can be characterized as that of *fear of failure* and *preservation of self-esteem*. Students used terms such as "afraid, scared, judged, stupid, boring, gullible" and "looked down upon" when they talked about their experiences in science class. In contrast, when describing their experiences in a classroom environment that paid special attention to students development and defending of science ideas, students used terms consistent with *integrity* and *self-expression* such as "comfortable, interested, evidence, it's okay, legit, help each other, share," and "we have the answers." In this paper we further explore these two extremes in efforts of developing a theoretical perspective on identity. We argue that the differences between these two extremes have to do with the extent to which the students feel in control of and integrated with their classroom science activities and their peers.

Ames (1992) outlined a set of students' classroom motivational dispositions similar to the findings of Ross (2013), in terms of students' *goals*. She argued that students either engage in classroom activities through externally motivated goals associated with one's self worth, which she referred as *performance goals* or out of internally motivated goals of self-expression and inventiveness, which she referred to as *mastery goals*. Ames used the idea of *performance* versus *mastery goals* to argue that it is the characteristics of classrooms, not characteristics of students, which increased the likelihood that students will engage in performance goals (protecting their self-worth) or mastery goals (self-expression and inventiveness). Using prior studies as examples, she demonstrates how these goals determine the quality of involvement of students in class, which greatly impacts students' efforts and outcomes. In our own terms, Ames (1992) showed that outcomes were largely attributable to whether the classroom context engendered an orientation toward *fear and self-preservation* or whether these contexts engendered an orientation toward *integrity and self-expression*—these are phrases we use throughout this paper. Ames emphasizes that the constructs of motivation and goals are determined by the nature of the context and how people see themselves in relation to that context and the people in it. We use Ames's work in relation to other work in the area of classic motivation theory (Ryan & Deci, 2000; Ryan, 1995) to develop a sociocultural perspective on identity and its relation to student performance in physics.

Other sociocultural researchers have worked toward perspectives on learning using terms such as "agency," "identity," and "culture," in attempts to establish a theory of student learning as a function of sociocultural factors (Barton & Tan, 2010; Basu, Barton, Clairmont, & Locke, 2008; Brown, 2006; Holland, Lachichotte, Skinner, & Cain, 1998; Penuel & Wertsch, 1995). However, in each of these cases crucial terms that were used, such as "identity" and "agency," were not operationalized or no mechanism for their development was provided to the extent that these perspectives on identity could be easily utilized in other contexts.

The challenge of characterizing how learning environments interact with the hearts, minds, and social and cultural histories of students has been the subject of investigation since physics entered the high school curriculum in the late 1800s. During the "New Movement Among Physics Teachers" that began in 1906, university physics professors and physics high school teachers worked together to change the way in which students' experienced physics in the high school classroom (C. R. Mann & Twiss, 1910; C. R. Mann, 1909a, 1909b; Millikan & Gale, 1906; Twiss, 1920). In 1914 physicist, Charles Mann concluded that the classroom had become a place in which teachers try to "impose [the scientific] order of thought on our pupils with the idea that we were thereby serving science." He went on to say that, "We have failed because the essence of the spirit we want is not of this sort. The essence of the scientific spirit is an emotional state, an attitude toward life and nature, a great instinctive and intuitive faith" (Mann, 1914, p. 518). Although physicists and physics teachers wanted to understand how to create an environment that fostered what Mann (1909) referred to as *the spirit of science,* what Ames (1992) referred to as mastery learning goals, or what we refer to as *integrity and self-expression,* the work of the early physicists did not quite achieve this—nor did the work of later physicists or science educators. We have also struggled with the terminology and methodologies for describing and studying students' interactions in learning contexts. The purpose of this paper is to establish a language and a theoretical perspective for research and implementation of learning contexts that engage the hearts and minds of students.

    Our own previous studies in this area have focused on defining the characteristics of specific learning environments that appear to lead to increased social connections and motivations. We have made conjectures about the mechanisms that create motivation, and measured improved outcomes that emerge from these settings. Van Dusen & Otero (2012) examined how an urban, high school physics class responded to the inclusion of a classroom set of iPads and associated applications, such as screencasting. This study was exploratory in nature

and was designed to identify how students integrated themselves, iPads, and the physics classroom. Ethnographic field notes and student survey responses led to the observation of three general trends in students interactions with iPads, screencasting, and physics: increased *social status*, opportunities to engage in *play*, and a sense of *agency* in their work. These findings led us to further examine how these three factors (social status, play, agency) were related to students' motivation to engage in physics. We hypothesized that the iPad's screencasting technology combined with the digital sharing of assignments with classmates allowed the students to blend their peer-cultural practices with the cultural practices of the physics classroom and the physics community. We further hypothesized that by drawing on students' "sense of self," the iPad/screencasting activity was increasing the motivation of students to solve physics problems and would therefore lead to higher scores. By comparing students' physics problem solutions using screencasts with their physics problem solutions using traditional pen and paper notebooks, we found that students' screencasting solutions were more complete and correct than their notebook problems. We also found that students' actions in screencasts exhibited significant opportunities for authorship and social interactions. These findings were bolstered by students' survey responses, which indicated a sense of autonomy and social connectedness when doing their screencasts. This led us to deepen our hypothesis, to include that it is students' sense of social belonging that leads to increased motivation, ultimately improving the quality of the their physics solutions.

These findings led us to further examine the literature on motivation. *Self-Determination Theory* (Deci & Eghrari, 1994; Ryan & Deci, 2000) emerged as a theory that offered a potential explanation of how our physics students' motivations and identities shifted over time. Self-Determination Theory states that changes in identity and motivation are linked to a person's sense of competence (feelings of success), autonomy (feelings of having choice), and relatedness (feelings of connection to one's peers). The constructs of competence, autonomy, and relatedness aligned with our earlier finding of play, agency, and social status being salient factors in the physics classroom (Van Dusen & Otero, 2012). Although Self-Determination Theory is rarely associated with a sociocultural perspective, like the work of Ames (1992), we interpret it through a sociocultural lens. Below we discuss how Self-Determination Theory can be viewed through a sociocultural lens and what implications this perspective has for understanding the classroom environments' role in shaping students' shifting identities and motivations.

## Literature Review

### *Self-Determination Theory*

Self-Determination Theory considers identity and associated motivations as properties of an individual in relation to social environments that shift over time and contexts (Deci & Eghrari, 1994; Ryan & Deci, 2000; Ryan, 1995). According to this perspective, as an individual engages within a particular context, she will come to see herself in various ways with respect to this context. As this changes, there are chances that the individual will become integrated (or alienated) from the activity/context and from the members of the community represented by the activity. The process of integration can shift one's engagement in an activity from being externally motivated to being internally motivated. We extend this perspective to say that, as one becomes increasingly interested in engaging, the goals and practices of the activity become difficult to distinguish from the goals and practices of the individual. As the difference between the goals of the activity and the goals of the individual become increasingly similar, the individual is said to "internalize" or identify with the activity. The model proposes a mechanism for identity development through the integration of goals and practices.

According to Deci and Ryan (Ryan & Deci, 2000), internalization can be thought of as the assimilation of behaviors that were once external to the self. Through the process of internalization, individuals come to feel that what makes them engage in an activity (their locus of causality) moves from the external to the internal. Based on the study of organismic integration, Self-Determination Theory states that internalization occurs as an activity fulfills an individual's *basic psychological needs* (Deci & Ryan, 1991; Ryan & Deci, 2000). Organismic integration states that, like all natural processes, development through integration must be nurtured by the fulfillment of basic needs. These basic needs are defined as the, "conditions that are essential to an entity's growth and integrity" (Ryan, 1995, p. 410). In the case of the human psyche, the basic conditions for growth (or basic psychological needs) are: a sense of *competence, autonomy,* and *relatedness* (Reis, Sheldon, Gable, Roscoe, & Ryan, 2000; Ryan, 1995). When people engage in activities which provide them the experiences of competence, autonomy, and relatedness (as contrasted with excessive controls, overwhelming challenges, and relational insecurity) they will be more likely to identify with and choose to engage in the same activities in the future.

Self-Determination Theory identifies six successive stages of internalization (*amotivation, external regulation, introjection, identification, integration,* and *intrinsic motivation*), shown in Figure 1. Each column in figure 1 represents one of the six stages of internalization, while the rows represent the corresponding *regulatory processes*, *mental and emotional processes*, *perceived locus of causality*, and *relative autonomy* for each stage (the minus and plus symbols in the figure show levels of absence and presence, respectively, of autonomy). A person's development of intrinsic motivation is developed in conjunction with the internalization of an activity.

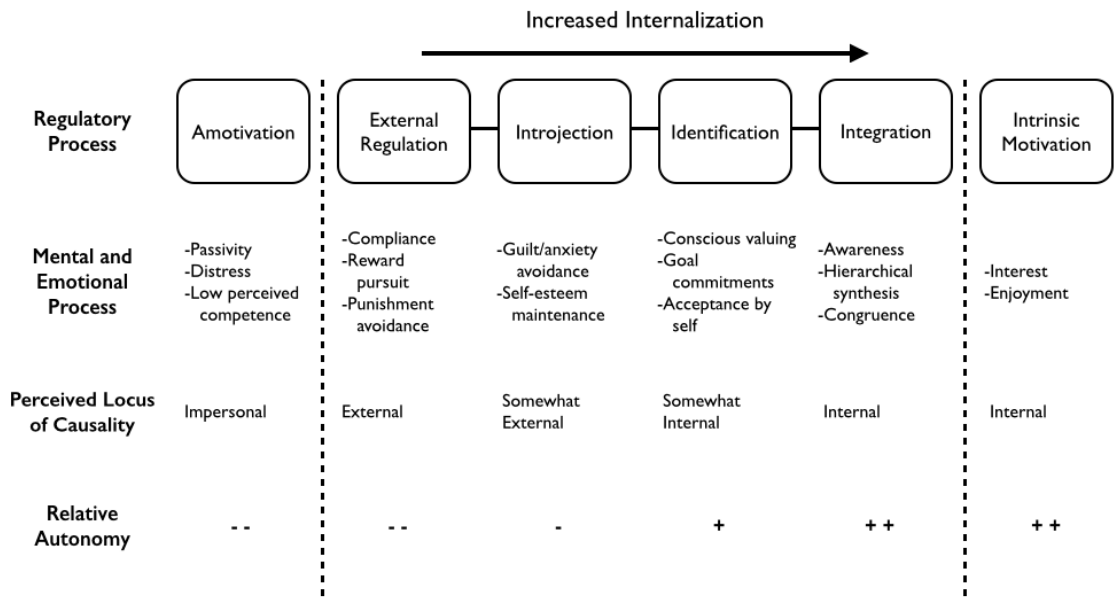

**Figure 1.** Internalizing an activity and developing intrinsic motivation (Ryan, 1995).

The extremes of the regulatory process scale in figure 1 (row 1 and 6) are *Amotivation* and *Intrinsic Motivation*. Amotivation actions are those that are seen as out of one's control. Intrinsic motivation, on the other hand, is the driver of actions that are done for their pure enjoyment and are perceived by the individual as being within one's control (Deci & Eghrari, 1994). Between the extremes of the regulatory scale, Self-Determination Theory identifies four regulatory process and their associated mental and emotional mechanisms that drive peoples' actions (Deci & Eghrari, 1994; Ryan & Deci, 2000; Ryan, 1995). *External Regulation* is the first stage of internalization and is represented in the second column of Figure 1. Activities that undergo External Regulation are done out of compliance and hold little to no personal meaning to the individual. These types of activities require external coercion or reward. *Introjection* is the second stage of internalization and is represented in the third column. While Introjection is not entirely based on external motivation, it is driven by a sense of guilt or anxiety avoidance about potential judgment by others. *Identification* is the third stage of internalization and is represented in the fourth column. When in the Identification stage, motivation to engage in an activity is based on a feeling of acceptance and personal valuing of the activity. While this regulation process is perceived as being internally motivated and autonomously driven, it lacks integration with other parts of the self. The Identification stage can be thought of as the "trying on" stage, in which a person values an activity but has not yet fully embraced it. *Integration* is the fourth stage of internalization and is represented in the fifth column. In the Integration stage of regulatory processes, various identifications are organized and brought into congruence with one's identity as a whole. Activities that have been assimilated to the Integration stage are those that people see as being central to their identities and are intrinsically motivated to engage in (i.e. the goals of the activity and the goals of the individual are essentially the same).

Figure 1 also shows the proposed *mental and emotional processes*, *perceived locus of causality*, and *relative autonomy*. The *mental and emotional processes* (row 2) are proposed psychological mechanisms that drive people to engage in activities. The *perceived locus of causalities* (row 3) determines whether a subject's engagement in an activity is experienced as being driven externally or internally. The *relative autonomy* (row 4) is the extent to which people feel that they have control over whether or not they engage in activities.

The inherent sociocultural nature of Self-Determination Theory's model for internalization becomes particularly apparent when viewed in light of Vygotsky's model of internalization. Vygotsky viewed learning as the internalization of social practices through imitation of a more experienced other (Vygotsky, 1978; Vygotsky, 1981). People engage in activities that are external to them long before they have internalized and understand the activities. For example, a child will learn how to brush his teeth by imitating his parents. Initially, the child does not know how to brush his teeth and has no understanding of why he should brush his teeth, but participates in order to make his parents happy and avoid punishment. Over time the child begins to brush his teeth without his parents prompting or knowledge. At this point, the child has internalized the social practice of brushing teeth, can perform the action on his own, and does so out of self-expression. Self-Determination Theory's model of internalization builds on Vygotsky's model (intentionally or not) by fleshing out the features of different stages of internalization and by proposing a set of social experiences (competence, autonomy, and relatedness) that are required for internalization to occur. Like Vygotsky, Deci and Ryan are interested in how individuals are shaped by their social environments.

*A Sociocultural Interpretation of Self-Determination Theory*

Integrating the work of Ames (1992), Ross (2013), and Ross & Otero (2012) with Self-Determination Theory (Ryan & Deci, 2000) leads to the "integration spectrum" shown in figure 2. We stress that the term motivation is not being used here with a strictly traditional attributions to internal characteristics of an individual. Instead, both "motivation" and "integrity/identity" are used in a sociocultural way—emphasizing the role of community and context in determining the extent to which one is integrated with it (or has integrity). "Integrity" in figure 2, is thus defined as the extent to which the goals and practices of an activity or community have become subjective, rather than objective, to the individual. That is, the individual can barely tell the difference between her own goals and the goals of a community. Figure 2 helps us to better understand how one's social environment acts to shape the emergence and expression of identities and their associated motivations.

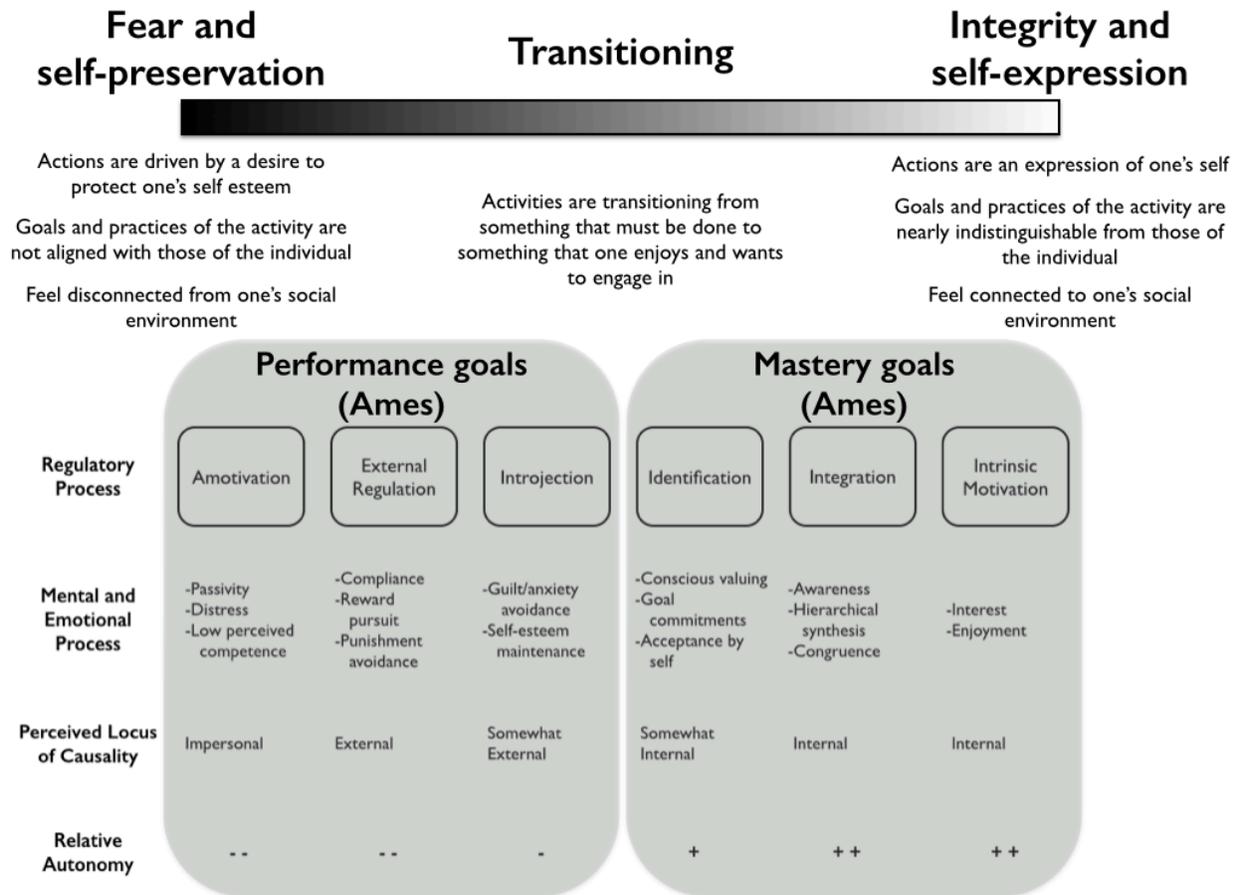

**Figure 2.** Integration Spectrum superimposed on classic Self Determination Theory.

As shown on the left side of the figure, when people do not feel connected to (or integrated with) their social environments (peers and community practices) the individual's goals and practices are not in alignment with those of the activity and they may act out of fear and self-preservation. As Figure 2 illustrates, the regulatory process that are on left end of the spectrum (Amovation, External Regulation, and Introjection) are driven by mental and emotional processes such as distress, compliance, punishment avoidance, anxiety avoidance, and self-esteem maintenance—in which the individual lacks control. We interpret this lack of control as characterized by what Ames (1992) refers to as *performance goals*.

> Central to a performance goal is a focus on one's ability and sense of self-worth (e.g., Covington, 1984; Dweck, 1986; Nicholls, l984b), and ability is evidenced by doing better than others, by surpassing normative-based standards, or by achieving success with little effort (Ames, l984b; Covington, 1984). Especially important to a performance orientation is public recognition that one has done better than others or performed in a superior manner (Covington & Beery, 1976; Meece et al., 1988). (Ames, 1992, p. 262)

The quotation above expresses that an individual is motivated by performance goals when her actions are geared toward appearing competent to her peers. In such a case, a person's senses of competence and autonomy are *socially-normed*, or normed against her interpretation of the practices of a broader, objective community.

Similarly, we interpret Ames's (1992) *mastery goals* as existing on the right end of the integration spectrum. When a person feels connected to, integrated with, and in control of her social environment her actions are driven by integrity. In such a case, the individual is authentically engaged in an activity such that it becomes subjective and difficult to distinguish from her self. As Figure 2 illustrates, the regulatory processes that are on the integrity end of the spectrum (Identification, Integration, and Intrinsic Motivation) are driven by mental and emotional processes that are characteristic of having control of one's environment, such as consciously valuing, congruence, interest, and enjoyment. Ames (1992) describes integrity as being driven by *mastery goals*.

> Central to a mastery goal is a belief that effort and outcome covary, and it is this attributional belief pattern that maintains achievement-directed behavior over time (Weiner, 1979, 1986)... The focus of attention is on the intrinsic value of learning (Butler, 1987; Meece & Holt, 1990; Nicholls, l984b), as well as effort utilization. One's sense of efficacy is based on the belief that effort will lead to success or a sense of mastery (see Ames, l 992a, Ames & Archer, 1988). With a mastery goal, individuals are oriented toward developing new skills, trying to understand their work, improving their level of competence, or achieving a sense of mastery based on self-referenced standards (Ames, l992b; Brophy, l983b; Meece, Blumenfeld, & Hoyle, 1988; Nicholls, 1989). (Ames, 1992, p. 262)

The quotation above highlights how personal goals have become integrated within a broader system of interactions. Terms such as "developing skills," "trying to understand," "improving competence" are used to illustrate that the purpose (or motivation) for engaging in the activity are indistinguishable from the activity itself. In such a case, a person's sense of competence is *self-normed,* as distinguished from *socially-normed*. The activity is subjective and integrated with the actor and surrounding community—this person has integrity and her identity is tied up in the activity.

We have defined *fear and self-preservation* and *integrity and self-expression* as representing the extremes of the integration spectrum, however we assume that individuals' motivations in various contexts typically fall somewhere within the spectrum, and evolve through engagement in activities. In some cases, such as those on the far left of the spectrum, the individual drops out completely or maintains the same level of detachment forever. In many cases however, identities reflexively evolve as an individual interacts with, and modifies behaviors as a result of feedback from the community. Figure 2 illustrates that there is a point at which an individual begins to engage in an activity for internal rather than external reasons. This transition in the valuing of an activity coincides with an individual beginning to identify with the activity's potential for positive affective or intellectual experiences.

Our theoretical model highlights the critical nature of sociocultural factors, in particular how one relates to peers, social norms, and cultural practices in the emergence of identity. For example, when an individual expresses a physics identity, she cannot tell the difference between her own goals and practices and those of physics—she is integrated with physics. As mentioned earlier, this has been the goal of physics classrooms since the early 1900s. Charles Mann (1914) expresses this *integrity* (or integrated-ness) in the quotation below.

> The essence of the scientific spirit is not, as has been generally supposed, a method of thinking. It is not the intellectual process that has been divided into the steps called observation, induction, hypothesis, and verification. This process, if it signifies anything real, is at best but one of the modes in which the presence of the scientific spirit inside is made manifest. Many of us have consciously tried, and as consciously failed, to impose this order of thought on our pupils with the idea that we were thereby serving science. We have failed because the essence of the spirit we want is not of this sort. The essence of the scientific spirit is an emotional state, an attitude toward life and nature, a great instinctive and intuitive faith. It is because scientists believe in their hearts that the world is a harmonious and well-coordinated organism, and that it is possible for them to find harmony and coordination, if only they work hard enough and honestly enough and patiently enough, that they achieve their truly great results. It is this faith inside them that inspires them to toil on year after year on one problem (Mann, 1914, p. 518).

In the quotation above, Mann calls for the building of learning environments in which students can embody, the goals of physics inquiry. He makes clear, that statements of scientific practices, such as observation, induction, hypothesis, and verification lead to a distraction from the central goal of student identity development and integration with physics.

Integration with social environments can be highly contextual. For example, a student may demonstrate a strong sense of integrity when doing physics in her normal classroom environment but revert to a state of fear and self-preservation when moved to a physics class in a new school. As the examples that we analyze later in this paper will show, there can be great variability in the levels of student integration between activities in a single class.

### *A Sociocultural Interpretation of Basic Psychological Needs expressed in SDT*

As described earlier, Self-Determination Theory is founded on the idea of *Basic Psychological Needs,* which are made up of the constructs *competence, autonomy,* and *relatedness*. Based on the findings from our exploratory factor analysis reported in our third manuscript, together with our sociocultural interpretation of self-determination theory presented above, we concluded that the construct of *relatedness* determines an individual's levels of social *integration* and determines how the constructs *competence* and *autonomy* are expressed in context. Figure 3 provides a toy model of how relatedness determines the ways in which competence and autonomy are expressed. This toy model was derived from the findings reported in our second manuscript, which is consistent with figure 2, representing our reworking of Ryan's (1995) model (figure 1).

Figure 3, shows how social relatedness (or levels of social integration) lead to different outcomes relevant to the basic psychological needs outlined by Deci & Ryan (1991). The constructs *competence* and *autonomy* will be expressed differently depending upon an individual's level of social integration. A context in which an individual feels highly integrated with their social environment (right side of figure) will lead to autonomy being experienced as self-expression/innovation, and competence being experienced as efforts for improvement of the self or environment. A context in which an individual feels like an outsider (left side of figure) will lead to autonomy being experienced as aloneness and alienation, and competence being experienced as efforts to preserve self-esteem. For example, students who are friends with their classmates and feel connections to the practices of their physics classrooms are willing to take risks and try out new ideas (autonomy) with the hope of better understanding physics (competence). Students who do not feel socially connected to other students, the teacher, or the goals and practices of physics are often afraid to express their ideas (competence), unlikely to take risks in problem solving, and often criticize the teacher and the content in efforts of not looking like the ones who are responsible for not fitting in (autonomy). Figure 3 illustrates that both competence and autonomy emerge in both types of settings, however ones integration within the social environment frames how these constructs are experienced and expressed. In order to have nurturing positive feelings of competence and autonomy emerge from a system

students must feel some connections to their social environment (peers and the cultural practices within the system).

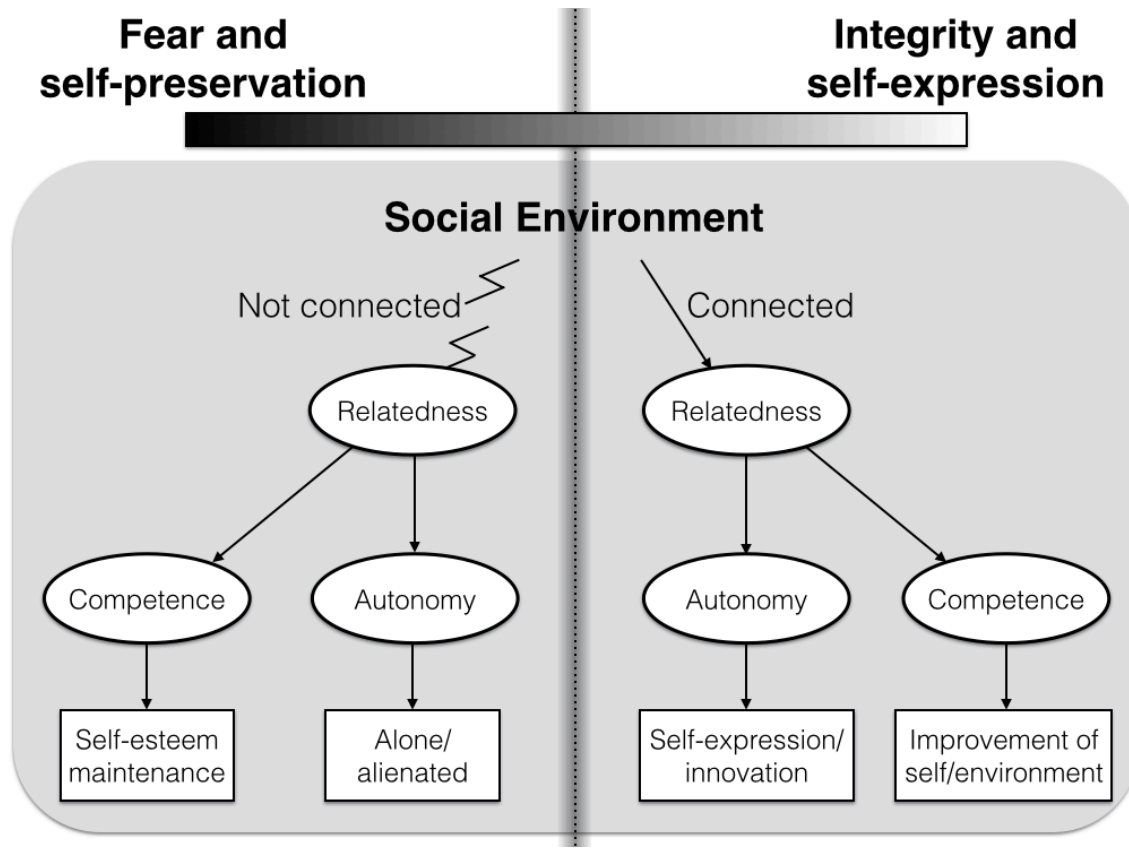

**Figure 3.** A toy model of the how "basic psychological needs" are expressed in context.

Differing perspectives on Self-Determination Theory allow for the construction of theoretical arguments for why, or why not, competence, autonomy, and relatedness should be distinguishable constructs. Ultimately, however, the constructs' distinguishability is an empirically measurable feature. In our third manuscript we examined the nature and distinguishability of each of these constructs using a traditional Self-Determination Theory style survey and exploratory factor analysis. Through factor analysis of the survey, we showed that while relatedness was a primarily independent construct competence and autonomy were not distinguishable from each other. The competence and autonomy questions loaded onto one of two emergent constructs. The two emergent constructs were primarily composed of either the competence and autonomy questions that were framed in a positive social setting or a negative social setting. For example, a positive and a negatively framed question about competence from the survey were, "people told me I was good at what I was doing," and, "I did not get a chance to show how capable I am." This finding is consistent with the sociocultural interpretation of Self-Determination Theory in which our experiences of competence and autonomy are ultimately driven by (and therefore are inseparable from) our experiences of social belonging and integration. Within this perspective, one should not be surprised that the constructs of *fear and self-preservation* and *integrity and self-expression*, rather than the constructs of competence and autonomy, emerged from the exploratory factor analysis.

It is important to note that our interpretation of these constructs is not in conflict with Ryan and Deci's (2000). While these founders of Self-Determination Theory do not go into detail about how relatedness (integration) drives experiences of competence and autonomy, they do give relatedness a special importance in facilitating the internalization process (or the process of developing intrinsic goals).

> Because extrinsically motivated behaviors are not typically interesting, the primary reason people initially perform such actions is because the behaviors are prompted, modeled, or valued by significant others to whom they feel (or want to feel) attached or related. This suggests that relatedness, the need to feel belongingness and connectedness with others, is centrally important for internalization. (Ryan & Deci, 2000, p. 73)

In this quotation Ryan and Deci (2000) state that the desire to experience integration (relatedness) is what ultimately drives the process internalization. They go on to discuss how experiences of competence and autonomy are dependent on first feeling connected to one's social environment.

The relative internalization of extrinsically motivated activities is also a function of perceived competence. People are more likely to adopt activities that relevant social groups value when they feel efficacious with respect to those activities... contexts can yield autonomous regulation only if they are autonomy supportive, thus allowing the person to feel competent, related, and autonomous. (Ryan & Deci, 2000, p. 73)

The first half of this quotation expresses that while competence is required for the internalization of activities, the internalization is facilitated by the activities relevance within a social group that one values. In the second half of this quotation, autonomy is said to be only experienced in settings that facilitate all three basic psychological needs. In our terms, autonomy can only be experienced in social environments that facilitate social integration.

## Classroom Applications

Our sociocultural interpretation of Self-Determination Theory highlights the importance that students' senses of social belonging and integration play in shaping their experiences and motivations in classroom environments. Our model also suggests that contexts can be changed to facilitate students' feelings of belonging. In this section we apply our proposed model to better understand students' actions in three classroom examples from an AP physics classes we documented in a previous study (Van Dusen & Otero, in review a).

*Example #1*

The first example took place an AP physics class in which students were utilizing screencasts to create and share solutions with their fellow students (Van Dusen & Otero, in review a). Three months prior to this event, the teacher had engaged the class in the creation of a rubric that the students could use to guide their creation of screencasts. The students created three categories that the rubric would evaluate (content, organization, and presentation & aesthetics) and determined what features a "proficient" or "advanced" screencast would have in each of these categories. While the teacher did not use the rubric to evaluate students' screencasts, the students used it to provide each other feedback. In this particular class session, the teacher asked the students how the screencasting process was going and if there were things that could be improved. One student, who we refer to as Juan, raised his hand and shared that he did not think that the screencasting rubric was fair. Juan took particular issue with the Presentation & Aesthetics category of the rubric that stated that to be classified as "advanced" a student had to incorporate humor into their screencast. Juan did not feel that the requirement was fair because he was, "not an entertainer," and was not comfortable, "including jokes or using funny voices," in his screencasts. Initially, some students felt that the rubric should not be changed because the inclusion of humor made the screencasts more enjoyable to watch. After some discussion, the students decided that the rubric should be altered such that in order to be classified as Advanced in the Presentation & Aesthetics category screencasts did not have to be "humorous" but that they did have to be "engaging."

In this example, we claim that Juan is internally motivated to exercise his agency by expressing disagreement with a previous class consensus. He was able to express aspects of his identity (not an entertainer) with the rest of the class, and to admit openly that he did not feel that he possessed something that the class openly valued (humor in screencasts). We interpret this as Juan being *integrated* with his class environment—he felt that he was a functioning part of the classroom system and acted out of a belief that his thoughts not only mattered, but that he would not be chastised for disagreeing with the entire community. Juan's belief that his opinion would be heard and honored is evidence of integration, or integrity, and is associated with risk-taking in a space that is open for innovation. This example brings to light how our model uses the term "integrity" to mean more than "wholeness," "completeness," "honesty," and "pureness." We also use the term in the engineering sense—"structural integrity" to maintain wholeness within context. We associate the term "integrity" with "integration"—a sense that one is a critical part of the context. In this example, Juan did not appear to see the classroom rules or participants as separate from his self. Instead, he experiences a sense of social belonging, his personal goals were in alignment with those of the class and through the act of improving the situation for himself, he also felt that he was improving the conditions of the broader classroom community. Juan experienced competence through improvement of his environment and experienced autonomy through self-expression and innovation.

While Juan appeared to feel comfortable and connected with his fellow classmates in this scenario, he may not have felt as connected when attempting to make "humorous" screencasts. Instead, we claim, he experienced autonomy as isolation and competence as performing well enough to protect his self-esteem. By proactively expressing his feelings in an environment in which he was integrated, Juan was able to convince his peers to explicitly alter the classroom goals and practices to value screencasts that were engaging in any form. The class's responsiveness to Juan's plea is indicative of a context in which many of the students felt connected to and identified with their social environment. The students meaningfully engaged in the innovative process of creating new classroom norms in order to improve the learning environment such that their fellow classmates would not have to engage out of fear and self-preservation.

While there are many facets of a learning environment, we claim that allowing students to develop their own screencasting rubric was central to creating a classroom context that fostered students' feelings of belonging and self expression. In creating their rubric, the students integrated practices that were relevant to them, such as "presenter incorporates humor, jokes, puns, accents," with practices that were important to learning physics, such as "clearly shows all work, equations, variables, and units." By creating classroom norms through the development of the screencasting rubric, students may have begun to feel a sense of ownership and connection to the classroom environment more broadly.

*Example #2*

The second example is a set of twenty physics problem solutions that another AP physics student (who we refer to as Lih-Hann) created in his notebook. The teacher did not formally grade the notebook solutions. After completing the notebook solutions at home, each group was randomly assigned a problem to share with the rest of the class using whiteboards. Lih-Hann's twenty solutions were completed in approximately half of a page in his notebook. The solutions had only 49% of the steps that the teacher had asked students to show. The final answers were correct on 58% of the problems. The percentages complete and correct are in line with the rest of the class's performance on the problem set (Van Dusen & Otero, in review a).

Because these problems were completed at home and were not generally shared with classmates there were not many opportunities for the activity to foster students' senses of social connection. Lih-Hann's notebook solutions show minimal effort. We claim that this is because Lih-Hann felt little connection to his social environment or integration with the activities goals and only a small amount of effort was required to maintain his self-esteem. Lih-Hann's actions are indicative of working towards a performance goal (finishing the assignment as quickly as possible).

We claim that the context in which students created their notebook problems was isolating for some students because it allowed minimal opportunities for social interaction or engagement in peer cultural practices. The students created their solutions separately from their peers, the solutions were usually only seen by the students who created them and the teacher, and the expectations for their final product were based off of the AP grading guidelines. While these features may be common to many physics classes, they are not particularly conducive to students feeling a sense of connection to their classroom environment. However, as the next example will illustrate, a small change the context (e.g. having students present some homework solutions to the class through screencasts) provided rich opportunities for students to build and maintain connection to a community.

*Example #3*

The third example is Lih-Hann's creation of a screencasting solution from the kinematics unit. The screencasting assignments were not formally graded by the teacher but were uploaded to the course website for fellow students to view. There are several features of Lih-Hann's screencast that indicate that he was engaging more out of integrity than out of fear. When working out his 4min 11sec long solution Lih-Hann used his normal voice but when reading the question he used a humorous high-pitch voice. Once Lih-Hann reaches his final answer he shares that it does not match the answer that the course website gives. Lih-Hann spends over a minute discussing the differences between the two answers and how they should be interpreted.

> It equals, 6,078N. But I actually checked on [the course website] to see what answer we're supposed to get and it says $6.8 \times 10^3$N. So I don't know about *that*. You just saw what I did here, so it's possible I messed something up. But all I know is that these are the numbers I got, I followed the equations, and this is the answer that I came up with here: 6,078N. But according to this, it's supposed to be $6.8 \times 10^3$N. So if you see a mistake in this process go ahead and change it to try to get this answer here, *I guess*. Like I said, I don't see why this is the answer. But try to get $6.8 \times 10^3$N using this method correcting wherever I made my mistake. And if my answer is the right answer, that's how you get it. But it's not though. So aim for this, but hopefully this process will help everyone. (Lih-Hann, 3/7/13)

In Lih-Hann's quotation, he expends significant effort trying to determine whether his answer is correct and how others might use his solution. It appears that Lih-Hann does not feel intimidated by the fact that he has the "wrong answer." While Lih-Hann may have never found out, in our investigation we discovered that the course website had a typo and Lih-Hann's answer was, in fact, correct.

Lih-Hann's ability to express is disagreement with the physics textbook solution is evidence of integrity and self-expression. Like Juan, Lih-Hann was not acting out of fear of rejection or self-preservation. He put himself out on a semi-public website in open disagreement with the textbook authors, in front of the whole class. Instead of choosing another problem to solve for the class, Lih-Hann invited the class into his confusion about the difference between his answer and the answer in the text. Lih-Hann's integrity, or integration with his social and physical environment can be seen in the way that he directly addresses his audience as if he was in an active conversation with them. Lih-Hann's use of humor in the tone of his voice is indicative of him experiencing autonomy through innovation and self-expression. The significant amount of time and effort that Lih-Hann put into creating and sharing his solution, even though it was in disagreement with the textbook, is indicative of experiencing competence as attempts to improvement his environment, or himself, by asking for help from his classmates.

In the three examples above, we illustrate how two of the students in the class expressed integrity and self-expression in physics-centric activity. In this classroom, (a) the teacher required that each student "publicly" share their homework solutions as screencasts on the class website, (b) these screencasts provided many opportunities for creativity, self-expression, and innovation, (c) the teacher required that the students assess one another on the basis of a class-generated rubric, (d) the rubric was established (with the teacher's guidance) and agreed upon by the students in the class. These factors may have all contributed to increased connections and social belonging among the students and their physics learning environment. This environment was shown throughout Van Dusen & Otero (in review a) to improve motivation through increased performance and self-reported connection to the class, the other students, and the physics.

The four points listed above, along with the three examples, suggest that *learning environments* or *learning contexts* can be modified in order to promote student motivation. Educators have classically thought of motivation as an intrinsic property of the individual, something that a teacher could do nothing about. Our research along with our model for motivation and identity demonstrates that although the individual is a critical player in motivation, so too is the context that is built and maintained in part by the teacher. Student motivation can be altered by setting up a context in which competence and autonomy are more likely to be *expressed* as improvement of the self/environment and self-expression/innovation rather than as self-esteem maintenance and alienation. Our proposed model can be used to interpret how the student experiences activities within learning environments. This is encouraging because it sets the stage for teachers having input and control on factors that can lead students to engage or disengage.

## Implications for Classroom Contexts

Our empirical and theoretical work in understanding the role integration of goals and practices have on students' classroom experiences suggests design principles for creating engaging and motivating physics learning environments. These principles are listed below.

**Students' cultural practices.** Facilitate integration between the students' social selves and the content of the course. Students should not be forced to abandon their peer cultural practices at the door of the classroom. If classroom practices and goals do not in some way align with those of the students' views of themselves, then they will lack senses of connection to their classroom activities. One way to facilitate the integration of students' peer cultural practices into the classroom is through the incorporation of *boundary objects*, tools that hold meaning within multiple sets of cultural practices (Buxton et al., 2005; Star & Griesemer, 1989). Boundary objects can act as bridges between distinct sets of cultural practices, creating a common space for them to blend. By incorporating tools in the physics class that a priori hold meaning in the students' peer communities, the students peer cultural practices can be meaningfully merged with the cultural practices of the physics classroom (Van Dusen & Otero, 2014). Another technique for incorporating students' peer cultural practices into the classroom to create space for students' peer languages (a specific type of tool) into the classroom discourse. By creating a common space for different discourses to exist and blend, a *third space* can emerge (Gutierrez et al., 1999; Gutiérrez, 2008; Gutierrez, 1997). By facilitating the interaction between the class's "script" and the students' "counterscripts" hybrid spaces, or third spaces, can be formed that value both systems of engagement and allow for unique practices to emerge.

**Physics cultural practices.** Facilitate the integration between the physics and the students' social selves. Cultural practices that are central to physics should not be marginalized in an attempt to appeal to students. The importance of embedding science learning in the practice of science are evident in the last several incarnation of a national science education standards (AAAS, 1989; NRC, 2011, 2013). The Next Generation Science Standards (NRC, 2013) goes as far as to create a three dimensional model for learning physics in which scientific knowledge is embedded in the practices of science such that science can only be understood through the practices themselves. The idea of embedding science learning in the physics community's cultural practices is not new. As early as 1925, Robert Millikan, a physics Nobel Lauriat and central member of the "new movement among physics teachers" (Mann, C.R., Smith, & Adams, 1906; National Education Association, 1920), stated that one can only embody science, "by getting in close touch with science itself; by *absorbing* is method and its spirit through *actual contact* with it and through *practice* in using it (Millikan, 1925, p. 972)." As Millikan's quotation emphasizes, it is only by engaging authentically in the practices of physics that one can come to embody the spirit of physics.

**Students are diverse.** Make a concerted effort to get to know your students. Be aware that while you may not ever understand the peer cultural practices of some of your students, students will appreciate your efforts in getting to know them. Having advocates who can help socialize students from non-dominant background into the practices of science has been shown to be one of the most important factors fostering students' persistence in science education (Aschbacher et al., 2009; Lee & Buxton, 2010). It is critical that physics teacher reach out to their students and get to know them as individuals.

**Ask students for help.** Explicitly engage your students in designing the classroom norms. Classroom environments must be flexible and dynamic enough that students have the opportunity to integrate their diverse and ever shifting peer cultural practices into their activities. By creating space for students to imprint their own interests onto activities, such as through the creation of rubrics, a teacher does not have to a priori know every student's needs. By creating an environment in which students feel a sense of ownership over their learning environment they will have increased opportunities to engage out of self-expression and experience autonomy as innovation (Nicholson-Dykstra et al., 2013). These sort of learner-centered environments have been shown to be effective in improving motivation and a sense of participant ownership a wide range of settings (Lee & Buxton, 2010; National Research Council, 2000; Ross, Van Dusen, &

Otero, 2014; Ross, Van Dusen, Sherman, & Otero, 2011; Van Dusen & Otero, 2011; Van Dusen, Ross, & Otero, 2012).

**Efforts over outcomes.** Promote student effort and growth rather than a fixed goal state. Rigor, students can handle, but fear of failure is what leads them to feeling isolated and ultimately rejecting physics. To alleviate students fear of failure, provide them feedback using language that reflects mastery rather than performance goals (Ames, 1992). For example, instead of grading something as simply incorrect, let the student know that their understanding is "not yet" where it needs to be. Encourage students to continually strive to master the practices of physics while listening to, and be true to, themselves. By fostering what Dweck refers to as a growth mindset over a fixed mindset (C. Dweck, 2010), students will be more likely to take risks, engage in productive failures (Kapur & Bielaczyc, 2012; Kapur, 2008), and form lasting physics educational trajectories.

**Students and systems.** Students' disinterest in physics classes may have little to do with their lack of abilities or their interests in physics. Confrontational or sullen students are often acting out of fear and self-preservation due to an all too common lack of connection to their social environments (Aschbacher et al., 2009; Garn, Matthews, & Jolly, 2010; National Governers Assocation, 2007; Yerrick & Roth, 2005). Rather than assume that these students are destined for failure, or that you can help them "get by," assume that you can make some environmental modifications that can be more inclusive. Try to consider what it is about the environment that is making them feel isolated from the activities.

**Traditional physics classes are preferentially selective.** Be aware that physics has traditionally served the population of students whose parents are already part of the scientific or academic community (Banilower et al., 2013; National Science Foundation, 2010, 2012). These students have been predisposed to language and practices that lead to success in traditional physics learning environments (Lemke, 1990). Many of the students from groups that are traditionally underrepresented in physics do not have examples that are consistent with the academic process at home (Rosebery, Warren, & Conant, 1992). However, they do have many home practices and everyday experiences that can be built upon (Basu & Barton, 2007; Rosebery, Ogonowski, DiSchino, & Warren, 2010; Suarez & Otero, 2013). It is critical that teachers do not assume that all students are the same and that all students are predisposed to the norms and practices of traditional physics classrooms. Like the students, teachers must feel that the classroom is a space for innovation, and should therefore draw on their creativity and good sense to do what is right for the students. This may not always look like traditional physics.

**Students like to feel good about themselves.** Always remember that you not only teach physics, more importantly, you teach *students*. These people want to learn, they want to feel good about themselves, they want to have fun, and they want you to like them. Teachers can use physics as a tool to help students learn, to help students feel good about themselves, and to help students perform well. By creating environments in which students can engage in scientific play (Rieber, 1996; L. Vygotsky, 1978) students can safely explore their environments without experiencing a significant fear of failure (Podolefsky, Rehn, & Perkins, 2012). Further, by creating environments in which students feel that the challenges are high, but that they are able to meet the challenges, leads students to enjoy participating, stretches their abilities, and increases their self-esteem (Csikszentmihalyi & LeFevre, 1989; Csikszentmihalyi, 1990). In physics classes, these types of optimal challenges can be created when students are asked to support their claims with evidence. Having one's idea taken up by a whole class is exciting. By providing a space in which students can engage in structured innovation, teachers can greatly increase student motivation.

While there are no surefire tricks to creating physics learning environments that will engage all students, but general guidelines such as those listed above can help physics teachers increase their students' uptake and embodiment of the goals and practices of the physics community. Through this process students will become more motivated to engage in physics and begin to form lasting physics identities.

## Conclusion

Our work highlights the importance of students' feelings of belonging in their physics learning environments. It is through this sense of belonging and acceptance that students will feel safe enough to engage in the creative and exciting process of learning physics. It is no wonder that when students go into a physics classroom in which the teacher talks to their students using Greek (literally) and only wants to hear correct answers from them that students retreat into a place of simply trying to look and sound smart so that they can protect their self-esteem. The theoretical perspective on identity and motivation presented in this paper, if valid, is encouraging. It implies that physics teachers can influence students' motivations. Teachers may not be in control of their students, but they are in control of their classrooms. By making seemingly small changes in classroom learning environments, teachers can radically alter their students' interactions with physics such that the students feel motivated and build positive relationships to physics.

# CHAPTER 7

# CONCLUSION AND IMPLICATIONS

Our examinations of students' engagements and behaviors in two AP physics classrooms have yielded several findings with both practical and theoretical implications. The first manuscript details a pilot study that examined how physics students responded to the inclusion of iPads in their class. Ethnographic field notes and student surveys showed the emergence of several positive classroom trends associated with the iPads and the way they were used. The analysis reported in the first manuscript resulted in the following findings:

1. iPads facilitated students' use of evidence.
2. iPads facilitated student play.
3. iPads increased student agency.
4. iPads appeared to impact student social status.

From these findings we inferred that the iPad and its associated applications were tools that were personally meaningful to students and acted as a conduit to increase the students' social status, play, and agency. By integrating the iPad into physics activities, we hypothesized that over time, physics itself may become personally meaningful to the students, such that they would value physics even without the iPads. These findings led us to more closely examine the constructs that might lead to shifts in students' identification with, and motivation to engage in, physics.

The second manuscript focused on students' engagements and motivations in solving physics problems using two different sets of tools (screencasting on the iPad and traditional pen and paper notebooks). To analyze students' behaviors we drew on a motivation theory (Self-Determination Theory) that utilized a similar set of constructs to those that emerged from our pilot study. Students' behaviors in the activities, self-reports, and performances were examined and findings include:

1. Students' screencasts demonstrated social activity.
2. Students exhibited opportunities for autonomy in their unique styles of creating screencasts.
3. Students' self-reported higher levels of social engagement, authorship, affect, and effort when creating solutions on screencasts versus pen and paper notebooks.
4. Students' solutions were, on average, more complete when created using screencasts rather than pen and paper notebooks.
5. Students' solutions were, on average, more correct when created using screencasts versus pen and paper notebooks.

From these findings we inferred that screencasts were acting as boundary objects that facilitated the blending of students' peer cultural practices with the physics classroom cultural practices. This process allowed students to feel more connected to classroom activities and to increase their motivations to engage in them. We claim that it was these increases in connections to the activities that led to the students improving their problem solving performances.

The findings from the second study led us to examine the specific mechanisms that might create shifts in students' identities and motivations. The use of Self-Determination Theory in our second study led us to investigate the role that the constructs of competence, autonomy, and relatedness played in improving students' outcomes. Students' surveys, class grades, and self-reported interests in physics were examined. The results from the analysis reported in our third manuscript indicate that:

1. A blend of competence, autonomy, and relatedness were predictive of students' final grades.

2. A blend of competence, autonomy, and relatedness were predictive of students' self-reported shifts in their interest in learning physics.

3. Most studies assume that competence, autonomy, and relatedness are distinguishable constructs that can be measured independently.

4. The survey items about relatedness emerged as a largely independent factor.

5. The survey items about competence and autonomy were blended into two emergent factors.

6. The two blended factors were primarily split between questions that were framed in positive or negative social settings.

From these findings we concluded that the constructs of competence, autonomy, and relatedness were important in shaping student motivations. Unlike most studies that draw on Self-Determination Theory we did not conclude, however, that the constructs of competence, autonomy, and relatedness could be disentangled. We concluded that the quality of students' social connections played a key role in shaping their experiences of competence, autonomy, and relatedness and that these constructs should be considered together as part of a more holistic picture of students' experiences in social environments. Theoretically, we determined that the constructs of competence, autonomy, and relatedness should not be thought of in terms of *whether* they exist in a context, but in terms of *how they are expressed* in various contexts.

Our findings about the interrelated nature of competence, autonomy, and relatedness led us to more fully explicate our interpretation of Self-Determination Theory in our fourth manuscript. Unlike traditional interpretations of Self-Determination Theory, we drew on the work of Ames (1992) , Ross (2013), and Ross & Otero (2012) to create a sociocultural model of motivation as it pertains to identity by highlighting the role that social belonging plays in shaping students' physics motivations and identities. The fourth manuscript provides a sociocultural interpretation of Self-Determination Theory and resulted in the following conjectures:

1. Identity and motivation are reflexively dependent on each other and emerge from students' interactions with their social contexts.
2. Feelings of social belonging in an environment are critical to the internalization of, or identification with, an activity.
3. When students' feel connections to their social environments they can be driven to act out of integrity and experience positive forms of competence and autonomy such as self-improvement and innovation.
4. When students' do not feel connections to their social environments they can be driven to act out of fear and experience negative forms of competence and autonomy such as self-esteem maintenance and alienation.

Our sociocultural perspective reframes students' motivations from being internal to the students to being emergent from the students' interactions with the classroom context. Seen in this way, teachers have significant opportunities to shape their students' motivations through the creation of their learning environments. By creating contexts in which students feel a sense of belonging they can begin to engage in physics out of integrity and self-expression rather than fear and self-preservation. We conclude our fourth manuscript with several recommendations that still need to be empirically tested. We discuss this briefly in the upcoming section on future work.

## Implications for Classroom Contexts

Our empirical and theoretical work on understanding students' integration of classroom goals and practices suggests a set of "design principles" that can help guide the creation of physics learning environments that could be more engaging and motivating for students. These principles, which are more fully explicated in our fourth manuscript, are all formulated to create environments that can simultaneously value the students, physics, and the interplay between them.

The first step in designing learning environments is recognizing that there is a need for contexts that allow students' peer cultural practices to blend with the practices of the physics classroom. In order to do this, the instructor must respect her students' peer cultural practices, even when these practices are are hard to understand. By actively engaging with students and getting to know them as people, teachers can begin to give them voice in the classroom and provide them the support they need to take chances and to be innovative. By reframing motivation as an emergent property of students' relationship to their environment, teachers can begin to move beyond the deficit model of "broken" students who are intrinsically unmotivated to engage in physics. It is the teachers' jobs to create environments where their students can feel safe and that their efforts are valued. It is only when students connect to their physics learning environments that they will begin to engage in classes out of integrity and to create lasting physics identities.

At the same time, teachers are working in complex district contexts themselves. It is important to note that the recommended teacher moves take place in a broader educational and political context, which often make it difficult for teachers to do what they know is right.

# Implications for Learning Theory

Much has been written about the apparent divide between theoretical perspectives that use the individual as the unit of analysis (cognitive) versus theoretical perspectives that use the context as the unit of analysis (situative). Many education and psychological researchers have either taken up the taken up the defense of one of the two general types of perspectives on learning (e.g. Anderson, Reder, & Simon, 1996; Greeno, 1997), while others have proposed a sort of détente in which both perspectives are allowed to co-exist despite inherent tensions between their underlying assumptions (e.g. Cobb, 2007; Greeno & van de Sande, 2007; Sfard, 1998). We propose that this tension is a manufactured divide that need not exist at all.

We see socioculturalism erasing this schism by viewing individuals as agents interacting in, and defining, learning contexts. This perspective acknowledges that integral to all human cognition are the very real electrical impulses that course through an individual's neural pathways, while simultaneously acknowledging that the development and activation of these neural networks are profoundly shaped by our engagement in current and past social environments. Within this perspective, human behaviors and cognition can only be understood by examining the dynamic interactions between an individual and their context.

Vygotsky, the progenitor of sociocultural theory, utilized a perspective that valued the importance of both the individual's mind and the social context in shaping learning. Whether writing about the role of more experienced others in creating optimal learning spaces (1978) or how students' socially developed academic concepts mediate their experience-based concepts (1962), Vygotsky explicitly addressed the role that social environments play in the process of learning and in individual cognition. At the same time, through writing about processes such as internalization of cultural practices (1981), expansion of the activities that an individual can accomplish unassisted, and the role of one's experiential (personal) knowledge in mediating the formation of academic (social) knowledge, Vygotsky also explicitly addressed the processes that are internal to an individual. We do not believe that it was Vygotsky's intention to ever have learning examined either as a purely individual or social phenomenon.

Cognitive Scientist, Andy Clark (1997) uses scientists' discovery of how a sponge breathes as an analogy to understand the importance of viewing human cognition in context.

> The simple sponge, which feeds by filtering water, exploits the structure of its natural physical environment to reduce the amount of actual pumping it must perform: It orients itself so as to make use of ambient currents to aid its feeding. The trick is an obvious one, yet not until quite recently did biologists recognize it. The reason for this is revealing: Biologists have tended to focus solely on the individual organism as the locus of adaptive structure. They have treated the organism as if it could be understood independent of its physical world. In this respect, biologists have resembled those cognitive scientists who have sought only inner-cause explanations of cognitive phenomena (Clark, 1997, p. 46).

In this quotation, Clark highlights the importance of examining cognition in context. Scientists could not fully understand a sponge's ability to breath in isolation from its environment, nor could they understand it from the environment alone. It is only by examining the interactions between the sponge and its environment that the scientists were able to discover how a sponge is able to effectively breath. Similarly, it is only by examining the reflexive interactions between a person and her social contexts that the processes of learning and cognition can be understood.

**Future Work**

The research reported in this dissertation leads to three different lines of future research. First, various recommendations or "design principles" were developed on the basis of the collective findings from the four studies reported. However, these design principles have not been empirically tested in terms of their actual effects on students. Therefore, studies are needed that test each design principle to determine the extent to which it truly impacts how students experience physics. Second, there is a need for careful qualitative investigations of the specific features of learning environments that lead to student expressions of competence and autonomy as self-esteem maintenance and alienation and which features tend to lead to lead to student expressions of competence and autonomy as self-expression/innovation and self/community improvement as proposed in the fourth manuscript (fig. 3). Such studies would focus on both students' talk about themselves and on ethnographic data of learning environments. It is likely that there will exist diversity among students, but that trends will be noticeable. Third, it is important to carefully catalog the types of student in-class and out-of-class behaviors that are associated with fear and self-preservation versus integrity and self-expression. Although these extremes are established in our model of motivation and identity reported in the fourth manuscript, they have not been empirically investigated. Finally, a next step in refining the sociocultural model of internalization and identity development proposed in my dissertation will be to increase the both the breadth and depth in which it is applied and to investigate the specific features. The data and the literature that were used in developing the model were largely focused on high school physics classes. Investigations will be designed in collaboration with teachers in a variety of settings to explicitly apply the model and test its robustness and generalizability. These investigations will also provide additional examples of effective practices that are grounded in existing classroom settings.

Specific learning contexts that will be useful in testing the generalizability of the model include in an elementary school science learning environment, a high school math class, a non-STEM high school class, a college physics class, and an informal science learning environment. In each of these environments, teachers will be identified that utilize innovative and effective techniques for fostering students' feelings of connection to their subject matters.

The design of each investigation will be tailored to the context in which it is set, but the general types of data collected will be static across the settings. In each setting, qualitative and quantitative evidence will be collected on how students express competence and autonomy in various contexts, and how this is connected to their sense of relatedness to various elements of the classroom context. Students will also be interviewed regarding their engagement and motivations and how they feel about themselves in relation to the class and topic. We hope to be able to make inferences about the alignment of students' goals with the goals of their learning contexts. Mixed methods analyses techniques will be used to examine the extent to which students' behaviors are consistent with the proposed model.

In addition to providing evidence for or against the robustness of the model, the investigation will also provide real-world examples of practices that foster students' sense of belonging or isolation. These examples will be used to refine the list of principles for designing effective learning environments, as well as offer teachers specific pedagogical tools and techniques that have been demonstrated to be effective.